\documentclass[prx,aps,letterpaper,twocolumn,allowfontchageintitle=true,unpublished]{quantumarticle}

\usepackage{bbm}

\usepackage{dcolumn}
\usepackage{bm}
\usepackage{graphicx}
\usepackage{color}
\usepackage{version}
\usepackage{mathtools}
\graphicspath{{figures/}}
\usepackage{natbib}
\usepackage{multicol}

\usepackage{float}
\usepackage{wasysym}
\usepackage{algorithm2e}

\SetAlgoCaptionSeparator{}
\SetAlCapSkip{0.5em}
\NoCaptionOfAlgo

\usepackage[table]{xcolor}
\usepackage{booktabs}

\usepackage{xcolor}
\usepackage{braket}

\begin{document}

\title{\centering \color{quantumviolet} Active volume: An architecture for efficient fault-tolerant quantum computers with limited non-local connections \newpage}
\author{Daniel Litinski}
\author{Naomi Nickerson \vspace{-2ex}}
\affiliation{PsiQuantum, Palo Alto}
\date{\vspace{-6ex}}
{\centering \maketitle}

\begin{abstract}

In existing general-purpose architectures for surface-code-based fault-tolerant quantum computers, the cost of a quantum computation is determined by the circuit volume, i.e., the number of qubits multiplied by the number of non-Clifford gates.
We introduce an architecture using non-2D-local connections in which the cost does not scale with the number of qubits, and instead only with the number of logical operations.
Each logical operation has an associated active volume, such that the cost of a quantum computation can be quantified as a sum of active volumes of all operations.
For quantum computations with thousands of logical qubits, the active volume can be orders of magnitude lower than the circuit volume.
Importantly, the architecture does not require all-to-all connectivity between $N$ logical qubits.
Instead, each logical qubit is connected to $\mathcal{O}(\log N)$ other sites.
As an example, we show that, using the same number of logical qubits, a 2048-bit factoring algorithm can be executed 44 times faster than on a general-purpose architecture without non-local connections.
With photonic qubits, long-range connections are available and we show how photonic components can be used to construct a fusion-based active-volume quantum computer.

\end{abstract}

\begin{figure}[t]
\centering
\includegraphics[width=\linewidth]{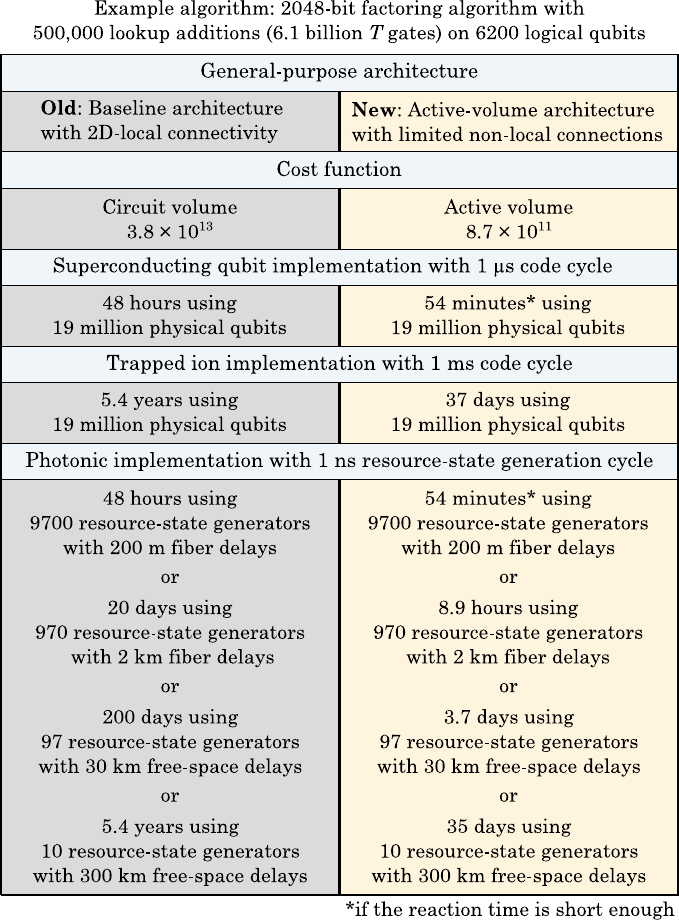}
\caption{Resource estimates for the 2048-bit factoring algorithm described in Ref.~\cite{Gidney2021} in a baseline architectures~\cite{Litinski2019,Fowler2018,Chamberland2022,Chamberland2022a,Bombin2021} and in the active-volume architecture described in this paper. More details are found in Appendix \ref{sec:resourceestimate}.}
\label{fig:examplealgorithm}
\end{figure}

How does one determine the cost of a quantum computation?
How does one quantify the performance of a quantum computer? 
These questions can only be answered with a specific quantum computer architecture in mind.
In the absence of error correction, it can be sensible to count the cost of algorithms in terms of qubits and two-qubit gates, and quantify the performance of quantum computers in terms of memory and two-qubit gate error rates. 
However, currently known commercially useful applications of quantum computers~\cite{vonBurg2020,Lee2020,Gidney2021,Campbell2021,Sanders2020,Chakrabarti2021} cannot be executed without error correction, so, e.g., CNOT counts of algorithms are irrelevant when quantifying the cost of useful quantum computations.
The introduction of error correction~\cite{Campbell2016,TerhalRMP} changes the relevant cost metrics of quantum algorithms.
With surface codes~\cite{Kitaev2003,Bravyi1998,Fowler2012,Fowler2018,Litinski2019}, non-Clifford gates such as $T$ gates or Toffoli gates require the preparation of magic states~\cite{Bravyi2005,Bravyi2012,Haah2018,Litinski2019a}, whereas Clifford gates do not. 
Therefore, the number of $T$ gates and Toffoli gates is often used as a guiding metric in determining the cost of a computation.

\textbf{Baseline architectures.}
An architecture for a general-purpose fault-tolerant quantum computer must be defined together with a compilation scheme to translate arbitrary quantum computations into instructions for that architecture.
In general-purpose surface-code architectures described in the literature~\cite{Litinski2019,Fowler2018,Chamberland2022,Chamberland2022a,Bombin2021}, which we refer to as \textit{baseline architectures}, the cost of a computation scales with its \textit{circuit volume}.
A computation that requires $n_Q$ qubits of memory and contains $n_T$ $T$~gates has a circuit volume of $n_Q \times n_T$. 
One example of a baseline architecture is the architecture in Ref.~\cite{Litinski2019} where each $T$ gate is implemented via a multi-qubit Pauli measurement involving some of the qubits and a magic state. 
It is compatible with a two-dimensional grid of physical qubits with physical two-qubit operations between nearest neighbors.
In this architecture, the \textit{spacetime volume cost} of a quantum computation, i.e., the number of logical qubits multiplied by the total number of logical cycles, is roughly twice its circuit volume.
For example, the 2048-bit factoring algorithm described in Ref.~\cite{Gidney2021} consists of $\approx 6.1\times 10^9$ $T$ gates on 6200 logical qubits, so the circuit volume is $3.8 \times 10^{13}$.
This implies that, in the aforementioned baseline architecture, it would take $6.1 \times 10^9$ logical cycles on a device with $2 \cdot 6200$ logical qubits to finish the computation.

These numbers can be used to perform resource estimates for various hardware implementations, as described in more detail in Fig.~\ref{fig:examplealgorithm} and Appendix \ref{sec:resourceestimate}.
With a code distance of $d=28$, this corresponds to a grid of 19 million physical qubits and a logical cycle of 28 code cycles.
In a grid of superconducting qubits with a 1-$\mu$s code cycle, the computation would finish after 48 hours. 
In an array of trapped-ion qubits with a slower 1-ms code cycle, the computation would finish after 5.4 years.
In addition, it is possible to perform linear space-time trade-offs, i.e., use approximately twice as many physical qubits to finish the computation twice as fast~\cite{Litinski2019}.
In a photonic fusion-based~\cite{Bartolucci2021} implementation, the quantum computer is not an array of physical qubits, but instead a network of so-called interleaving modules~\cite{Bombin2021}, each module consisting of a resource-state generator (RSG) producing one photonic resource state every $\tau_{\rm RSG}$, and a number of additional switches, delay lines and single-photon detectors. Out of these components, the RSG is the most complex one, so the total number of RSGs is the relevant metric for the physical cost of the device. With $\tau_{\rm RSG} = 1~\mathrm{ns}$ and a maximum fiber delay length of 200 m, 9700 RSGs can be used to finish the computation in 48 hours. 
Interleaving modules also allow for the opposite space-time trade-off, i.e., the use of fewer hardware modules with longer delay lines, but a slower computation. 970 RSGs with a maximum fiber delay length of 2 km can finish the computation in 20 days. In the extreme case of very long 300-km free-space delay lines, the computation can be executed by a network of only 10 RSGs, but it will take over 5 years to execute.

\begin{figure*}
\centering
\includegraphics[width=0.98\linewidth]{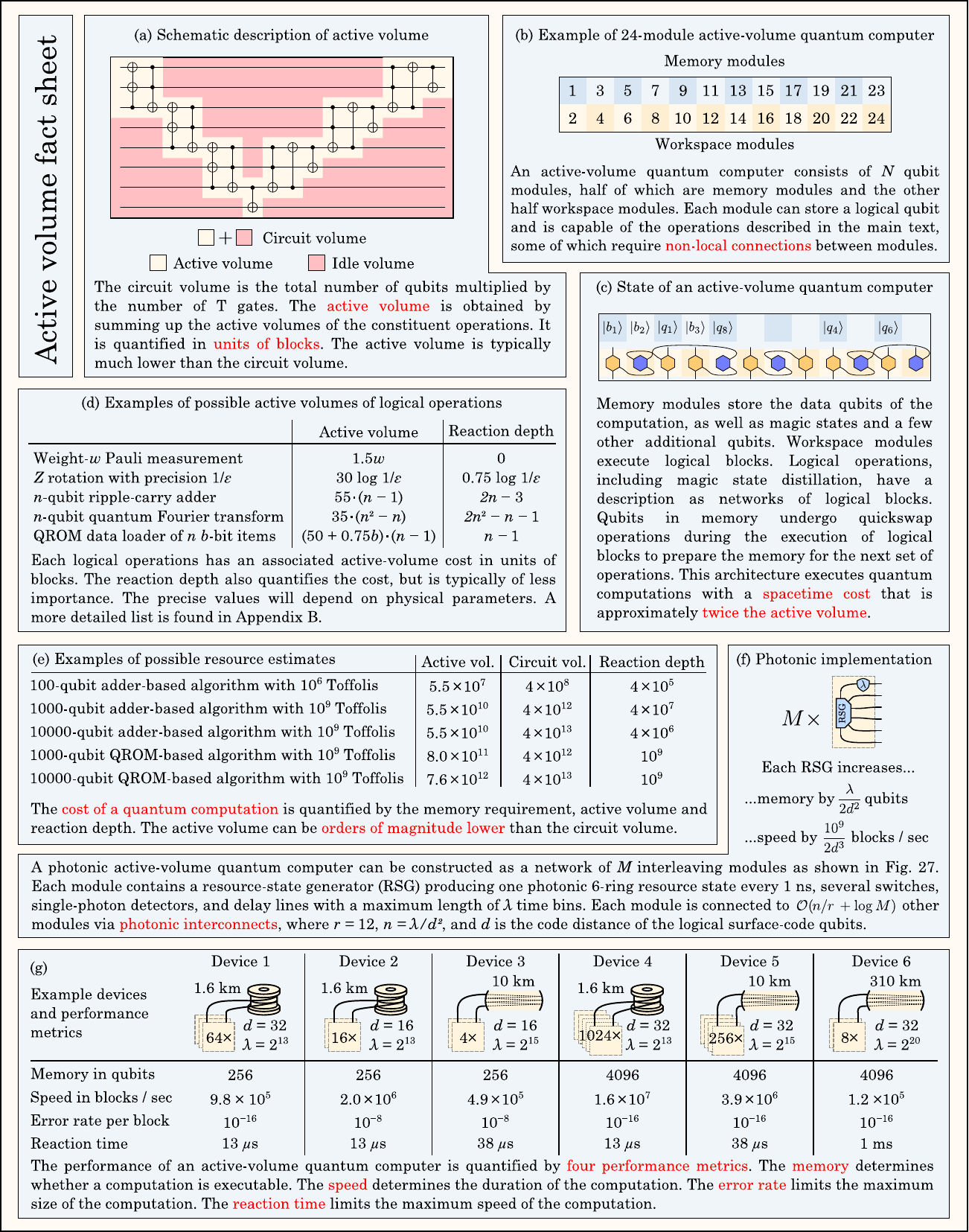}
\caption{Active volume fact sheet. The numbers in (d) assume active-volume costs of CCZ states and $T$ states of around 35 and 25, respectively. In (e), the considered algorithms are layers of disjoint 20-qubit adders, and $n$-qubit QROMs loading $n$-bit numbers. In (g), a per-block error rate of $p(d) = 10^{-d/2}$ and a reaction time of $\tau_r = 5~\mu\mathrm{s} + \lambda \cdot \mathrm{ns}$ is assumed.}
\label{fig:factsheet}
\end{figure*}

\textbf{Active volume.} 
In a baseline architecture, each multi-qubit operation may involve only a small subset of all $n_Q$ qubits, whereas the remaining logical qubits are idling during the execution of the operation.
Because idling logical qubits have the same cost as logical qubits that participate in a logical operation when using surface codes, the cost of each operation in a baseline architecture is identical, scaling with the total number of qubits $n_Q$.
In this sense, a large portion of the circuit volume may be \textit{idle volume}, i.e., volume attributed to idling logical qubits (see Fig.~\ref{fig:factsheet}a).
For example, in a 2000-qubit quantum computation consisting of a sequence of few-qubit non-Clifford operations such as Toffoli gates, more than 99\% of the circuit volume may be idle volume.
Qubits that are idling do not contribute to progressing the quantum computation, so we should be interested in generating as little idle volume as possible.
We refer to the remaining volume as \textit{active volume}, i.e., volume corresponding to logical operations that are required to progress the quantum computation.
We measure the active volume in units of \textit{logical blocks} (or simply \textit{blocks}).
Since the active volume depends on the cost of both Clifford and non-Clifford gates, it differs from operation to operation.
An adder will have a different active volume than a multi-qubit Pauli rotation, but, typically, the active volume will \textit{not} scale with the total number of qubits in the computation.
Instead, the active volume of a full quantum computation is obtained by summing over the active volumes of its constituent logical operations.
Due to the different scaling in the number of qubits, the difference between circuit volume and active volume becomes more pronounced for computations with more qubits. 
For example, the active volume of a 10,000-qubit quantum computation primarily consisting of adders is more than 100 times lower than the circuit volume.

\textbf{An active-volume architecture.} In this work, we construct a general-purpose architecture that can execute quantum computations with a spacetime volume cost of roughly twice the active volume instead of the circuit volume. This can significantly reduce the cost of quantum computations, as, e.g., the active volume of the aforementioned 2048-bit factoring algorithm is 44 times lower than the circuit volume. This enables a much faster execution of this computation, such that 19 million physical ion-trap qubits now finish the computation in 37 days instead of 5.4 years. 970 RSGs with 2-km-long fiber delays finish the computation in 8.9 hours instead of 20 days. And even a small network of only 10 RSGs with long 300 km free-space delays can finish the computation in 35 days instead of 5.4 years. However, the architecture relies on non-local connections between physical components. It can be thought of as a collection of surface-code patches with transversal physical two-qubit operations between a limited set of patch pairs. While the architecture can be implemented in any hardware platform with the required connectivity, it is primarily motivated by photonic qubits, where these connections are readily available, rather than matter-based ones. 

Note that surface-code methods to execute logical operations with a lower-than-baseline cost have been studied in the past, e.g, by Gidney and Eker\aa{}~\cite{Gidney2021} for the execution of the aforementioned factoring algorithm via the use of optimized adders~\cite{Gidney2019} which lead to a 6x lower spacetime volume cost compared to the baseline architecture.
However, these are special-purpose approaches and can only be applied to the execution of specific algorithms.
This also complicates resource estimates, as existing special-purpose methods rely on customized surface-code layouts that depend on the subroutines that need to be executed.
There exists no general-purpose architecture that can be used to execute arbitrary computations with a lower-than-baseline cost and a corresponding framework to perform resource estimates in a simplified manner.
This paper introduces such a general-purpose architecture.

We first define the architecture using a hardware-agnostic description in Sec.~\ref{sec:overview}. We describe on a high level how to perform resource estimates of quantum computations for an active-volume architecture, and how to quantify the performance of an active-volume quantum computer.
We then show how this architecture can execute logical operations in a cost-efficient manner and compute the active volume of several example logical operations in Secs.~\ref{sec:zmeasurements}-\ref{sec:compilation}, including multi-qubit Pauli measurements and rotations, adders, QROM reads and magic-state distillation protocols.
Finally, we demonstrate how an active-volume architecture can be implemented as a silicon-photonic fusion-based quantum computer in Sec.~\ref{sec:photonicmodules}.

\begin{figure*}
\centering
\includegraphics[width=0.72\linewidth]{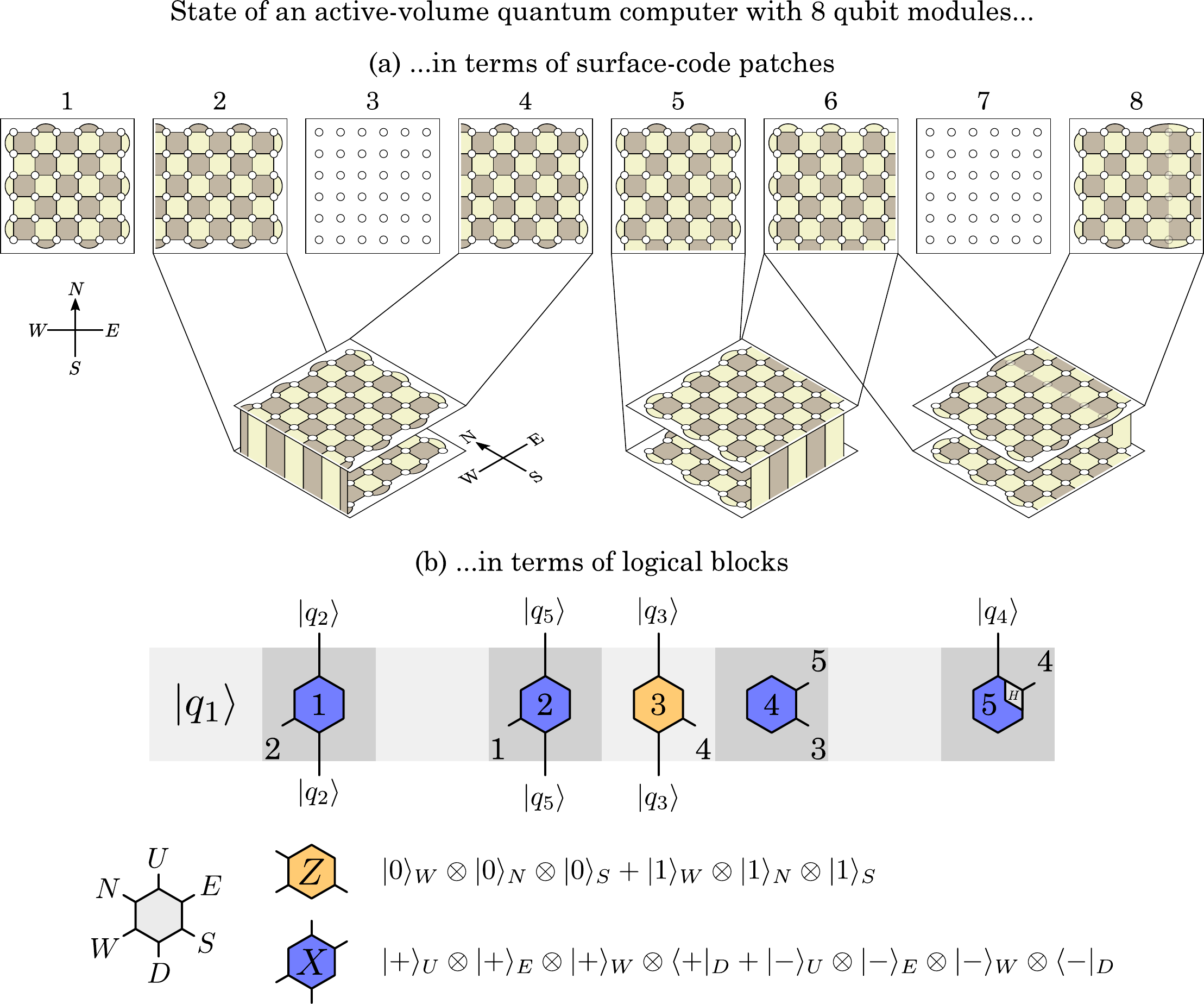}
\caption{Description of the active-volume architecture in terms of (a) surface-code patches and (b) logical blocks.}
\label{fig:architecturedef}
\end{figure*}

\section{Overview}
\label{sec:overview}

This section provides an overview of concepts that are explained in more detail in later sections. We first present a definition of the active-volume architecture in terms of surface-code patches that may be more familiar to the reader. Afterwards, we provide a more general definition in terms of logical space-time blocks that will be used throughout the paper.

\textbf{Active-volume architecture in terms of surface-code patches.} 
An active-volume quantum computer is a network of $N$ \textit{qubit modules} labeled 1 to $N$. Additional parameters defining the architecture are the \textit{range} $r$ and the \textit{code distance} $d$. Relevant time scales are the duration of a \textit{code cycle}, a \textit{logical cycle} corresponding to $d$ code cycles, and the \textit{reaction time} $\tau_r$.
Each qubit module can store a $d \times d$ surface-code patch encoding a logical qubit or a $d \times d$ ancilla patch~\cite{Litinski2019} facilitating multi-patch operations. Each patch has four boundaries, which we associate with the directions north (N), east (E), south (S) and west (W), as shown in Fig.~\ref{fig:architecturedef}a. A qubit module can also be empty, storing no qubits at all. Pairs of patch modules with labels $i$ and $j$ are referred to as \textit{in range}, if $|i-j| \leq r$, and as \textit{quickswappable} if $|i-j| = 2^k$. Here, $k$ is an integer between 0 and $\lfloor \log N \rfloor$ and all logarithms are binary logarithms.

The elementary unit of time of the quantum computation is a code cycle. In each code cycle, it shall be possible to perform all standard surface-code check measurements within each qubit module, including boundary checks, twist defects~\cite{Bombin2010} and lattice dislocations in at least one direction, single-qubit measurements and physical T gates for state injection. In addition, it shall be possible to execute all of the following operations in each code cycle:
\begin{enumerate}
\item Quickswaps: The qubits stored in pairs of quickswappable modules can be swapped via transversal physical SWAP gates.
\item Bell-state preparation: A logical Bell state $(|00\rangle + |11\rangle)/\sqrt{2}$ encoded in two surface-code patches can be prepared in two empty modules that are in range. This is achieved by transversal physical Bell-state preparations between physical data qubits, e.g., via the preparation of $|+\rangle$ states in one module and $|0\rangle$ states in the other, followed by transversal physical CNOT gates.
\item Lattice-surgery check measurements: If two surface-code patches are stored in two qubit modules that are in range, it shall be possible to measure surface-code checks between pairs of patch boundaries associated with the \textit{same} direction. In other words, it shall be possible to execute lattice-surgery operations~\cite{Horsman2012} between surface-code patches within a range $r$, where the lattice surgery merges boundaries associated with the same direction, rather than opposite directions. For example, Fig.~\ref{fig:architecturedef}a shows a west-to-west merge of patches stored in qubit modules 2 and 4, a south-to-south merge between patches in modules 5 and 6, and an east-to-east merge between patches in modules 6 and 8.
\item Bell measurements: A logical Bell-measurement can be performed between two logical qubits stored in two modules that are in range. This can be done via transversal physical Bell measurements between physical data qubits, i.e., transversal measurements of the two-qubit Pauli operators $X \otimes X$ and $Z \otimes Z$.

\end{enumerate}

\textbf{Logical blocks.} Before providing an alternative definition of the active-volume architecture, let us first define the notion of \textit{logical blocks}. These are operations described by ZX-calculus~\cite{Coecke2011} spiders, where we restrict ourselves to spiders with at most four legs in the context of this work. Logical blocks have a \textit{type} and \textit{orientation}. In the context of this paper, we define $Z$-type blocks as linear maps of the form 
\begin{equation}
|0\rangle^{\otimes m}\langle 0 |^{\otimes n} + |1\rangle^{\otimes m}\langle 1 |^{\otimes n} 
\end{equation}
and $X$-type blocks are linear maps of the form 
\begin{equation}
|+\rangle^{\otimes m}\langle + |^{\otimes n} + |-\rangle^{\otimes m}\langle - |^{\otimes n} \, ,
\end{equation}
where $2 \leq m+n \leq 4$, and $m$ and $n$ are integers with $0 \leq n \leq 1$ and $1 \leq m \leq 4$. In other words, they are operations with zero or one input qubits, and up to 4 output qubits. Each of the $n$ input and $m$ output qubits are referred to as \textit{ports}. Each port is associated with one of the six directions west (W), east (E), north (N), south (S), up (U) and down (D). The three possible orientations of logical blocks are east, north and up. E-oriented blocks are those that do not contain ports in the W or E direction. N-oriented blocks are those that do not contain ports in the S or N direction. U-oriented blocks are those that do not contain ports in the D or U direction.  Input qubits are always associated with D-ports. An example of a U-oriented $Z$-type block without input qubits is
\begin{equation}
|0\rangle_W \otimes |0\rangle_N \otimes | 0\rangle_S + |1\rangle_W \otimes |1\rangle_N \otimes | 1\rangle_S \, .
\end{equation}
In other words, this operation prepares a three-qubit GHZ state. Input or output qubits in the west or east port direction may also be affected by a Hadamard gate. (Note that this is by convention and, in principle, one may allow for Hadamarded ports in other directions.) An example of such an operation is this $Z$-type N-oriented block with one input qubit:
\begin{equation}
|0\rangle_U \otimes | +\rangle_E \otimes \langle 0|_D + |1\rangle_U \otimes | -\rangle_E \otimes \langle 1|_D \, .
\end{equation}
We then refer to the corresponding port (in this case the E-port) as \textit{Hadamarded}. Furthermore, we refer to pairs of blocks as \textit{commensurate}, if they have the same type and same orientation, or different types and different orientations. Pairs of blocks that have the same type and different orientation, or different types and the same orientation, are referred to as \textit{incommensurate}. We use hexagonal symbols to represent logical blocks. The color of the hexagon indicates the block type, with orange (blue) hexagons corresponding to $Z$-type ($X$-type) blocks, as shown in Fig.~\ref{fig:architecturedef}b. Each port of a logical block is represented by an edge radiating outwards from the hexagon in one of six different directions, where the direction corresponds to the direction of the port.

We use logical blocks as building blocks of logical operations by combining them to \textit{logical-block networks}. Such networks are collections of logical blocks in which some pairs of ports are connected. Each port connection implies that the corresponding pair of qubits is projected onto the Bell state $(|00\rangle + |11\rangle)/\sqrt{2}$. The motivation behind logical-block networks is that they describe logical operations that can be implemented with surface codes and lattice surgery~\cite{deBeaudrap2020,Bombin2021a}. (Readers familiar with the spacetime picture of surface codes may find it helpful to take a glance at Fig.~\ref{fig:logicalblockintro} to see the correspondence between logical blocks and 3-dimensional pieces of spacetime diagrams.) We only allow port connections between ports pointing in the same direction. Furthermore, two connected blocks must be commensurate, unless exactly one of the two ports connecting them is Hadamarded, in which case they must be incommensurate. In a logical operation described by a logical-block network, all ports in the W, S, E and N direction must be connected.  Unconnected D (U) ports correspond to input (output) qubits of the logical operation. A pair of connected D ports indicates that the input to the logical operation is a Bell pair.

\begin{figure*}
\centering
\includegraphics[width=\linewidth]{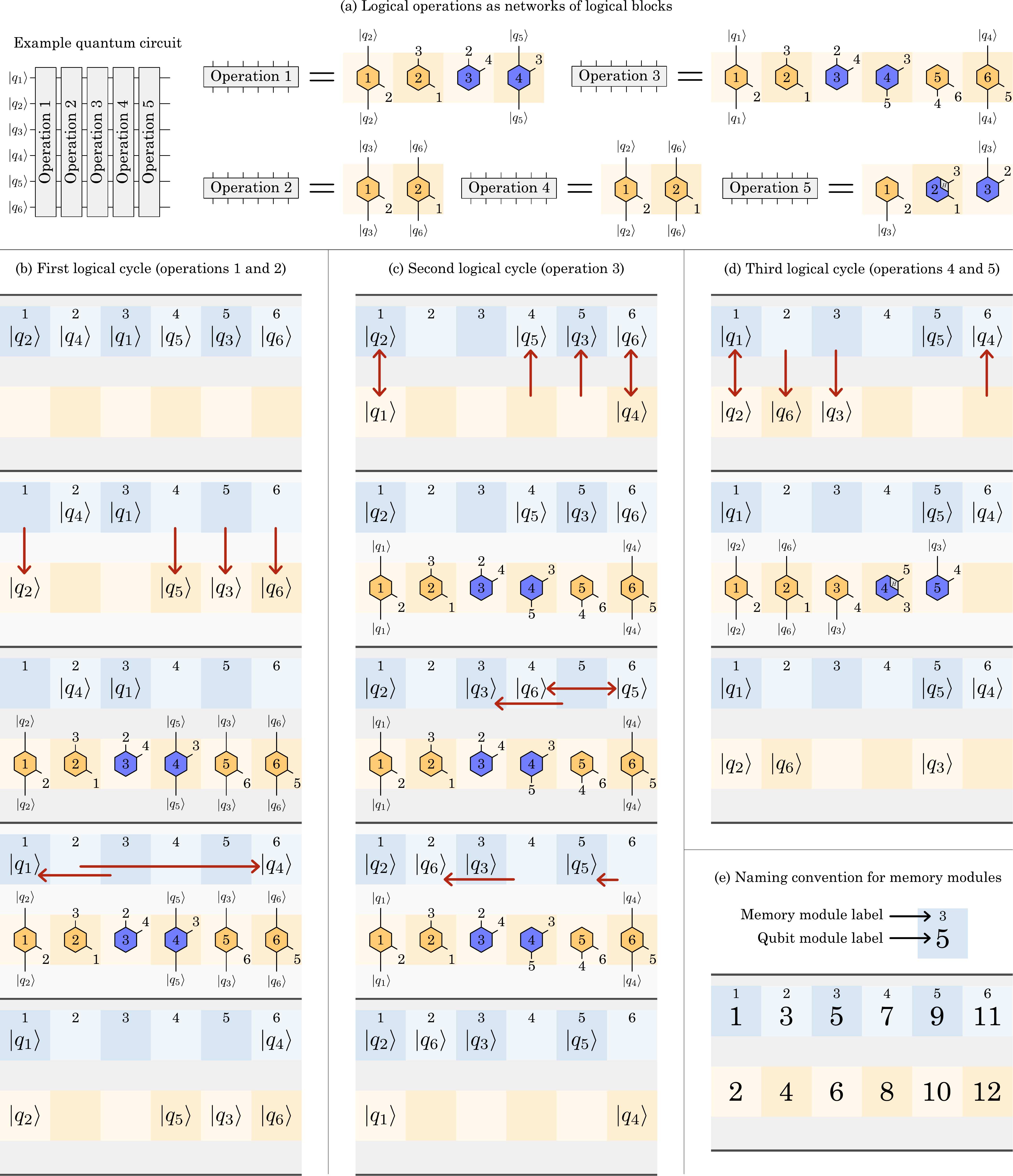}
\caption{Example demonstrating the structure of a quantum computation executed on an active-volume quantum computer with 12 qubit modules. (a) The quantum computation corresponds to a sequence of 5 operations on 6 qubits, where each operation has a representation as a network of logical blocks. (b) In the first logical cycle, operations 1 and 2 are executed. (c) In the second logical cycle, only operation 3 is executed. (d) In the third logical cycle, operations 4 and 5 are executed.}
\label{fig:compstructure}
\end{figure*}

When drawing networks of logical blocks, we label each block with a number which is drawn inside the hexagon. Numbers next to the edges representing ports indicate which block the port is connected to. For example, the network of 5 logical blocks shown in Fig.~\ref{fig:architecturedef}b is a logical operation with three input qubits ($|q_2\rangle$, $|q_3\rangle$ and $|q_5\rangle$) and four output qubits ($|q_2\rangle$, $|q_3\rangle$, $|q_4\rangle$ and $|q_5\rangle$). There are port connections between the W-ports of blocks 1 and 2, the S-ports of blocks 3 and 4, and the E-ports of blocks 4 and 5, where the E-port of block 4 is Hadamarded. Other examples of logical block networks are shown in Fig.~\ref{fig:compstructure}a, where the connection between the D-ports of blocks 4 and 5 in operation 3 indicates that the two input qubits of these logical blocks correspond to a Bell pair $(|00\rangle + |11\rangle)/\sqrt{2}$. While these are just meant as introductory examples of logical block networks, later sections show how to represent various logical operations using logical blocks, including Pauli rotations, adders, QROM circuits and magic state distillation protocols.

Let us briefly reiterate the motivation behind logical block networks and their connectivity rules, as these may seem very unintuitive to readers who are unfamiliar with the spacetime picture of surface codes. Logical blocks are operations that can be executed fault-tolerantly with surface codes. However, these are operations with at most one input qubit (the one associated with the D-port), but multiple output qubits. The connectivity rules are motivated by the problem that, when a logical qubit stored in a qubit module and a logical block operation is executed, then there cannot be suddenly two or three output qubits stored in the qubit module, since it can only store one. Therefore, the convention is that all output qubits, except for potentially (but not necessarily) the output qubit associated with the U-port, will be projected onto Bell states immediately after the operation is executed. With this construction, a logical block network with $n$ unconnected D-ports and $m$ unconnected U-ports describes a logical operation with $n$ input qubits and $m$ output qubits.

\textbf{Active-volume architecture in terms of logical blocks.} We now present a definition of the active-volume architecture in terms of logical blocks that is equivalent to the previous definition in terms of surface-code patches. It is this definition that will be used in the remainder of the paper. We define an active-volume architecture as a collection of $N$ qubit modules with labels from 1 to $N$. Qubit modules can each store a logical qubit or be empty, and are capable of the following operations:
\begin{enumerate}
\item Quickswaps: The contents of quickswappable qubit modules $i$ and $j$ can be swapped within one code cycle. Quickswaps between disjoint pairs of qubit modules can be performed simultaneously.
\item State preparations: Logical qubits can be initialized in the Pauli-$X$ and $Z$ eigenstates $|0\rangle$ and $|+\rangle$ in empty qubit modules within one code cycle. Furthermore, pairs of qubits $i$ and $j$ can be prepared in a Bell state $(|00\rangle + |11\rangle)/\sqrt{2}$ within one code cycle, if qubit modules $i$ and $j$ are empty and in range.
\item Measurements: Qubits can be measured in the $X$ or $Z$ basis within one code cycle. Furthermore, pairs of qubits stored in modules $i$ and $j$ can participate in a Bell-basis measurement within one code cycle, if the modules are in range. This is a measurement of the two-qubit Pauli operators $X_i \otimes X_j$ and $Z_i \otimes Z_j$.
\item Reactive measurements: If the choice of measurement basis ($X$, $Z$ or Bell measurement) depends on the outcome of previous measurements, it shall be possible to complete this measurement within a time $\tau_r$ after the completion of all previous relevant measurements, where $\tau_r$ is the reaction time.
\item Logical blocks: Each qubit module shall be capable of executing a logical-block operation within a logical cycle, i.e., $d$ code cycles. Port connections between simultaneously executed logical blocks are only allowed, if the logical blocks are executed by qubit modules that are in range. Qubit modules executing logical blocks with unconnected D-ports use the logical qubit that was previously stored in the module as the input qubit. Qubit modules executing logical blocks with unconnected U-ports will contain the output qubit of the logical block at the end of the logical cycle.
\item Magic state distillation: The qubit modules shall be capable of the execution of specific magic state distillation protocols, i.e., the use of a certain number of logical qubits for a fixed number of logical cycles to produce a fixed number of magic states such as $T$ states ${|T\rangle = (|0\rangle + e^{i\pi/4}|1\rangle)/\sqrt{2}}$ or CCZ states ${|\mathrm{CCZ}\rangle = \mathrm{CCZ}|+\rangle^{\otimes 3}}$.
\end{enumerate}

The correspondence between this definition and the definition in terms of surface-code patches is that quickswaps, Bell-state preparations and Bell measurements can be implemented via transversal physical two-qubit gates. Logical blocks with port connections in the N, E, S and W directions correspond to lattice-surgery operations via boundaries in the same direction, which are implemented by $d$ rounds of check-operator measurements. Port connections in the U and D direction correspond to transversal Bell measurements and Bell-state preparations, respectively. For example, Fig.~\ref{fig:architecturedef}b shows the state of 8 qubit modules in terms of logical blocks corresponding to the lattice-surgery operations of 8 surface-code patches shown in Fig.~\ref{fig:architecturedef}a. When drawing the state of qubit modules, each module is represented by a square, with a ket inside a square indicating that the module is storing a qubit, a hexagon indicating that it is executing a logical block, and an empty square representing an empty module. Modifications of this architecture are possible, e.g., by allowing for different code distances, additional non-Clifford resource states, additional transversal gates, or different connectivity.

\textbf{Photonic implementation.} While the implementation of such non-local connections may be challenging for some types of qubits, they can be implemented in a fusion-based photonic interleaving architecture with remarkably small modifications. As described in Ref.~\cite{Bombin2021}, in an interleaving architecture, a quantum computer is a network of interleaving modules. Each module consists of a resource-state generator (RSG) producing one photonic resource state (a fixed-size entangled state) in periodic time intervals of size $\tau_{\rm RSG}$, and a number of additional switches, delay lines and single-photon detectors. As we describe in Sec.~\ref{sec:photonicmodules}, a photonic active-volume quantum computer can be constructed as a network of modules with a maximum delay length of $\lambda = n \cdot d^2$ time bins. We describe two variants: one in which each interleaving module corresponds to $n$ qubit modules (Fig.~\ref{fig:module2}), and one in which each interleaving module corresponds to one qubit module, but contains a slowed down RSG with an $n$ times longer RSG cycle such that each full-speed RSG can supply $n$ interleaving modules (Fig.~\ref{fig:module1}). In any case, it is the total resource-state generation rate that determines the number of qubit modules in the quantum computer. Each 1 gigahertz of RSG rate (as provided by one full-speed RSG) adds $n$ qubit modules to the quantum computer. The main difference between these active-volume modules and the baseline modules in Ref.~\cite{Bombin2021} is the increased size of switches, requiring up to $\mathcal{O}(r + \log N)$ switchable options, where $N$ is the total number of qubit modules. However, since $r=12$ is sufficient for all operations described in this paper, the switch size is not expected to be a limiting factor.

\textbf{Active-volume compilation.} We refer to the odd-labeled qubit modules as \textit{memory modules} and the even-labeled qubit modules as \textit{workspace modules}, see Fig.~\ref{fig:factsheet}b. Even though their functionality is identical, they are envisioned to play different roles in the active-volume quantum computer. Memory modules store the data qubits of the quantum computation, whereas workspace modules are responsible for the execution of logical operations including magic state distillation (Fig.~\ref{fig:factsheet}c). Each logical operation can be expressed as a network of logical blocks, and therefore has an associated active volume in units of blocks. As we describe in the following sections in more detail, every workspace module will, ideally, execute one logical block in every logical cycle. Thereby, the quantum computer will be executing as many operations in parallel as possible. These operations do not need to commute and do not need to act on distinct sets of qubits, as explained in Sec.~\ref{sec:zmeasurements}. While the workspace qubits are executing logical blocks, memory qubits will be performing layers of quickswaps to rearrange the stored qubits such that they are in the right memory locations for the next set of logical blocks. As discussed in Sec.~\ref{sec:compilation}, a small number of quickswap layers is sufficient to rearrange large memories with thousands or even millions of logical qubits.

The structure of a quantum computation executed in an active-volume architecture is illustrated in Fig.~\ref{fig:compstructure}. The example shows the execution of a quantum circuit consisting of 5 logical operations on 6 logical qubits. Each logical operation can be represented as a network of logical blocks, as shown in Fig.~\ref{fig:compstructure}a. The number of blocks of an operation is its \textit{active volume}. While these are just small example operations, in practice, these logical operations may be any one of the subroutines found in Tab.~\ref{tab:subroutines}, such as an adder or a Pauli rotation. Note that each operation only affects a subset of qubits, e.g., operation 2 is a two-qubit operation on qubits $|q_3\rangle$ and $|q_6\rangle$. The figure shows how this circuit can be executed in a network of 12 qubit modules, where the 6 odd-labeled memory modules are drawn in blue and the 6 even-labeled workspace modules in orange. Note that we also refer to the $(2n+1)$-th qubit module as the $(n+1)$-th memory module, as shown in Fig.~\ref{fig:compstructure}e.

The 5 logical operations are executed in three logical cycles. For each logical cycle, our goal is to execute as many logical blocks as possible (i.e., ideally 6 logical blocks using 6 workspace modules), thereby executing multiple logical operations in parallel. Since operation~1 has an active volume of 4 blocks, and operation 2 an active volume of 2 blocks, we can execute them in parallel in the first logical cycle (Fig.~\ref{fig:compstructure}b). Qubits $|q_2\rangle$, $|q_3\rangle$, $|q_5\rangle$ and $|q_6\rangle$ are quickswapped from memory into workspace and a network of 6 logical blocks is executed, loading the pattern described by the networks of operations 1 and 2 in Fig.~\ref{fig:compstructure}a into the workspace. During the execution of logical blocks, up to $d$ layers of quickswap operations can be performed to prepare the state of the memory for the next logical cycle. In this example, we only perform one layer of quickswaps, moving $|q_1\rangle$ into the first memory module and $|q_4\rangle$ into the sixth memory module (i.e., qubit module 11). Remember that the locations connected by quickswaps must be quickswappable, which they are in this example. At the end of the logical cycle, the output qubits of the logical operation are found in the workspace.

In the second logical cycle (Fig.~\ref{fig:compstructure}), only one logical operation is executed, as operation 3 is particularly costly with an active volume of 6 blocks. Qubits $|q_1\rangle$ and $|q_4\rangle$ are quickswapped into the workspace, while the other qubits are quickswapped back into the memory. The purpose of the quickswaps in the previous logical cycle was to move $|q_1\rangle$ and $|q_4\rangle$ into the right locations such that the logical block network corresponding to operation 3 can now be executed in the second logical cycle after a single quickswap layer. In preparation for the third logical cycle, two layers of quickswaps are performed, moving $|q_3\rangle$ into memory module~3, $|q_5\rangle$ into memory module 5 and $|q_6\rangle$ into memory module 2. Finally, in the third logical cycle, operations 4 and 5 are executed, whose combined active volume is 5 blocks. Qubits are quickswapped between memory and workspace and the pattern of logical blocks is loaded into the workspace. Note that, in this cycle, not all workspace qubits are used to generate logical blocks, as the last workspace module is left idling. This may happen whenever the total active volume of the logical operations that are executed in parallel is slightly below the number of available workspace modules.

Note that, in this particularly simple example, consecutive operations act on disjoint sets of qubits. However, even if the same qubits are used by consecutive operations, these operations can be executed in parallel in an active-volume architecture, as discussed in Sec.~\ref{sec:zmeasurements}. This is also true even if consecutive operations do not commute. Such an architecture will therefore execute a quantum computation with a spacetime volume cost equivalent to approximately twice the active volume of the quantum computation, as $N$ qubit modules are used to execute $N/2$ logical blocks in every logical cycle. Since memory modules and workspace modules are identical, it is possible to use more memory in exchange for a slower quantum computer and vice versa. In addition to data qubits, the memory will also store distilled magic states, \textit{stale} magic states awaiting reactive measurements and \textit{bridge qubits} that connect concurrent operations accessing the same qubits. Therefore, quantum computations with memory requirements close to the maximum capacity of $N/2$ qubits may run more slowly than quantum computations using less memory.

\textbf{Resource estimates.} To quantify the cost of a quantum computation, it is sufficient to compute its active volume. While arbitrary quantum circuits consisting of $n_r$ single-qubit rotations on $n$ qubits and an arbitrary number of Clifford gates have an active volume of at most $\mathcal{O}(n^2 + n\cdot n_r)$ blocks, it is usually possible to significantly reduce the active volume of specific logical operations. Optimizations of several commonly used subroutines are shown in Secs.~\ref{sec:zmeasurements}-\ref{sec:distillation}. Their costs are summarized in Appendix \ref{sec:table}, and some examples are shown in Fig.~\ref{fig:factsheet}d. If a quantum computation consists exclusively of such optimized subroutines, its active volume is computed by simply summing over the active volumes of each constituent subroutine. If a quantum computation contains custom subroutines in addition to optimized ones, these can either be optimized using the methods shown in this paper, or treated as unoptimized blocks that are executed using the costly generic prescription in Sec.~\ref{sec:compilation}.

In addition to the active volume, a secondary quantity determining the cost of a quantum computation is the \textit{reaction depth}. Each logical operation has an associated reaction depth, which is the number of \textit{reaction layers} (i.e., consecutive layers of reactive measurements) that are part of the operation. Since reactive measurements are inherently sequential, as the choice of measurement bases of a layer of reactive measurements depends on the measurement outcomes of the previous layer, the reaction depth determines the minimum runtime of the quantum computation. However, computing the reaction depth of a full quantum computation is not as simple as adding the reaction depths of the constituent subroutines, as logical operations on disjoint groups of qubits typically can be performed in parallel, leading to a lower reaction depth. The output of a resource estimate of a quantum computation then consists of three numbers: the memory requirement in number of qubits, the active volume in units of blocks, and the reaction depth, as shown in Fig.~\ref{fig:factsheet}e.

\textbf{Performance metrics.} In order to determine how well a specific active-volume quantum computer can execute a quantum computation, we need to characterize its performance. The performance of an active-volume quantum computer is determined by four key performance metrics. The \textit{memory} corresponds to half the number of qubit modules and determines which quantum computations can be executed in the first place, as the memory needs to exceed the algorithmic memory requirement. The \textit{speed} is quantified in blocks per second, which is obtained by dividing half the number of qubit modules by the duration of a logical cycle. The duration of the quantum computation can be estimated by dividing the active volume of the quantum computation by the speed of the device. The \textit{error rate} is quantified as the logical error rate per block. It is governed by the (fixed) code distance of the logical qubits and the physical error rate of the device, and determines the maximum size of a computation that can be executed on the device. The total error probability of a computation with an active volume of $n_b$ blocks with a per-block error rate of $p_b$ is $1-(1-p_b)^{n_b}$. Finally, the \textit{reaction time} determines the maximum speed of the computation, as a computation cannot be executed in less time than its reaction depth multiplied by the reaction time, even if the speed in blocks per second would otherwise allow it.

\begin{figure*}[t]
\centering
\includegraphics[width=0.8\linewidth]{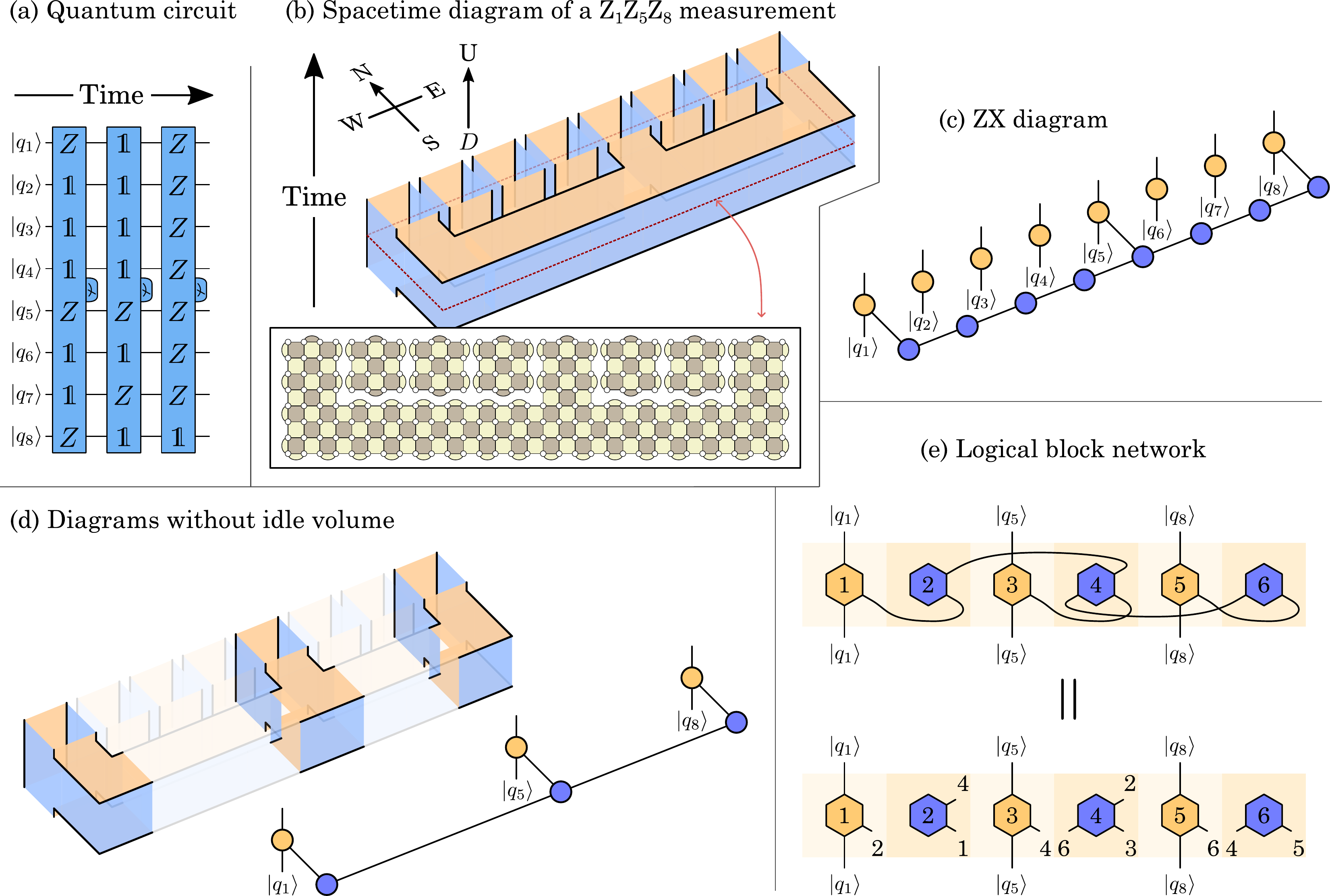}
\caption{(a) Quantum circuit of a sequence of $Z$-type measurements. In a baseline architecture, the first measurement may be executed via a lattice-surgery operation connecting qubits 1, 5 and 8. This operation can be described by (b) a spacetime diagram or (c) the corresponding ZX diagram. (d) The diagrams contain redundant volume corresponding to idle operations that do not contribute to the execution of the logical operation. (e) An active-volume quantum computer can execute the $Z$-type measurement generating only the volume relevant to the logical operation, i.e., the active volume of 6 logical blocks.}
\label{fig:zmeasexample}
\end{figure*}

\begin{figure*}[t]
\centering
\includegraphics[width=0.9\linewidth]{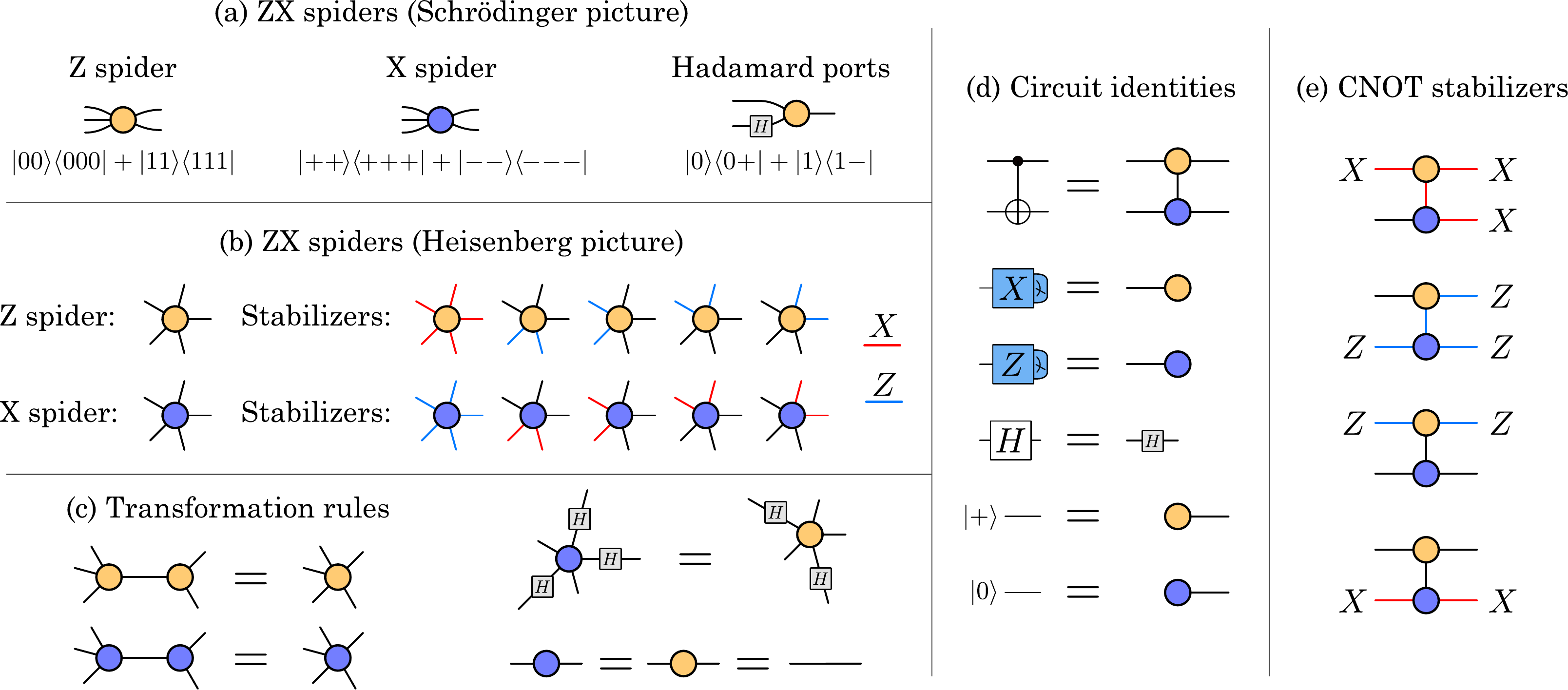}
\caption{(a) Definition of ZX spiders in the Schr\"odinger picture. (b) In the Heisenberg picture, $Z$ and $X$ spiders with $n$ legs are defined by the $n$ stabilizer generators of a $Z$-type or $X$-type $n$-qubit GHZ state. (c) Transformation rules can be used to manipulate ZX diagrams. (d) Circuit identities can be used to convert quantum circuits into ZX diagrams. Note that these identities are only valid up to Pauli corrections, as measurements are treated as projections. (e) Composite ZX diagrams with $n$ outgoing (unconnected) legs describe $n$ stabilizer generators, as shown for the example of the stabilizer generators of a CNOT gate consisting of two spiders.}
\label{fig:zxintro}
\end{figure*}

As shown in Fig.~\ref{fig:factsheet}f, the photonic implementation corresponds to a network of $M$ interleaving modules with a maximum delay length $\lambda$ and code distance $d$. Each module contains one RSG producing one resource state per nanosecond, such that each RSG increases the memory by $\lambda/(2d^2)$ and the speed by $10^9/(2d^3)$ blocks per second. A few example devices are shown in Fig.~\ref{fig:factsheet}g to visualize the effect of $M$, $\lambda$ and $d$ on the performance metrics. With a delay length of $\lambda \approx 8000$, as provided by 1.6 km of low-loss optical fiber, a quantum computer with 256 distance-32 logical qubits of memory can be constructed from 64 RSGs (device 1). With a speed of around $10^6$ blocks per second, it can execute the 100-qubit computation in Fig.~\ref{fig:factsheet}e in one minute, while the memory requirements of the other computations in the table exceed the memory of the device. Since a small $10^7$-block computation may tolerate a higher error rate, the physical footprint of the device can be reduced by using a smaller code distance (device 2). In addition, the use of longer delays can reduce the footprint. Each RSG adds four times as many qubits when using a 10-km free-space delay (device 3) instead of a 1.6-km fiber delay. The memory and speed of device 1 can be increased by adding more modules, e.g., device 4 with 4096 qubits and 1024 RSGs which can finish a $10^{11}$-block computation in 100 minutes. A 10-km free-space delay (device 5) reduces the footprint to 256 RSGs, but increases the computational time to 7 hours. A 1-million-photon 310-km free-space delay line (e.g., realized by two mirrors with 310 m separation and 1000 reflections) reduces the footprint to 8 RSGs, but increases the computational time to 10 days. However, to illustrate the limited usefulness of absurdly long delay lines and the importance of speed as a performance metric, consider that a single RSG with a 10,000-km delay line can provide 16,000 qubits of memory, but the slow speed may render the device useless for computations that require this much memory. Also note that the reaction time increases with the delay length in this particular construction.

\textbf{Summary.} Active volume is a cost function that assigns a cost in units of logical blocks to a quantum computation. The active volume of a quantum computation can be orders of magnitude lower than the circuit volume (number of qubits times number of $T$ gates). It is possible to construct a surface-code-based architecture that can execute computations with a spacetime cost proportional to the active volume, but this requires non-2D-local connections between surface-code patches. One possible implementation is via photonic interleaving modules with non-local photonic interconnects. In the following sections, we show how this architecture executes various logical operations.

\begin{figure*}[t]
\centering
\includegraphics[width=0.9\linewidth]{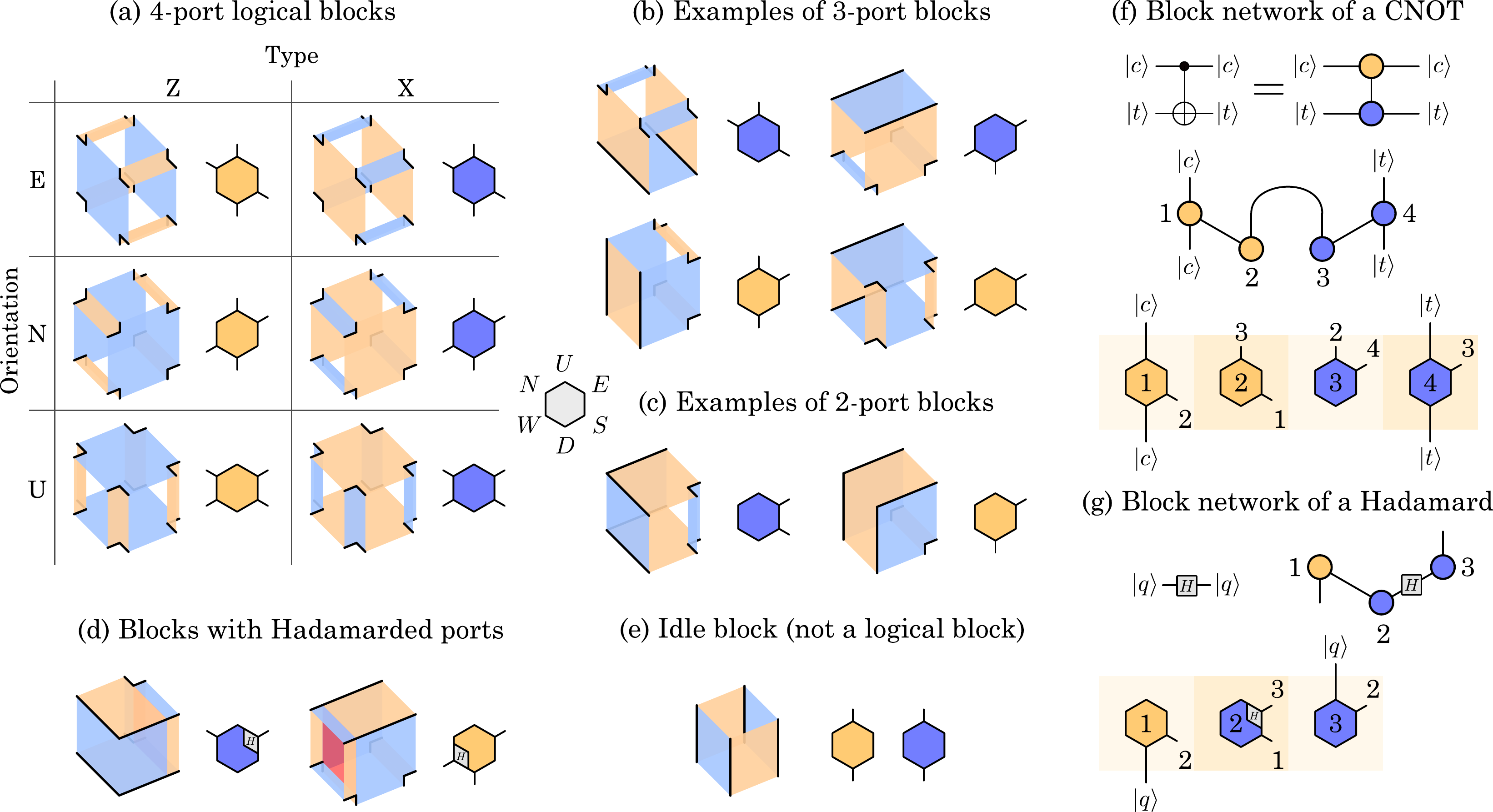}
\caption{Logical blocks describe segments of surface-code spacetime diagrams. They are similar to ZX spiders with 2 to 4 legs, except that logical blocks have an associated orientation. (a) There are six different 4-port logical blocks. Closing some of the ports of these six blocks generates (b) 3-port and (c) 2-port blocks. (d) In addition, ports of logical blocks may be Hadamarded, corresponding to a surface-code lattice dislocation. (e) 2-port blocks with ports pointing in opposite directions are idle blocks, so they are never used in logical operations. We use the convention that the $Z$ edges ($X$ edges) of idling logical qubits are located on the north and south (west and east) side of logical qubits. Quantum circuits can be translated into ZX diagrams, transformed to satisfy connectivity constrains, and converted into a network of logical blocks, as shown for (f) a CNOT gate with 4 blocks and (g) a Hadamard gate with 3 blocks.}
\label{fig:logicalblockintro}
\end{figure*}

\section{Introductory example: Z-type measurements}
\label{sec:zmeasurements}

As an introductory example, consider a quantum circuit consisting of a sequence of $Z$-type measurements. An example of an 8-qubit quantum circuit with three measurements is shown in Fig.~\ref{fig:zmeasexample}a. While no useful quantum computation consists exclusively of $Z$-type measurements, this will be our guiding example to introduce the different types of diagrams used in the following chapters. We will show how to execute this circuit on an active-volume quantum computer with 22 qubit modules and a cost proportional to the active volume of these operations.

In a baseline architecture such as the \textit{intermediate block} of Ref.~\cite{Litinski2019}, each measurement can be executed as shown in Fig.~\ref{fig:zmeasexample}b for the example of the first measurement. The figure shows a spacetime diagram, which is a diagram describing how the configuration of surface-code patches changes over time. Using the convention of Fig.~6 in Ref.~\cite{Bombin2021}, each line tracks a corner of a surface-code patch through spacetime, each orange surface a $Z$ boundary, and each blue surface an $X$ boundary. Each slice through the spacetime diagram then corresponds to a configuration of surface-code check operators. The slice shown in Fig.~\ref{fig:zmeasexample}b is a 3-qubit $Z$-type lattice surgery~\cite{Horsman2012,Fowler2018} where all logical qubits are encoded in distance-4 surface-code patches in this example. We will refer to the in-plane directions as north (N), east (E), south (S) and west (W), and to the perpendicular time directions as up (U) and down (D).

Two additional types of diagrams that will be used in the following chapters are ZX-calculus diagrams~\cite{Coecke2011} and logical-block diagrams. Before properly introducing these diagrams, note that ZX diagrams can be used to describe surface-code operations~\cite{deBeaudrap2020} as shown in Fig.~\ref{fig:zmeasexample}c. In a baseline architecture, this diagram will contain many degree-2 vertices which turn out to be redundant, such that many components of spacetime diagrams and ZX diagrams can be removed, as shown in Fig.~\ref{fig:zmeasexample}d. The remaining part is what we refer to as the active volume, with each segment of the spacetime diagram referred to as a logical block. Our goal will be to convert logical operations into chains of as few logical blocks as possible. In this example, an active-volume quantum computer may be able to execute the operation by executing 6 logical blocks, as shown in Fig.~\ref{fig:zmeasexample}e. \linebreak

\textbf{ZX diagrams.} The ZX calculus~\cite{Coecke2011} is an immensely useful graphical language for linear maps between qubits that can be used to describe and optimize quantum circuits~\cite{Duncan2020,Kissinger2020,deBeaudrap2020a} as well as surface-code operations~\cite{deBeaudrap2020,Gidney2018a}. We will use only a subset of the tools available in the construction and simplification of ZX diagrams, specifically the phase-free ZX spiders and Hadamard ports shown in Fig.~\ref{fig:zxintro}a. A ZX diagram consists of vertices, edges connecting pairs of vertices and edges that connect to only one vertex. Each vertex is referred to as a \textit{spider}, and each of the edges connected to a spider can be thought of as a qubit. We will refer to the edges as \textit{ports}. A spider with $n+m$ ports can be thought of as a linear map with $n$ input qubits and $m$ output qubits. $Z$ spiders are drawn as orange circles and correspond to linear maps
\begin{equation}
|0\rangle^{\otimes m}\langle 0|^{\otimes n} + |1\rangle^{\otimes m}\langle 1|^{\otimes n} \, .
\end{equation}
Similarly, $X$ spiders correspond to linear maps
\begin{equation}
|+\rangle^{\otimes m}\langle +|^{\otimes n} + |-\rangle^{\otimes m}\langle -|^{\otimes n} \, ,
\end{equation}
where $|\pm\rangle = (|0\rangle \pm |1\rangle)\sqrt{2}$.
Some ports may be \textit{Hadamarded}, corresponding to the application of Hadamard gates to the corresponding qubits.

An alternative definition of ZX spiders that relies on operators rather than states is shown in Fig.~\ref{fig:zxintro}b. Spiders can be thought of as describing stabilizer-state projections, with each $n$-port spider describing $n$ stabilizer generators on $n$ qubits. The stabilizer generators described by $Z$ ($X$) spiders are $X^{\otimes n}$ ($Z^{\otimes n}$) and all pairwise $Z^{\otimes 2}$ ($X^{\otimes 2}$) operators. Multiple spiders can be connected to form composite ZX diagrams, e.g., the two-spider diagram in Fig.~\ref{fig:zxintro}e describing a CNOT gate. Each connection between two spiders corresponds to a Bell-state projection, i.e., the two qubits $i$ and $j$ corresponding to the two ports are identified via the projection $Z_iZ_j = X_i X_j = +1$. A composite ZX diagram with $n$ unconnected ports also describes $n$ stabilizer generators on $n$ qubits, with the 4 stabilizer generators of a two-spider diagram shown in Fig.~\ref{fig:zxintro}e. When translating quantum circuits to ZX diagrams we will rely on the circuit identities in Fig.~\ref{fig:zxintro}d. When simplifying ZX diagrams, we will use the identities shown in Fig.~\ref{fig:zxintro}c.

In the operator picture of phase-free ZX diagrams, composite diagrams describe Clifford gates and Pauli measurement. Each stabilizer generator $P_i \otimes P_o$, where $P_i$ and $P_o$ are multi-qubit Pauli operators supported on the input and output qubits (ports), respectively, describes a map $P_i \rightarrow P_o$. For example, the CNOT gate in Fig.~\ref{fig:zxintro}e maps $X_c \rightarrow X_cX_t$, $Z_t \rightarrow Z_cZ_t$, $Z_c \rightarrow Z_c$ and $X_t \rightarrow X_t$, where $c$ and $t$ label the control and target qubits. Stabilizer generators supported only on input or output qubits can be thought of as multi-qubit Pauli measurements and preparations of Pauli eigenstates.

\textbf{Logical blocks.} The operations described by the ZX diagrams that can be constructed from the building blocks in Fig.~\ref{fig:zxintro} are identical to the operations implementable by surface-code qubits. ZX spiders have a one-to-one correspondence to segments of surface-code spacetime diagrams. However, ZX spiders do not have a notion of orientation, whereas segments of spacetime diagrams need to be oriented in 3D space and satisfy certain constraints on connectivity and number of ports. For this reason, we introduce \textit{logical blocks} as hexagons with up to four connected edges in Fig.~\ref{fig:logicalblockintro} to describe segments of surface-code spacetime diagrams. The six corners of each hexagon are identified with the six directions of spacetime diagrams (U, E, S, D, W and N).

\begin{figure}[b!]
\centering
\includegraphics[width=\linewidth]{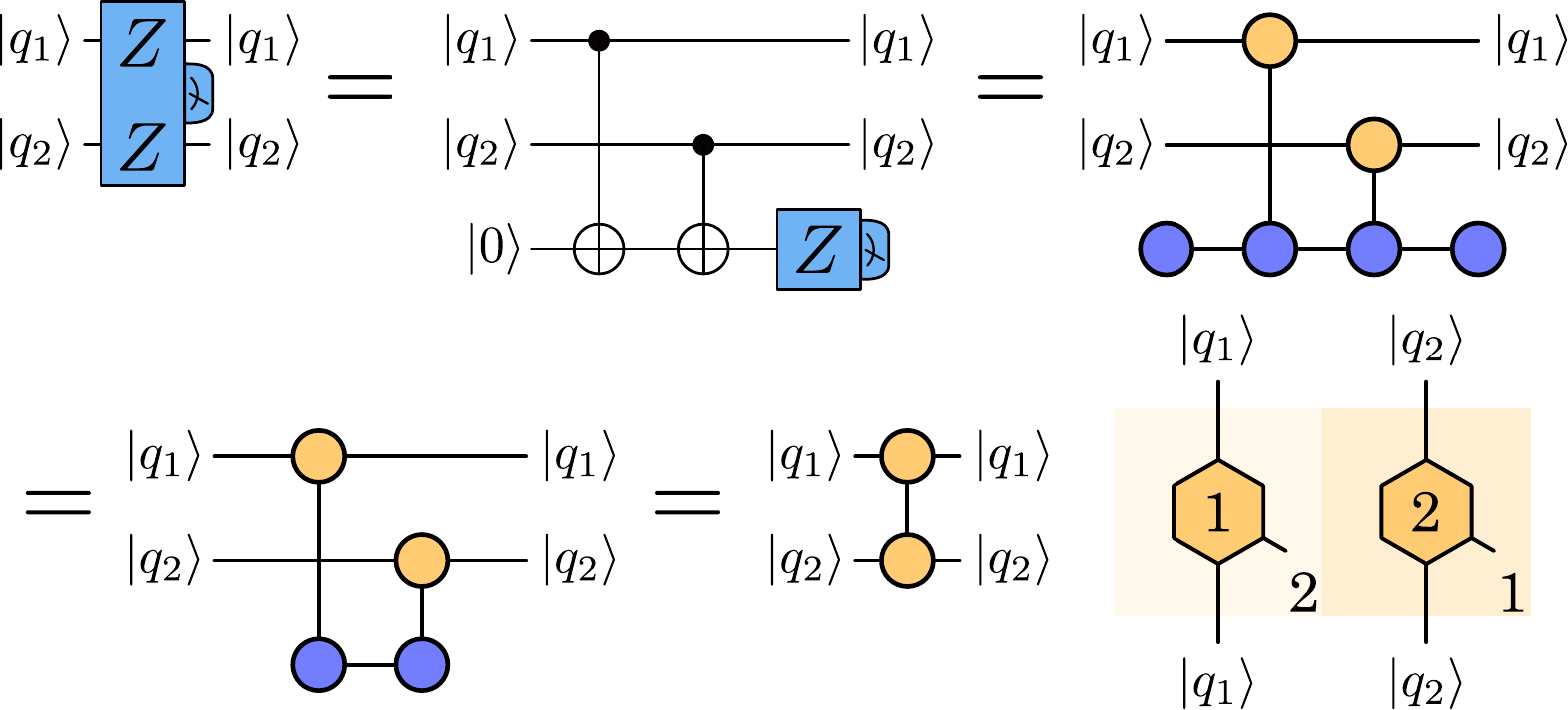}
\caption{A 2-qubit $Z \otimes Z$ measurement has an active volume of 2 blocks. Changing $Z$-type to $X$-type blocks implements a 2-qubit $X \otimes X$ measurement with a volume of 2 blocks.}
\label{fig:weighttwozmeas}
\end{figure}

\begin{figure*}[t]
\centering
\includegraphics[width=\linewidth]{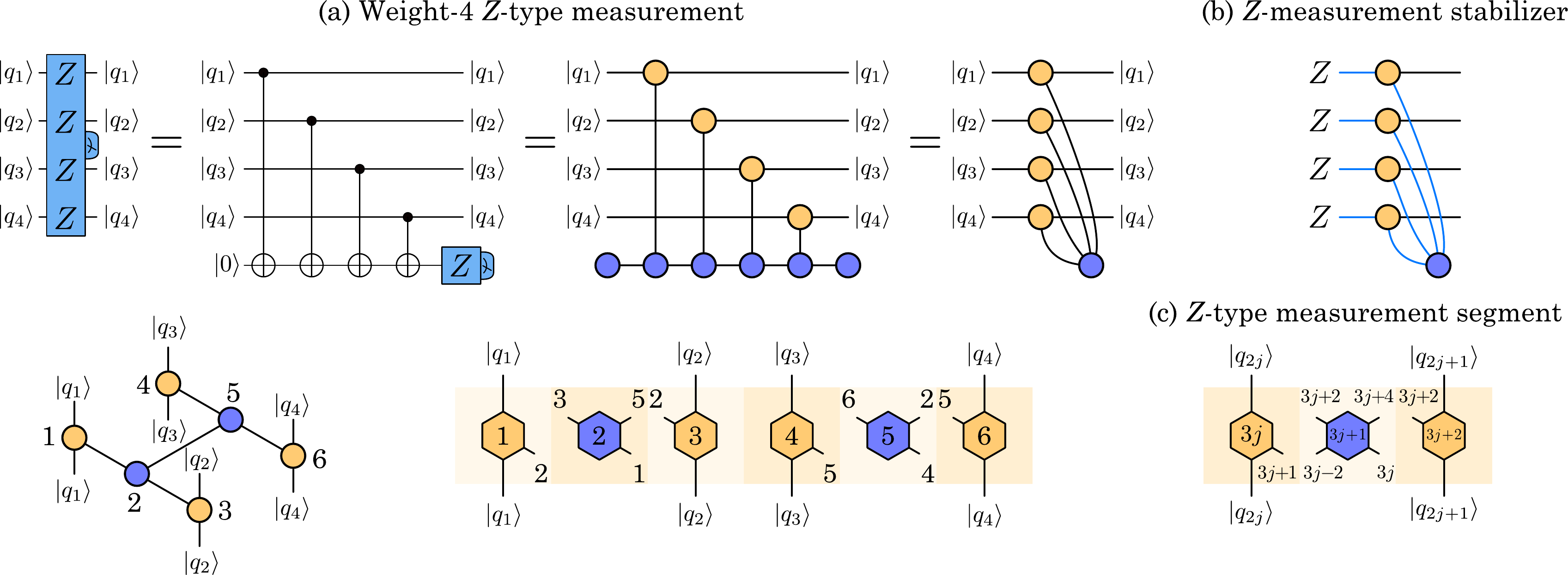}
\caption{(a) A weight-4 $Z$-type measurement has an active volume of 4 blocks. (b) The ZX diagram can be used to verify that the implemented operation is indeed a measurement of the 4-qubit $Z$ operator. (c) A weight-$w$ $Z$-type (or $X$-type) measurement has an active volume of $\lceil \frac{3}{2}w \rceil$ blocks for $w > 2$.}
\label{fig:zmeasurements}
\end{figure*}

Logical blocks can have 2, 3 or 4 ports. There are six types of 4-port blocks, which are shown in Fig.~\ref{fig:logicalblockintro}a. These are oriented in one of three different directions (E, N and U) and are either $Z$-type (orange) or $X$-type (blue). These blocks can either be interpreted as 4-port Z-spiders and X-spiders, or as surface-code spacetime diagrams. Each stabilizer generator of the ZX diagram corresponds to one logical membrane (or correlation surface) of the surface-code in the corresponding spacetime diagram~\cite{Bombin2021a}.
 The spacetime diagrams corresponding to $Z$-type blocks can be converted to $X$-type blocks by replacing primal ($Z$-type) boundaries with dual ($X$-type) boundaries and vice versa. 3-port and 2-port blocks are generated by closing some of the ports of the 4-port logical blocks, as shown in Fig.~\ref{fig:logicalblockintro}b/c. Ports can be Hadamarded, corresponding to surface-code lattice dislocation shown in red in Fig.~\ref{fig:logicalblockintro}d. Note that we will only use those 2-port blocks that correspond to corner blocks, i.e., we will not use 2-port blocks with ports pointing in opposite directions. 2-port blocks in the up-down direction correspond to idling qubits, i.e., qubits stored in memory. Since the assignment of primal and dual boundaries is ambiguous in this case, we will adopt the convention in Fig.~\ref{fig:logicalblockintro}e that the $Z$ boundaries of all idling qubits stored in memory point in the north and south directions, whereas the $X$ boundaries point west and east.

Similar to ZX spiders, 2/3/4-port logical blocks can be connected into networks of logical blocks. However, because surface-code boundaries must connect to boundaries of the same type (primal or dual), there are certain constraints that must be satisfied when assembling networks of logical blocks. The first constraint is that we only allow connections via ports pointing in the same direction, e.g., east ports can only be connected to east ports. This is different than in conventional spacetime diagrams as in Fig.~\ref{fig:zmeasexample} where east ports connect to west ports, but such a connectivity is allowed due to the reflection symmetry of even-distance surface-code patches. Secondly, connected logical blocks must either have the same type and orientation, or different types and different orientations. If one connecting port (but not both) is Hadamarded, then the connected blocks must have the same type and different orientations, or different types and the same orientation.

Figure \ref{fig:logicalblockintro}f shows the process that we will repeat for various logical operations to translate them into networks of logical blocks. We first convert the circuit to a ZX diagram. Next, we apply transformation rules to convert the ZX diagram into an \textit{oriented} ZX diagram, such that each spider can be replaced by a logical block. The oriented ZX diagram is not directly a logical block network, as it may contain connections between, e.g., east and west ports. Instead, it is meant as a guide to the eye with numbers next to the ZX spiders indicating the corresponding logical block. We use this to construct a chain of logical blocks, where each block is described by a hexagon. The numbers inside the hexagons label the blocks, while the numbers next to the ports indicate the connected blocks. For example, block 2 in Fig.~\ref{fig:logicalblockintro}f is connected to block 1 via the S-port and to block 3 via the U-port. Some D-port and U-port labels are qubits instead of numbers. These correspond to input and output qubits of the logical operation, respectively. The construction in Fig.~\ref{fig:logicalblockintro}f shows that a CNOT gate has an active volume of 4 blocks, even though the ZX diagram consists of only two spiders. Due to the orientation of idle blocks in Fig.~\ref{fig:logicalblockintro}e, an additional constraint is that blocks with input or output qubits must either be E-oriented $Z$-type blocks or N-oriented $X$-type blocks. Therefore, a Hadamard gate has an active volume of 3 blocks as shown in Fig.~\ref{fig:logicalblockintro}g, even though it corresponds to a single 2-port spider.

\begin{figure}[t!]
\centering
\includegraphics[width=\linewidth]{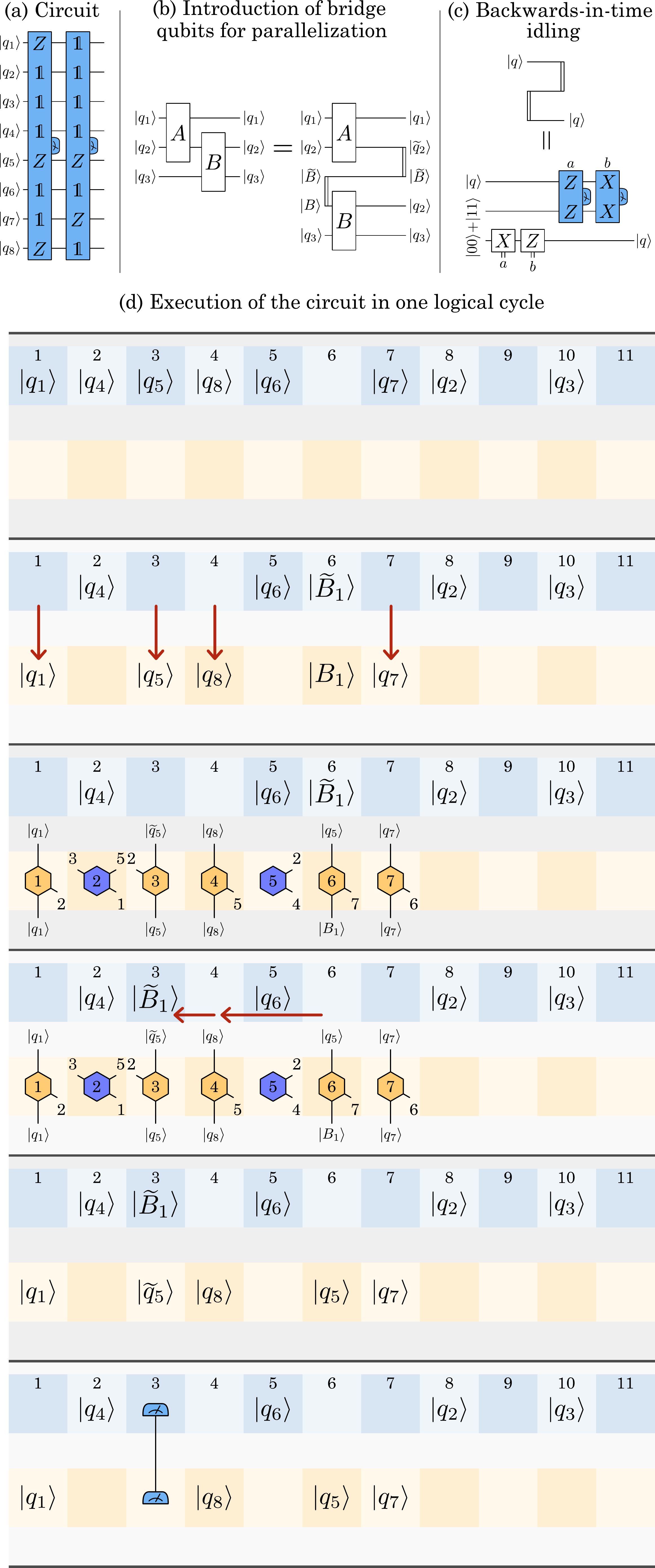}
\caption{Example of the execution of an 8-qubit circuit with two $Z$-type measurements using a 22-qubit active-volume quantum computer with 11 memory qubits and 11 workspace qubits.}
\label{fig:memstate1}
\end{figure}

\textbf{Two-qubit $Z \otimes Z $ measurements.} As a slightly more advanced example, consider the $Z \otimes Z$ measurement shown in Fig.~\ref{fig:weighttwozmeas}. We can express this measurement as two CNOT gates and a $Z$ measurement between the two qubits and an ancilla qubit initialized in $|0\rangle$. We translate the circuit into a ZX diagram and apply the simplification rules of Fig.~\ref{fig:zxintro}c to obtain a diagram consisting of two 3-port spiders. This diagram can be straightforwardly converted into a network of two logical blocks. Note that the ancilla qubit is only used in the construction and does not appear as an actual qubit in the final network of blocks. Furthermore, note that the active volume of large circuits can be smaller than the total active volume of the composite operations. While the active volume of two CNOT gates is 8 blocks, the active volume of a $Z \otimes Z$ measurement is only 2 blocks.

\textbf{Weight-$w$ $Z$-type measurements.} Next, consider the example of a 4-qubit $Z$-type measurement in Fig.~\ref{fig:zmeasurements}a. We can convert this into a ZX diagram with four Z spiders and one X spider. When constructing the oriented ZX diagram, we split the 4-port X spider into two 3-port X spiders. The operation has an active volume of 6 blocks. We can verify that this ZX diagram indeed implements a measurement, since the $Z^{\otimes 4}$ operator on the four input qubits is a stabilizer generator of the ZX diagram, as shown in Fig.~\ref{fig:zmeasurements}b. This construction generalizes to the measurement of arbitrary weight-$w$ multi-qubit $Z$ operators. Using the three-block segment in Fig.~\ref{fig:zmeasurements}c, a weight-$w$ measurement has an active volume of $\lceil \frac{3}{2}w \rceil$ for $w > 2$.

\textbf{Execution on an active-volume quantum computer.} We will now describe how to execute the example 8-qubit computation in Fig.~\ref{fig:zmeasexample} using an active-volume quantum computer with 22 qubit modules, i.e., 11 memory modules and 11 workspace modules. First, consider only the first two operations, i.e., the circuit in Fig.~\ref{fig:memstate1}a. The active volume of these operations is $5+2=7$, so we should be able to execute the circuit in one logical cycle by executing 7 logical blocks using 7 workspace modules. However, one complication is that qubit $|q_5\rangle$ participates in both operations. Whenever we need to execute multiple operations simultaneously that access the same qubits, we can introduce \textit{bridge qubits} using the construction shown in Fig.~\ref{fig:memstate1}b. A bridge qubit is initialized by initializing a Bell pair $(|00\rangle + |11\rangle)/\sqrt{2}$, which is a fast operation in an active-volume quantum computer. One half of the Bell pair $|B\rangle$ is used as an input qubit for the second logical operation, while the other half $|\widetilde{B}\rangle$ is stored in memory and referred to as a bridge qubit. At the end of the logical cycle, this bridge qubit needs to be destroyed via a Bell-basis measurement with the output qubit of the first logical operation, effectively teleporting the qubit back in time to be used as an input to the second operation, as shown in Fig.~\ref{fig:memstate1}c. Note that we will be referring to the two qubits of the Bell pair as $|B\rangle$ and $|\widetilde{B}\rangle$, even though the 2-qubit Bell state is not separable, so cannot be written as  $|B\rangle \otimes |\widetilde{B}\rangle$.

\begin{figure*}[]
\centering
\includegraphics[width=\linewidth]{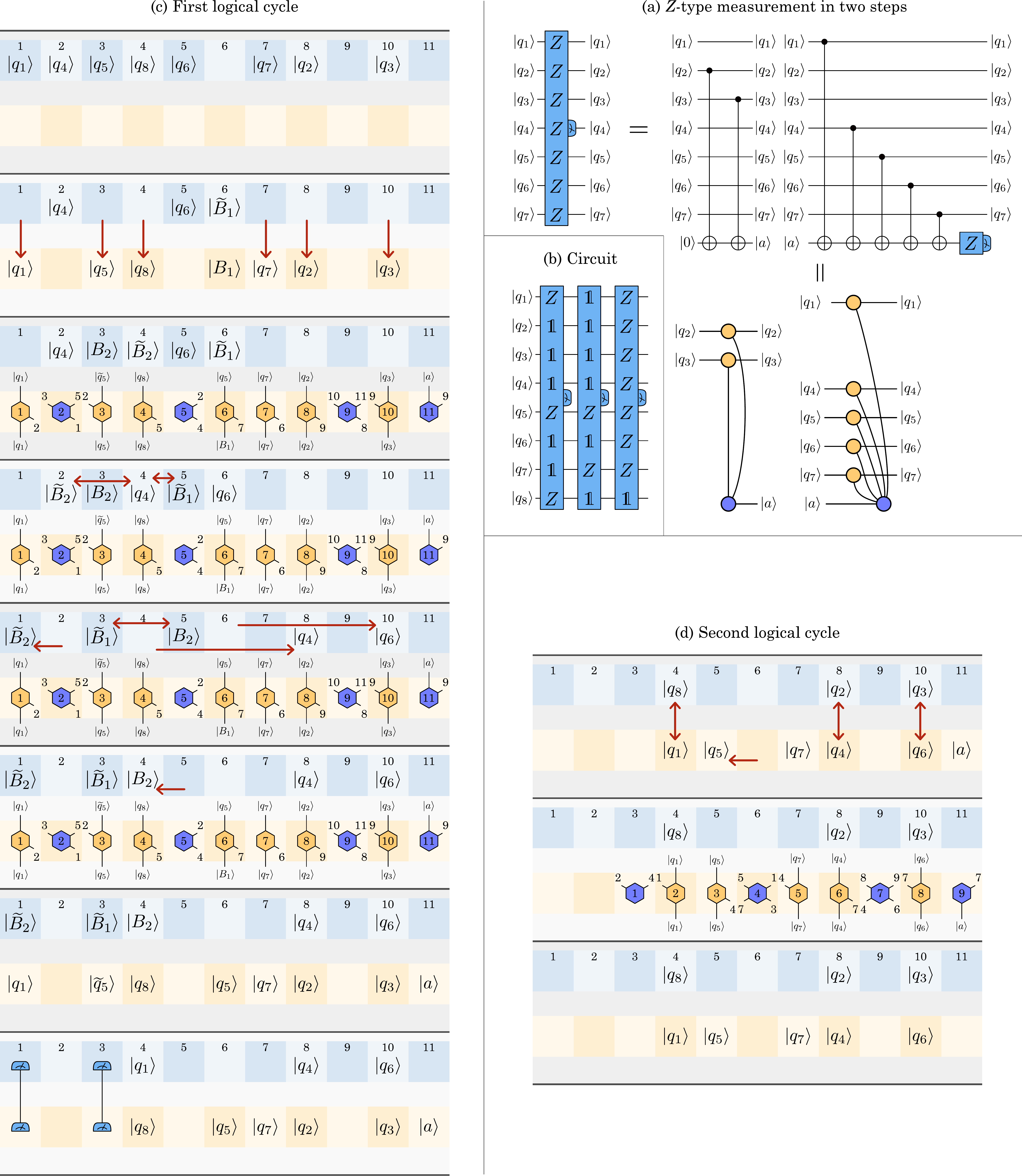}
\caption{Example of the execution of an 8-qubit circuit with three $Z$-type measurements using a 22-qubit active-volume quantum computer with 11 memory qubits and 11 workspace qubits.}
\label{fig:memstate2}
\end{figure*}

\begin{figure*}[t]
\centering
\includegraphics[width=\linewidth]{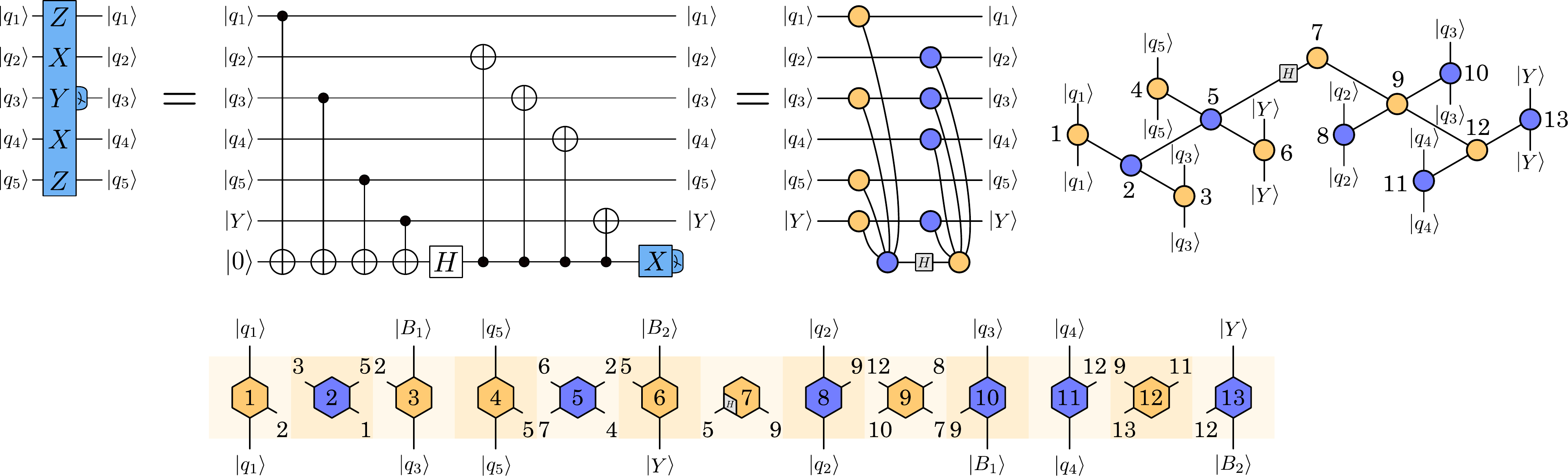}
\caption{Weight-$(w_x,w_z)$ Pauli product measurements have an active volume of $\lceil \frac{3}{2}w_x \rceil + \lceil \frac{3}{2}w_z \rceil + 1$. $X$ operators increase $w_x$ by 1. $Z$ operators increase $w_z$ by 1. $Y$ operators increase both $w_x$ and $w_z$ by 1. Additionally, if the total number of $Y$ operators is odd, $w_z$ and $w_x$ are increased by 1. Bridge qubits are used for qubits that participate in both the $X$ and $Z$ part of the measurement, i.e., $|q_3\rangle$ and $|Y\rangle$ in the example.}
\label{fig:ppm}
\end{figure*}

Figure~\ref{fig:memstate1}d shows all the steps required to execute the two logical operations in one logical cycle. Qubits $|q_1\rangle$, $|q_5\rangle$, $|q_7\rangle$ and $|q_8\rangle$ are quickswapped from memory into workspace. Simultaneously, a Bell pair consisting of two qubits $|B_1\rangle$ and $|\widetilde{B}_1\rangle$ is initialized. Next, 7 logical blocks corresponding to two logical operations are executed using 7 workspace modules, where $|B_1\rangle$ is used as an input qubit for the second logical operation instead of $|q_5\rangle$. Note that the output qubit $|q_5\rangle$ of the first logical operation is relabeled to $|\widetilde{q}_5\rangle$. During the execution of the logical-block operations, the bridge qubit $|\widetilde{B}_1\rangle$ is quickswapped into memory location 3 in two code cycles. This is done to move the bridge qubit to a location that is in range of the location where $|\widetilde{q}_5\rangle$ will emerge at the end of the logical cycle. Finally, the qubits $|\widetilde{B}_1\rangle$ and $|\widetilde{q}_5\rangle$ are removed via a two-qubit Bell measurement, implementing the bridge-qubit protocol of Fig.~\ref{fig:memstate1}b.

Now consider the full circuit in Fig.~\ref{fig:memstate2}b. The third operation has an active volume of 11 blocks, but there are only 4 unused workspace modules left in the first cycle. Since a $Z$-type measurement consists of multiple segments, we can split this operation into two steps, as shown in Fig.~\ref{fig:memstate2}a, where the first step uses 4 blocks and produces one ancilla qubit $|a\rangle$ as an output, and the second step uses 9 blocks and uses $|a\rangle$ as an input. Again, we start in the first logical cycle in Fig.~\ref{fig:memstate2}c by swapping qubits from memory into workspace. Looking ahead to the second logical cycle in Fig.~\ref{fig:memstate2}d, we can see that we will be generating 9 logical blocks, 6 of which have input qubits. If we want to avoid executing quickswap operations between logical cycles (i.e., additional operations to rearrange the memory after the end of the first logical cycle, but before the beginning of the second logical cycle, which would slow down the computation), these input qubits must be located in a quickswappable location at the end of the first logical cycle. For qubits $|q_4\rangle$, $|q_5\rangle$, $|q_6\rangle$, $|q_7\rangle$ and $|a\rangle$, this will be satisfied, as only one quickswap is required to move them into the right workspace locations in the second logical cycle.

However, block 2 in the second cycle requires qubit $|q_1\rangle$ as an input, but this qubit will be output in block 1 in the first logical cycle, which is not a quickswappable location. In this situation, we have a second use case for bridge qubits. Whenever the output qubit $|q\rangle$ of a logical operation in cycle $j$ is used as an input qubit in the subsequent logical cycle $j+1$, and the block using this input qubit is executed in a non-quickswappable location, we can generate a Bell pair in cycle $j$. Both Bell-pair qubits $|B\rangle$ and $|\widetilde{B}\rangle$ are stored in memory. The output of the block in cycle $j$ will participate in a Bell measurement with the bridge qubit $|\widetilde{B}\rangle$ at the end of the cycle, instantly teleporting $|q\rangle$ to the location of  $|B\rangle$. In Fig.~\ref{fig:memstate2}d, we generate a Bell pair $|B_2\rangle$ and $|\widetilde{B}_2\rangle$ in memory locations 3 and 4 at the beginning of cycle 1. The bridge qubit $|\widetilde{B}_2\rangle$ needs to be moved to memory location 1 to participate in a destructive Bell measurement with $|q_1\rangle$. This teleports $|q_1\rangle$ to the location of $|B_2\rangle$, so we rename $|B_2\rangle$ to $|q_1\rangle$ at the end of the first logical cycle. $|q_1\rangle$ can now be quickswapped into the workspace location where it is needed at the beginning of the second logical cycle (Fig.~\ref{fig:memstate2}d).

\textbf{Summary.} To summarize, in each cycle, we execute as many logical blocks as we can, ideally executing one logical block per cycle in each workspace location. The logical blocks executed in cycles $j$ and $j+1$ impose certain conditions on the state of the memory at the end of cycle $j$, as some memory locations must be empty, whereas others must store specific qubits. The conditions are the following: Input memory qubits to logical blocks must be located in quickswappable locations at the beginning of the cycle. Input bridge qubits require an empty memory location within range, such that a Bell pair can be generated, with one half of the Bell pair used as an input qubit for the logical block, and the other stored as a bridge qubit in memory. Output memory qubits either require an empty memory slot at a quickswappable position at the end of the operation or must be swapped with an input qubit for the logical block in the next cycle. Output qubits participating in a Bell measurement at the end of the cycle require the corresponding bridge qubit in memory to be located within range at the end of the cycle. In the example of Fig.~\ref{fig:memstate2}c, the memory can be rearranged within 3 code cycles using 3 layers of quickswap operations. As we show in Sec.~\ref{sec:compilation}, even very large memories can be rearranged within a sufficiently low number of quickswap cycles.

Provided that there is sufficient memory available and that the memory can be rearranged quickly enough, an active-volume quantum computer with $n$ qubit modules ($n/2$ of which are workspace modules) will finish a quantum computation with an active volume of $m$ blocks in approximately $2m/n$ logical cycles, i.e., an overall spacetime cost of $2m/n \cdot n = 2m$. Therefore, we should decompose logical operations into networks of as few logical blocks as possible in order to optimize them for an active-volume quantum computer, which is what we do in the following sections for various commonly used subroutines. We are also interested in keeping the degree of non-locality required to implement these logical block networks as low as possible, as ports of logical blocks $i$ and $j$ can only be connected, if they are in range $r$, i.e., if $|i-j| \leq r/2$ (since only every second logical qubit is a workspace qubit). For all operations described in the following sections, a range of $r\geq 12$ will be sufficient.

\section{Pauli product rotations and measurements}
\label{sec:pprs}

Having discussed $Z$-type Pauli measurements, we now consider arbitrary multi-qubit Pauli product measurements (PPMs). First, note that weight-$w$ $X$-type Pauli measurements also have an active volume of $\lceil \frac{3}{2}w \rceil$, as they can be obtained by replacing $Z$-type blocks in Fig.~\ref{fig:zmeasurements} with $X$-type blocks and vice versa. Using a circuit that is referred to as \textit{fast PPMs} in Ref.~\cite{Kim2022} or \textit{twist-free lattice surgery} in Ref.~\cite{Chamberland2022}, an arbitrary PPM can be decomposed into a $Z$-type measurement followed by an $X$-type measurement. As shown in Fig.~\ref{fig:ppm}, an ancilla qubit is initialized in $|0\rangle$, each qubit contributing a $Z$ or $Y$ is part of the first set of CNOTs, a Hadamard is applied to the ancilla, and then each qubit contributing an $X$ or $Y$ is part of the second set of CNOTs. The ancilla is measured in the $X$ basis, yielding the PPM outcome. This construction only works for PPMs with an even number of $Y$ operators. If the PPM contains an odd number of $Y$ operators, as in the example of Fig.~\ref{fig:ppm}, a $|Y\rangle = (|0\rangle + i |1\rangle)/\sqrt{2}$ state can be used as a catalyst state (i.e., it is a resource state that is not consumed by the operation). Since $Y=1$ for the $Y$ state, it can contribute an extra $Y$ to the PPM without changing the measurement outcome.

\begin{figure*}[t!]
\centering
\includegraphics[width=\linewidth]{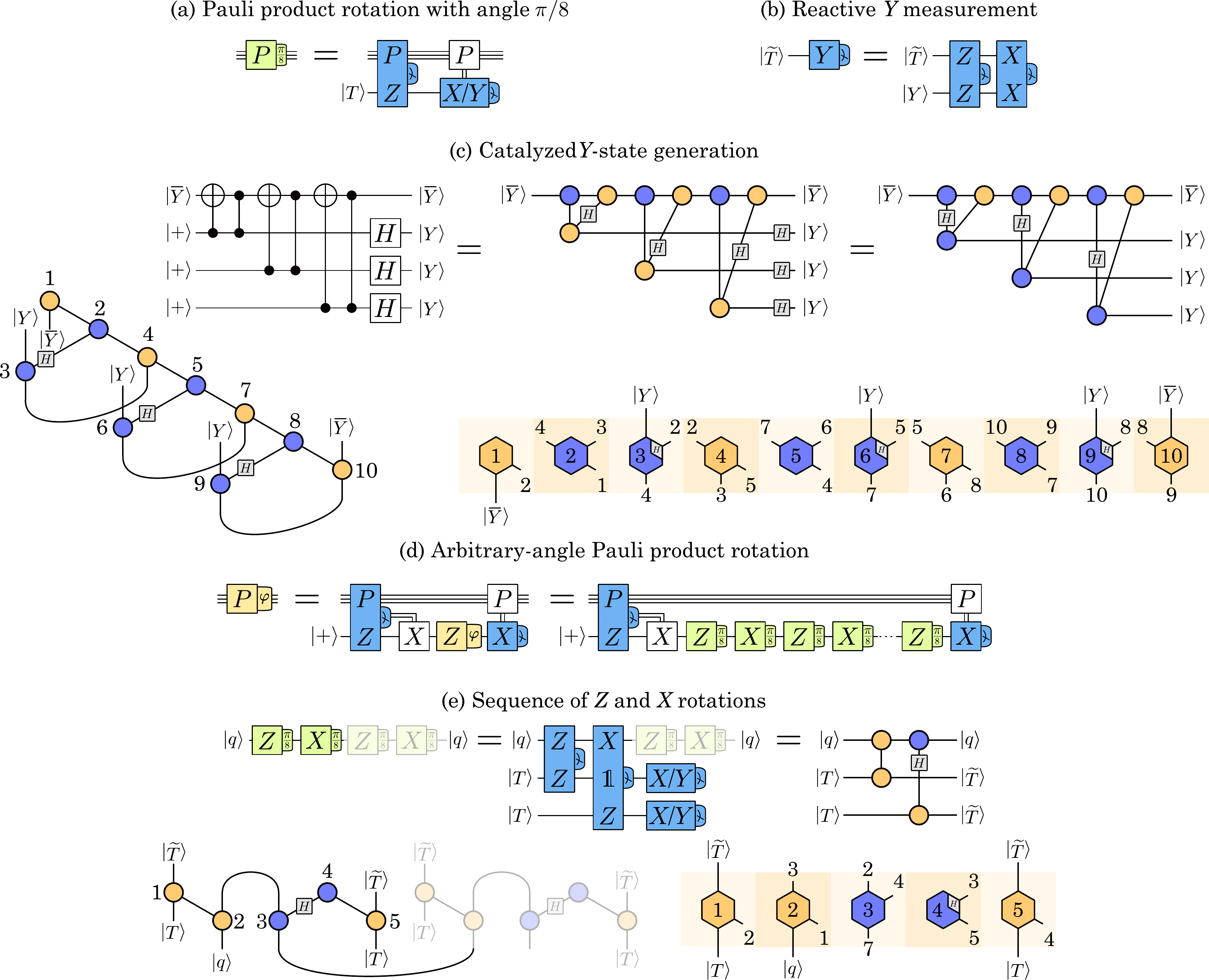}
\caption{(a) A distilled $T$ state can be consumed via a $P \otimes Z$ measurement to execute a generalized $T$ gate. This converts the distilled $T$ state into a stale $T$ state that needs to be removed via a single-qubit $X$ or $Y$ measurement, depending on the outcome of the $P \otimes Z$ measurement. The single-qubit measurement generates a Pauli correction. (b) A fast $Y$ measurement of a stale magic state can be performed by consuming a $Y$ state via a reactive Bell-basis measurement. (c) Using a $|\bar{Y}\rangle$ state as a catalyst, new $Y$ states can be generated with a cost of 3 blocks per state. (d) An arbitrary-angle $Z$ rotation with an error of $\varepsilon$ can be implemented using a sequence of $3\log 1/\varepsilon$ $X$ and $Z$ rotations with an angle of $c \cdot \pi/8$, where $c$ is an odd integer~\cite{Ross2014}. (e) Each pair of $X$ and $Z$ rotations has an active volume of 5 in addition to the cost of two $T$ states and one reactive Y measurement.}
\label{fig:pprs}
\end{figure*}

As shown in Fig.~\ref{fig:ppm}, the active volume of an arbitrary PPM is $\lceil \frac{3}{2} w_x \rceil + \lceil \frac{3}{2} w_z \rceil + 1$, where each $X$ ($Z$) operator in the PPM increases $w_x$ ($w_z$) by 1. Each $Y$ operator in the PPM increases both $w_x$ and $w_z$ by 1. If a $Y$ state is required to turn an odd number of $Y$ operators into an even number, $w_x$ and $w_z$ are again increased by 1.

\textbf{Pauli product rotations.} Next, we consider Pauli product rotations (PPRs), i.e., operations $P_\varphi = e^{-i P \varphi}$, where $P$ is a multi-qubit Pauli operator and $\varphi$ is a rotation angle. First, we consider PPRs with an angle $\varphi = \pi/8$. These are generalizations of $T$ gates, which are $Z_{\pi/8}$ rotations. As shown in Fig.~\ref{fig:pprs}a, such a PPR can be executed by consuming a $T$-gate magic state $|T\rangle = (|0\rangle + e^{i\pi/4}|1\rangle)/\sqrt{2}$ via a $P \otimes Z$ measurement involving the data qubits and the $T$ state. The measurement is non-destructive, so it leaves behind a qubit which we will refer to as a \textit{stale magic state}. Each $\pi/8$ rotation uses a \textit{distilled} $T$ state and turns it into a stale $T$ state. This qubit needs to be removed via a destructive single-qubit measurement. The basis of this measurement depends on the outcome of the $P\otimes Z$ measurement: If the outcome is $P \otimes Z = +1$, the stale $T$ state needs to be measured in the $X$ basis, otherwise in the $Y$ basis. Depending on the outcome of the single-qubit measurement, there can be a corrective $P$ Pauli operation on the qubits. Note that Pauli gates are not logical operations that require quantum hardware operations, but merely influence the interpretation of future PPM outcomes. Also note that rotations with $\varphi = \pm \pi/8$ and $\varphi = \pm 3\pi/8$ can all be executed by the circuit in Fig.~\ref{fig:pprs}a, differing only in the classical logic determining the basis of the single-qubit measurement and the presence of the Pauli correction.

\begin{figure*}[t!]
\centering
\includegraphics[width=0.8\linewidth]{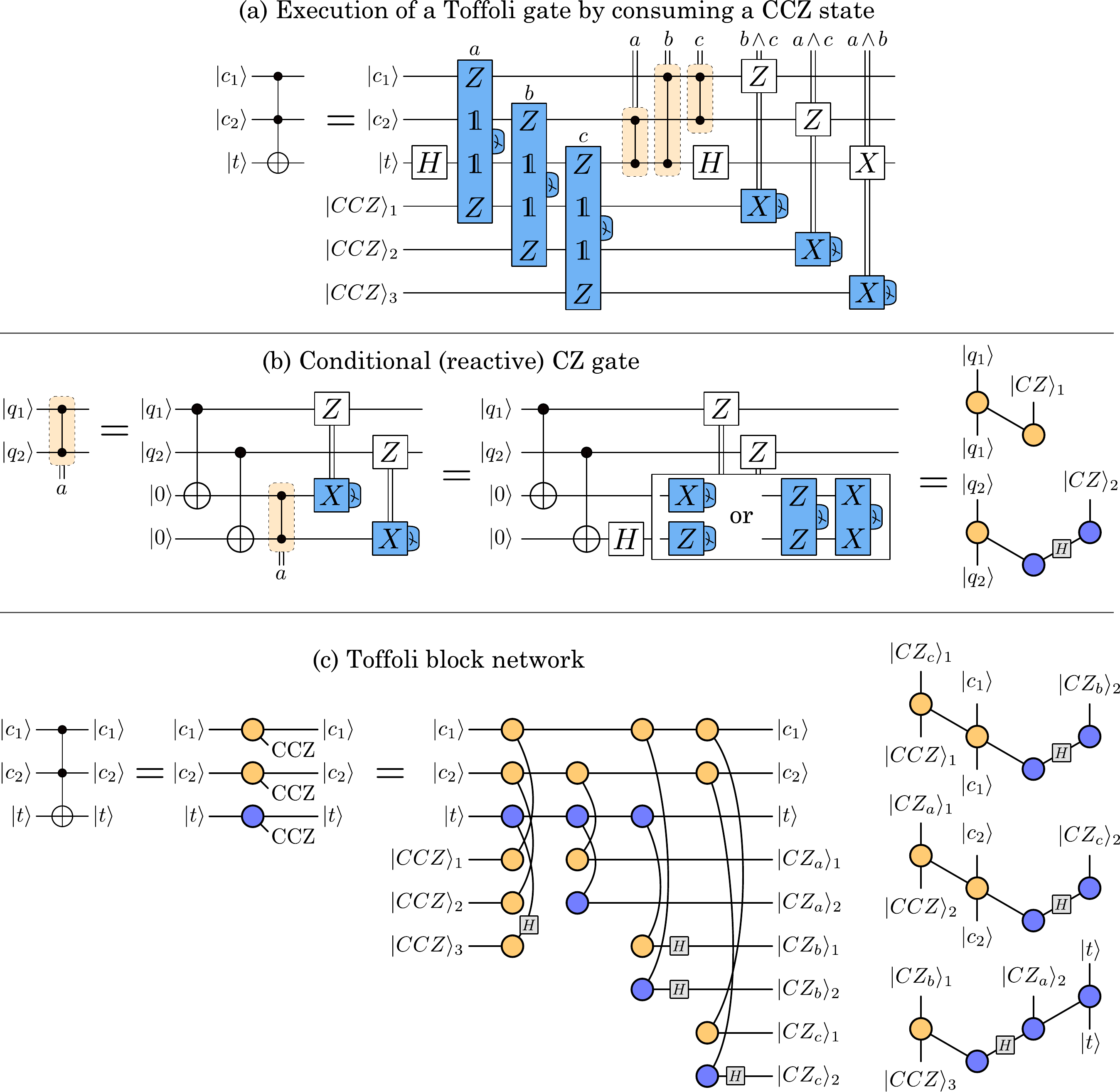}
\caption{(a) CCZ states can be consumed to execute Toffoli gates via 3 two-qubit measurements. Each measurement outcome determines the presence or absence of a subsequent CZ gate. The same measurement outcomes as well as the single-qubit measurement outcomes determine a set of Pauli corrections. (b) Such conditional CZ gates can be performed in a reactive manner (i.e., using only fast reactive measurements) by generating a pair of qubits that are referred to as a CZ state. If these qubits are destroyed by single-qubit $X$ and $Z$ measurements, no CZ gate is generated. If, instead, they are destroyed by a Bell-basis measurement, a CZ gate is teleported into the circuit. A reactive CZ gate has an active volume of 5 blocks. (c) Consequently, a Toffoli gate has an active volume of 12 blocks.}
\label{fig:toffoli}
\end{figure*}

\begin{figure*}[t]
\centering
\includegraphics[width=0.98\linewidth]{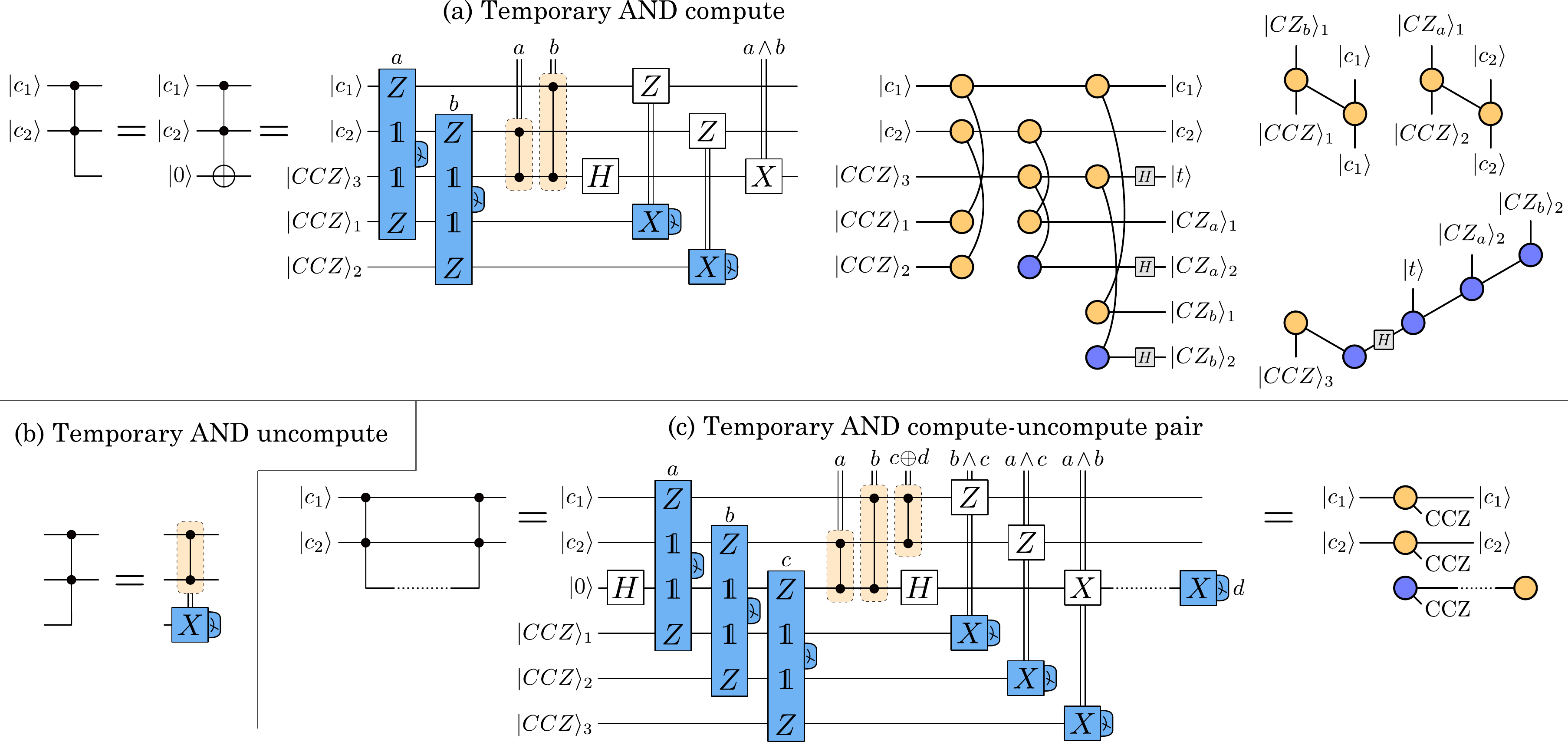}
\caption{(a) Toffoli gates corresponding to the first half of a temporary AND have a reduced active volume of 9 blocks. (b) The uncompute portion has an active volume of 5 blocks. (c) If the conditional CZ of the uncompute portion of the temporary AND can be commuted to the first half of the temporary AND, a compute-uncompute pair has the same active volume as a Toffoli gate, i.e., 12 blocks.}
\label{fig:tempand}
\end{figure*}

We refer to measurements whose basis depends on the outcome of previous measurements as \textit{reactive measurements}. Because the speed of reactive measurements ultimately dictates how fast a quantum computation can be executed, we should always execute reactive measurements using operations that can be performed in a single code cycle, rather than a logical cycle. This limits the allowed reactive measurements to single-qubit $X$ and $Z$ measurements and to two-qubit Bell-basis ($X\otimes X$ and $Z \otimes Z$) measurements. Notably, single-qubit $Y$ measurements are not fast measurement operations with surface codes. A reactive $Y$ measurement of a stale $T$ state can be performed using a Bell-basis measurement between a stale $T$ state and a $Y$ state, consuming the $Y$ state in the process, as shown in Fig.~\ref{fig:pprs}b. The $Y$ outcome is given by the outcome of $Y \otimes Y = -(Z \otimes Z )(X \otimes X)$. New $Y$ states can be prepared using a $|\bar{Y}\rangle = (|0\rangle - i|1\rangle)/\sqrt{2}$ catalyst as shown in Fig.~\ref{fig:pprs}c. Preparing $n$ $Y$ states costs $3n+1$ blocks, so the cost of a $Y$ state can be assumed to be 3 blocks, if $Y$ states are generated in batches. The initial $|\bar{Y}\rangle$ catalyst can be prepared at the very beginning of the quantum computation either via twist defects~\cite{Bombin2010,Brown2017,Litinski2018} or via magic state distillation.

Since a reactive $Y$ measurement only happens with a probability of 50\%, the cost of a $\pi/8$ rotation is $C_m + 1.5 + C_{|T\rangle}$, where $C_m = \lceil \frac{3}{2} w_x \rceil + \lceil \frac{3}{2} (w_z+1) \rceil + 1$ is the cost of the initial PPM, and 1.5 is half the cost of a $Y$ state. For every two $\pi/8$ rotations, we need to generate a $Y$ state. A stockpile of sufficiently many $Y$ states should be kept in memory, so that one does not run out of $Y$ states whenever many such states are needed at the same time due to unfavorable random measurement outcomes. $C_{|T\rangle}$ is the cost to prepare a $|T\rangle$ state. These states need to be prepared via magic state distillation, the cost of which depends on physical error rates and target logical error rates. In Sec.~\ref{sec:distillation}, we estimate that $C_{|T\rangle} \approx 25$ for reasonable error parameters.

Arbitrary-angle PPRs can be decomposed into sequences of $\pi/8$ rotations, e.g., using the methods in Ref.~\cite{Ross2014}. Here, each $Z_\varphi$ rotation can be approximately synthesized with an error $\varepsilon$ as a sequence of $3\log 1/\varepsilon$ rotations with angles $\varphi_c = c\cdot \pi/8$, where $c$ is an odd integer. The bases of these rotations alternate between $X$ and $Z$, as shown in Fig.~\ref{fig:pprs}d. For a $P_\varphi$, rotation, the $Z$ operator is copied onto an ancilla qubit via a $P \otimes Z$ measurement, and a sequence of single-qubit rotations is executed. Each pair of rotations has an active volume of $8 + 2C_{|T\rangle}$, as shown in Fig.~\ref{fig:pprs}e. Therefore, the active volume of an arbitrary-angle PPR using the method of Ref.~\cite{Ross2014} is $C_m + 3\log 1/\varepsilon \cdot (4 + C_{|T\rangle})$. Since consecutive $X$ and $Z$ rotations anticommute, the reaction depth of this operation is $3\log 1/\varepsilon$. In other words, the $3\log 1/\varepsilon$ stale $T$ states generated by the PPR need to be reactively measured sequentially, as the outcome of a reactive measurement generates a Pauli correction that is required to determine the basis of subsequent reactive measurements.

\begin{figure*}[t!]
\centering
\includegraphics[width=\linewidth]{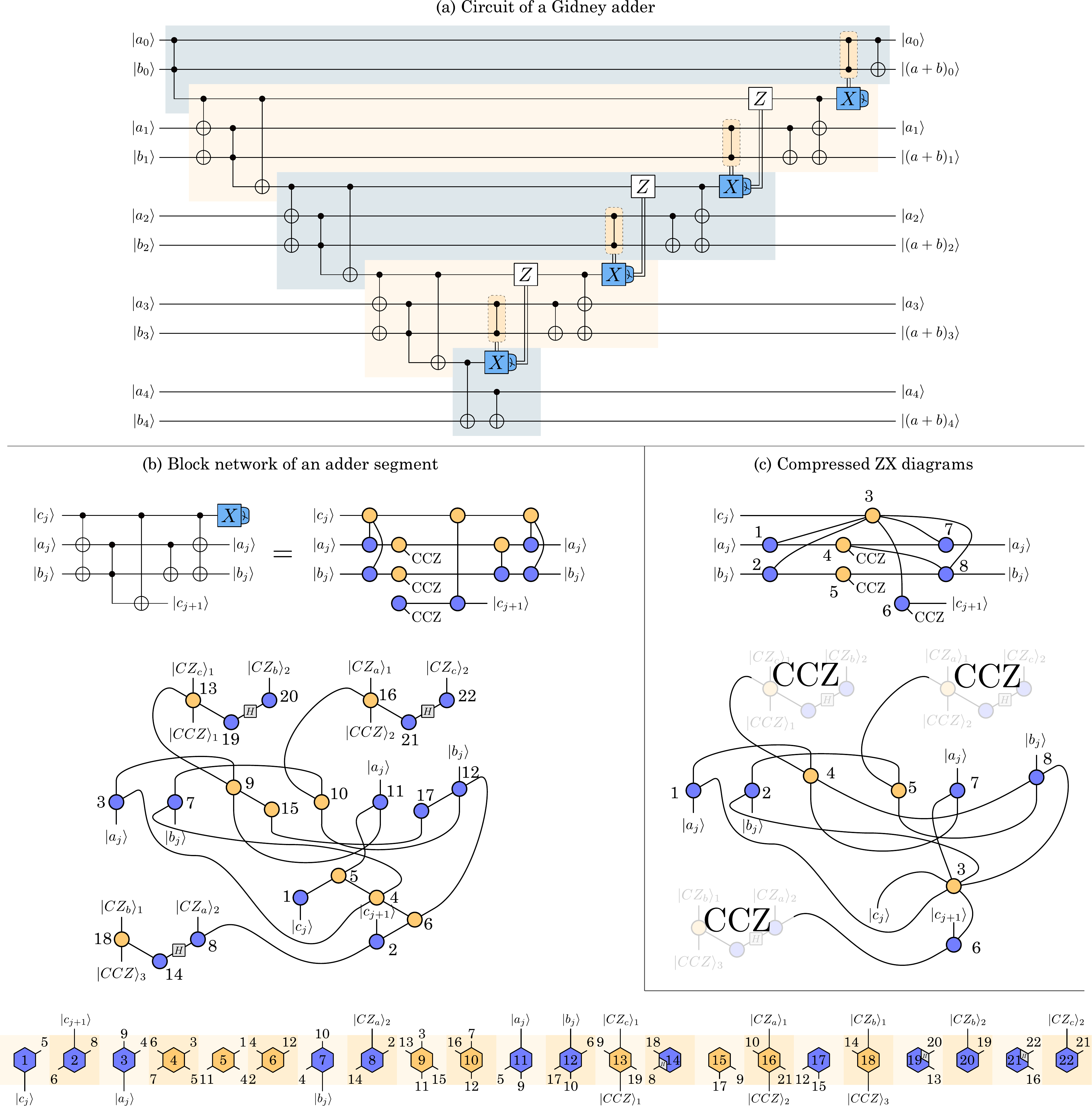}
\caption{(a) Modified version of the ripple-carry adder circuit shown in Ref.~\cite{Gidney2018}. (b) Each repeating segment of this adder has an active volume of 22 blocks in addition to the volume required to generate a CCZ state. (c) The compressed ZX diagrams can be used to verify that both diagrams describe the same operation.}
\label{fig:addercircuit}
\end{figure*}

\begin{figure*}[t]
\centering
\includegraphics[width=0.65\linewidth]{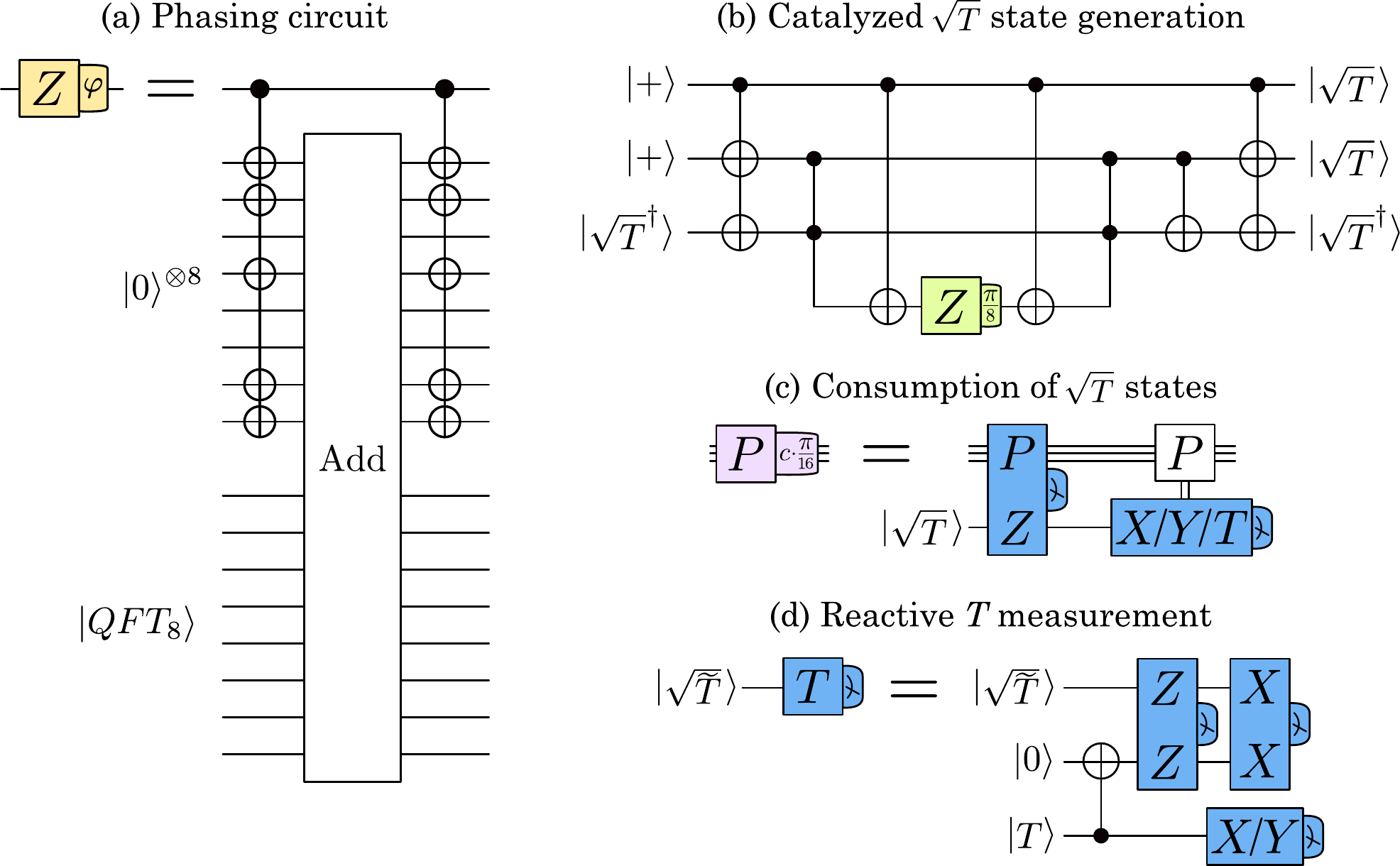}
\caption{(a) An arbitrary-angle rotation with the angle specified using $b$ bits of precision can be executed via an addition into a phase-gradient state. The example shows a rotation with an angle $\varphi = (0.11001011)_2 \cdot \pi$. (b) Adders can also be used to generate $\sqrt{T}$ states using a $\sqrt{T}^\dagger$ state as a catalyst~\cite{Gidney2018a}. (c) Such states can be consumed to execute $\pi/16$ rotations. (d) Depending on the outcome of the measurement used to consume the $\sqrt{T}$ state, a reactive $T$ measurement may be required.}
\label{fig:adderpprs}
\end{figure*}

\begin{figure*}[t]
\centering
\includegraphics[width=0.9\linewidth]{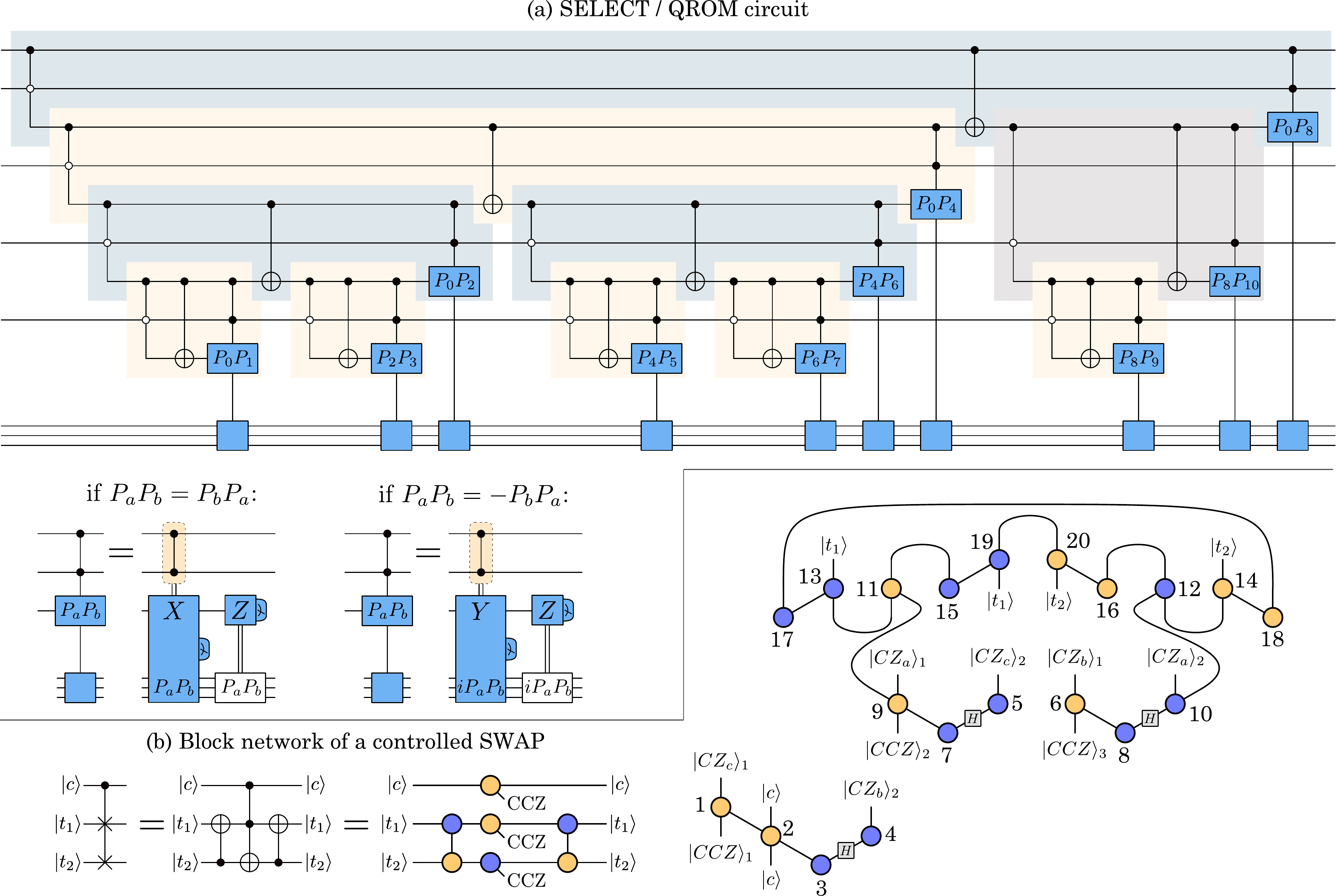}
\caption{(a) Modified version of the SELECT circuit from Fig. 7 of Ref.~\cite{Babbush2018}. Each segment consists of a temporary AND compute-uncompute pair, a CNOT, and a Pauli product measurement. A QROM circuit corresponds to a SELECT with $X$-type Pauli measurements. (b) A controlled SWAP has an active volume of 20 blocks.}
\label{fig:select}
\end{figure*}

\section{Toffoli gates, adders and data loaders}

Next, we consider circuits containing Toffoli gates. While it is possible to decompose Toffoli gates into four $T$ gates~\cite{Jones2013}, it can be cheaper to execute Toffoli gates by consuming $|CCZ\rangle$ resource states instead of $T$ states. These are three-qubit states $\mathrm{CCZ}|+\rangle^{\otimes 3}$, where CCZ is a controlled-controlled-$Z$ gate. Such states can be consumed to execute a Toffoli gate via the circuit shown in Fig.~\ref{fig:toffoli}a. The outcomes of the three PPMs used to consume the CCZ state determine the presence or absence of a CZ Clifford gate. Such a conditional CZ gate can be converted into a reactive measurement using the circuit in Fig.~\ref{fig:toffoli}b in a construction similar to AutoCCZ states~\cite{Gidney2019}. This 5-block operation generates a pair of qubits $|CZ\rangle_1$ and $|CZ\rangle_2$ that are stored in memory. These qubits can be used to retroactively teleport a CZ gate into the circuit. If a CZ gate needs to be generated, the qubit pair is removed via a Bell-basis measurement, otherwise via two single-qubit $X$ and $Z$ measurements. This converts the decision about the conditional CZ gate into a reactive measurement. Therefore, the full circuit for the execution of a Toffoli gate in Fig.~\ref{fig:toffoli}c generates 6 output qubits that are used for reactive measurements. The active volume of a Toffoli gate is 12 blocks with a reaction depth of 1.

\textbf{Temporary-AND ancilla qubits.} In many circuits, the target qubit of a Toffoli gate is an ancilla qubit initialized in the $|0\rangle$ state. Such temporary-AND Toffolis~\cite{Gidney2018} can be executed with a reduced cost of 9 blocks, as shown in Fig.~\ref{fig:tempand}a. Typically, temporary-AND Toffolis come in compute-uncompute pairs. The uncomputation of the Toffoli can be performed via a single-qubit measurement and a conditional CZ gate, see Fig.~\ref{fig:tempand}b. In many situations, the conditional CZ commutes with all operations between the two Toffoli gates of the compute-uncompute pair. The entire compute-uncompute pair can then be treated as a standard Toffoli gate, except that one of the conditional CZs requires the outcome of the $X$ measurement used to uncompute the Toffoli gate, as shown in Fig.~\ref{fig:tempand}c. Therefore, the compute-uncompute pair has an active volume of 12 blocks.

\textbf{Ripple-carry addition.} Such compute-uncompute pairs were used by Gidney in Ref.~\cite{Gidney2018} to construct an $n$-qubit in-place ripple-carry adder using $n-1$ Toffoli gates. A slightly modified version of this circuit is shown in Fig.~\ref{fig:addercircuit}a for the example of $n=5$. The circuit consists of $n-2$ identical segments, and a different first and last segment. Each of the $n-2$ segments can be converted into a network of 22 blocks as shown in Fig.~\ref{fig:addercircuit}b. The ZX diagram looks complicated, but we can confirm that it is identical to the depicted circuit by comparing the compressed ZX diagrams in Fig.~\ref{fig:addercircuit}c and verifying that they are indeed identical. Each adder segment inputs a carry qubit $|c_j\rangle$ that is destroyed, and generates a different carry qubit $|c_{j+1}\rangle$. This qubit may be the input of a different segment. If these segments are generated simultaneously, bridge qubits (Fig.~\ref{fig:memstate1}b) can be used to connect the different segments. Note that the labels of connected blocks differ by at most 6, so that a range of $r = 12$ is sufficient to implement this network of logical blocks.

The first and last segment of an adder have an active volume of $15 + C_{|CCZ\rangle}$ and 4, respectively, as shown in Fig.~\ref{fig:adderfirstlast}. Here, $C_{|CCZ\rangle}$ is the cost to distill a CCZ state. In Sec.~\ref{sec:distillation}, we estimate that $C_{|CCZ\rangle} \approx 35$ for reasonable error parameters. The total active volume of an $n$-qubit Gidney adder is therefore $(n-1)(22 + C_{|CCZ\rangle})-3$ with a reaction depth of $2n-3$.

\textbf{PPRs via addition.} The active volume of an adder also has implications for the cost of arbitrary-angle PPRs. Using a phase-gradient state as a catalyst, adders can be used to perform single-qubit rotations~\cite{Gidney2018}. A phase-gradient state is an $n$-qubit state 
\begin{equation}
|QFT_n\rangle = \bigotimes_{j=0}^{n-1} \frac{|0\rangle + e^{-i\pi/2^j} |1\rangle}{\sqrt 2} \, .
\end{equation}
Since these qubits are catalysts, these states only need to be prepared at the beginning of the quantum computation (e.g., via the methods in Fig.~\ref{fig:pprs}) and are then stored in memory until the end of the computation. A single-qubit $Z$-rotation with an angle $\varphi$ specified by $b$ bits of precision can be executed by performing a $b$-qubit addition, as shown for the example of $b=8$ and $\varphi = (0.11001011)_2 \cdot \pi$ in Fig.~\ref{fig:adderpprs}a. The initial CNOT copies the $Z$ observable onto a subset of the 8 ancilla qubits. Since these operations can be realized with two-qubit $Z$ measurements as in Fig.~\ref{fig:weighttwozmeas}, they have an active volume of $b/2+1$ for a random $b$-bit number with a Hamming weight of $b/2$. The CNOTs for the uncomputation are free, as they can be realized by single-qubit $X$ measurements. The total cost of a $b$-bit precision PPR is therefore $C_m + (b-1)(22.5 + C_{|CCZ\rangle})-3.5$ with a reaction depth of $2b-3$. With $b \approx \log 1/\varepsilon$, this has a lower depth compared to the sequence of $\pi/8$ rotations in Fig.~\ref{fig:pprs}d and, with $C_{|T\rangle} \approx 25$ and $C_{|CCZ\rangle} \approx 35$, a lower active volume of $\approx 57.5b$ (compared to $\approx 87b$).

\textbf{PPRs via $\sqrt{T}$ gates.} Adder circuits can be used to construct even cheaper PPRs by using the methods introduced in Ref.~\cite{Kliuchnikov2022}. Here, each arbitrary-angle single-qubit rotation with an error $\varepsilon$ can be decomposed into a sequence of $0.6\log 1/\varepsilon$ single-qubit $X/Y/Z$ rotations, half of which are rotations with an angle $\varphi = c \cdot \pi/8$, and the other half with $\varphi = c \cdot \pi/16$, where $c$ is an odd integer. The $\pi/16$ rotations can be executed using $|\sqrt{T}\rangle = (|0\rangle + e^{i\pi/8}|1\rangle)/\sqrt{2}$ states. Such states can be generated in pairs using a $|\sqrt{T^\dagger}\rangle = (|0\rangle + e^{-i\pi/8}|1\rangle)/\sqrt{2}$ catalyst state via an adder-type circuit~\cite{Gidney2018a} as shown in Fig.~\ref{fig:adderpprs}b. This is an adder segment and a $T$ gate, and therefore has an active volume of $25.5 + C_{|CCZ\rangle} + C_{|T\rangle}$, producing two $\sqrt{T}$ states. 

A $\pi/16$ rotation can be executed by consuming a $\sqrt{T}$ state as shown in Fig.~\ref{fig:adderpprs}c. Depending on the outcome of the $P\otimes Z$ measurement, we may need to apply a $T$ gate to the consumed (stale) $\sqrt{T}$ state. We refer to this as a \textit{reactive $T$ measurement}. As shown in Fig.~\ref{fig:adderpprs}d, it can be performed in two steps using a $T$ state encoded in a two-qubit repetition code, which can be prepared via a $Z \otimes Z$ measurement with a volume of 2 blocks. The stale $\sqrt{T}$ state is Bell-measured with one half of the repetition code. Based on the measurement outcome, the remaining qubit is measured in the $X$ or $Y$ basis. Since a reactive $Y$ measurement is needed with a 50\% probability, and a reactive $T$ measurement with a cost of $2 + C_{|T\rangle}$ is needed with a 50\% probability, the total cost of a $\pi/16$ rotation is $C_m + 15.25 + \frac{1}{2}C_{|CCZ\rangle} + C_{|T\rangle}$. The reaction depth is 1.5, as it is 2 if a $T$ measurement is required, and 1 otherwise.

For single-qubit rotations, we can set $C_m = 2$ for $Z$ rotations, $C_m = 3$ for $X$ rotations, and $C_m = 8$ for $Y$ rotations. For uniformly random $X$, $Y$ and $Z$ rotations, $C_m = 13/3$ on average. The average cost of each $\pi/8$ rotation is therefore $13/3 + 3/2 + C_{|T\rangle} = 35/6 + C_{|T\rangle}$. Similarly, the average cost of each $\pi/16$ rotation is $235/12 + \frac{1}{2}C_{|CCZ\rangle} + C_{|T\rangle}$. With $0.3\log 1/\varepsilon$ $\pi/8$ rotations and $0.3\log 1/\varepsilon$ $\pi/16$ rotations, the total cost of an arbitrary-angle PPR is $C_m + \frac{1}{40}\log 1/\varepsilon \cdot (305 + 6 C_{|CCZ\rangle} + 24 C_{|T\rangle})$  with a reaction depth of $0.75 \log 1/\varepsilon$. For $C_{|T\rangle} \approx 25$ and $C_{|CCZ\rangle} \approx 35$, this method is significantly cheaper than the previously mentioned methods, with a cost of $\approx 28 \log 1/\varepsilon$ per PPR.

\textbf{Controlled adders.} Using the construction of Ref.~\cite{Gidney2018}, a controlled adder uses twice as many Toffoli gates as an uncontrolled adder. The segments are shown in Fig.~\ref{fig:ctrladder}a-c. The active volume of a controlled adder is $(n-1)\cdot(30 + 2 C_{|CCZ\rangle}) + 9 + C_{|CCZ\rangle}$. Controlled adders can be used to construct a quantum Fourier transform (QFT). As shown in Fig.~\ref{fig:ctrladder}d, a QFT is a sequence of Hadamard gates and controlled rotations with angles $\pi/2^n$. An entire set of $n$ controlled rotations can be performed via a controlled addition into an $n+1$-qubit phase-gradient register, as shown in Fig.~\ref{fig:ctrladder}e. The active volume of an $n$-qubit QFT is therefore $(n^2-1)\cdot (15 + C_{|CCZ\rangle}) - 3n + 1$.

\textbf{Out-of-place adders.} As shown in Fig.~\ref{fig:outofplaceadder}, the cost of an out-of-place Gidney adder~\cite{Gidney2018} is $21 + C_{|CCZ\rangle}$ the compute block, and $18$ for the uncompute block. Such out-of-place adders can be used to efficiently execute sets of $n$ commuting PPRs with identical angles by performing $\approx n$ out-of-place additions and $\log n$ arbitrary-angle $Z$ rotations using a technique called Hamming weight phasing~\cite{Gidney2018,Kivlichan2020}. Therefore, the active volume of $n$ commuting equiangular PPRs is \linebreak $\approx (C_m + 39 + C_{|CCZ\rangle})\cdot n + \mathcal{O}(\log n \cdot C_{\rm rot})$, where $C_{\rm rot}$ is the cost of an arbitrary-angle single-qubit $Z$ rotation.

\textbf{SELECT and QROM.} Other circuits that can be constructed from temporary-AND Toffolis are data loaders which are widely used in various algorithms, e.g., in block-encoding circuits~\cite{Low2017,Low2019,Babbush2018}. The first type of data loader is a SELECT operation, where
\begin{equation}
\mathrm{SELECT} = \sum\limits_{k=1}^n |k\rangle \langle k| \otimes P_k \,
\end{equation}
applies one of $n$ Pauli operators $P_k$ to a target register controlled on a $\log n$-qubit control register. Using a slightly modified version of the circuit constructed in Fig. 7 of Ref.~\cite{Babbush2018}, a SELECT operation can be implemented as shown in Fig.~\ref{fig:select} for the example of $n=11$. It consists of $n-1$ segments, each containing a temporary-AND compute-uncompute pair, a CNOT, and a PPM. We can treat these as individual operations, such that the active volume of a SELECT operation is $(n-1)\cdot (13 + C_m + C_{|CCZ\rangle})$, where $C_m$ is the average cost of the PPMs.

If the Pauli operators $P_k$ are $X$-type operators acting on a $b$-qubit register, the same circuit can be used as a ``QROM read'' loading $n$ $b$-bit numbers into the quantum computer. The weight of the $X$-type operators is the Hamming weight of the $b$-bit numbers, so the PPMs will be weight-$b/2$ measurements on average. With $C_m \approx \frac{3}{4}b + 2$, the cost to load $n$ $b$-bit numbers via QROM is $(n-1)\cdot (15 + \frac{3}{4}b + C_{|CCZ\rangle})$. Using the construction in Ref.~\cite{Low2018}, it is possible to reduce the number of Toffoli gates by increasing the number of $b$-bit numbers that are loaded simultaneously. Effectively, the circuit in Ref.~\cite{Low2018} is a QROM loading $n/\lambda$ different $\lambda b$-bit numbers, preceded by a circuit of $b\cdot(\lambda - 1)$ controlled SWAP gates, where $\lambda$ is a tunable integer parameter. As shown in Fig.~\ref{fig:select}b, the active volume of a controlled SWAP gate is $20 + C_{|CCZ\rangle}$. Therefore, the active volume of a QROM read with the construction of Ref.~\cite{Low2018} is $(n/\lambda-1)\cdot (15 + \frac{3}{4}b\lambda + C_{|CCZ\rangle}) + b \cdot (\lambda - 1)\cdot (20 + C_{|CCZ\rangle})$.

Note that, regardless of $\lambda$, the active volume always contains a term proportional to $n \cdot b$, i.e., the total number of classical bits loaded into the quantum computer. While this contribution is due to large PPMs that do not consume non-Clifford resource states, and would be considered cheap in baseline architectures where the cost is primarily determined by the total number of $T$ gates and Toffoli gates, the scaling with $n \cdot b$ can make QROMs considerably more expensive than arithmetic circuits with the same number of Toffoli gates. For example, for $C_{|CCZ\rangle} = 35$, the per-Toffoli cost of an adder is 57 blocks, whereas the per-Toffoli cost of a 1000-bit QROM read is 800.

\begin{figure*}[t]
\centering
\includegraphics[width=\linewidth]{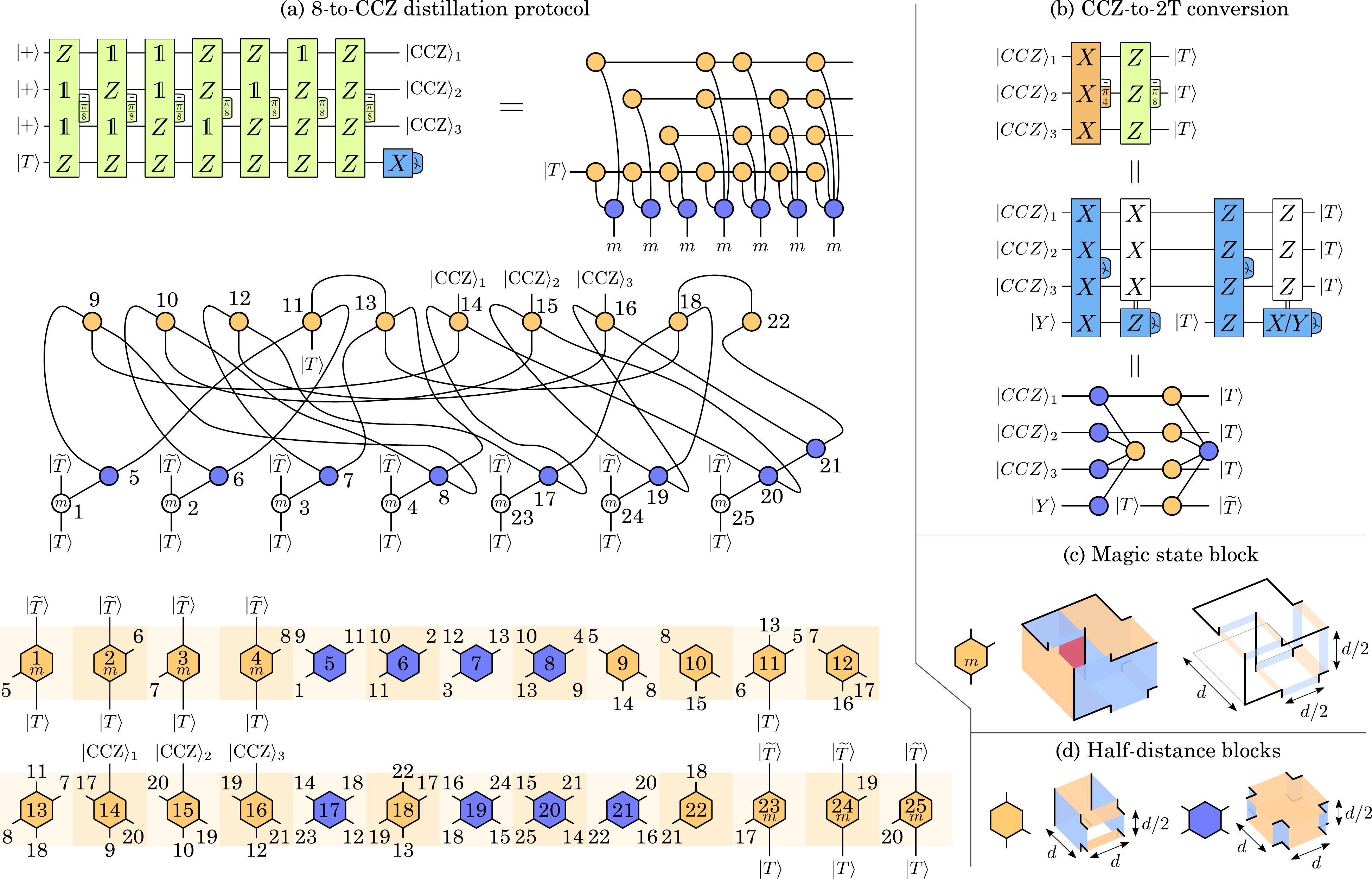}
\caption{(a) An $\text{(8-to-CCZ)}_{d,d,d/2}$ distillation protocol~\cite{Litinski2019a} has an active volume of 25 half-blocks, i.e., 12.5 blocks. It uses 8 half-distance $T$ states as inputs. Distilled CCZ states can be converted to two distilled $T$ states using 12 blocks and a $T$-state catalyst. This is a modified version of the circuit in Fig. 10 in Ref.~\cite{Gidney2018a}. (c) The distillation protocol uses magic state blocks that consume a magic state and apply an $X_{\pi/4}$ rotation, turning the reactive $X/Y$ measurements into $X/Z$ measurements. (d) All blocks are half-distance blocks with a reduced code distance in the time (up-down) direction.}
\label{fig:8toccz}
\end{figure*}

\begin{figure*}[t!]
\centering
\includegraphics[width=0.98\linewidth]{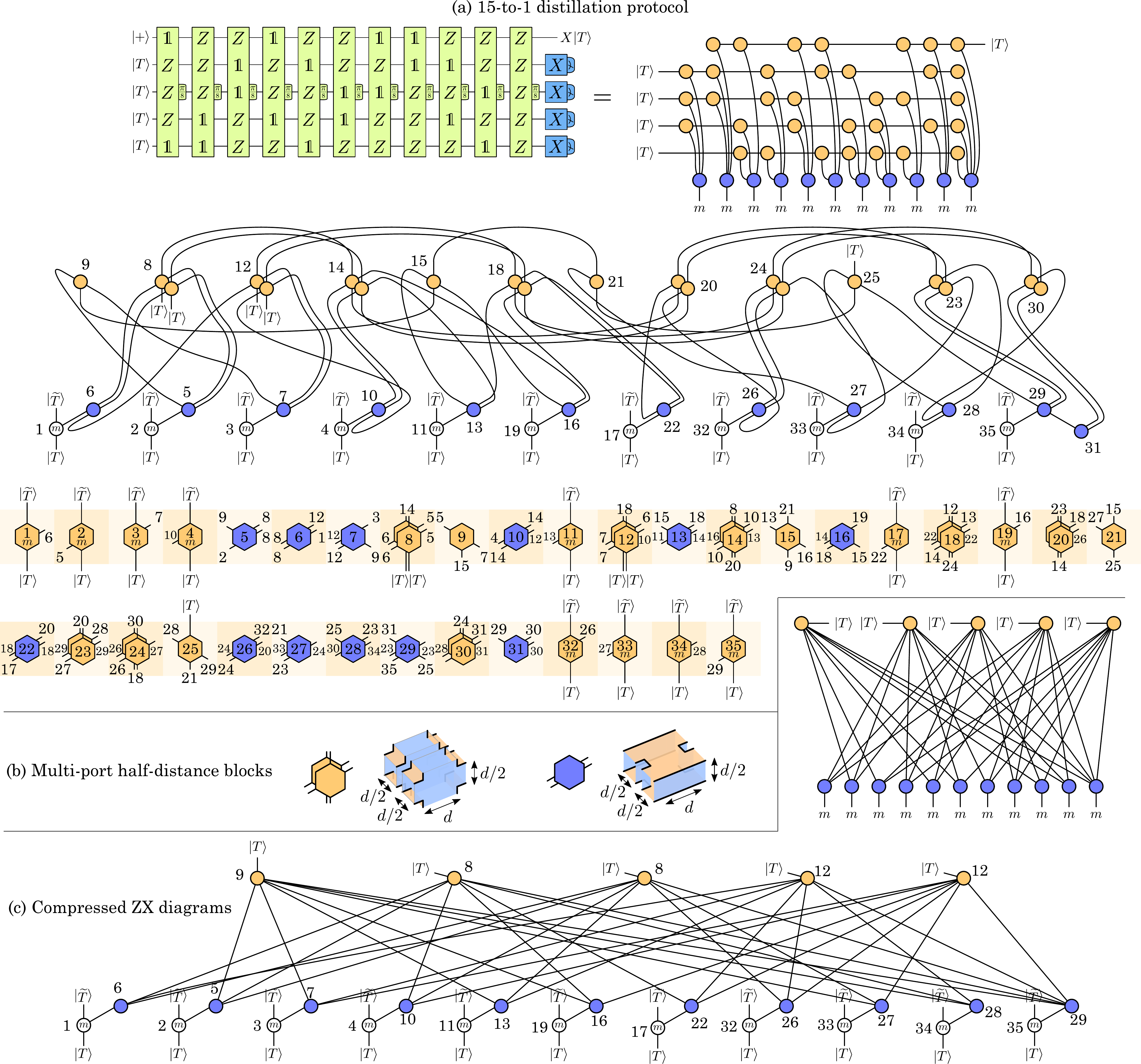}
\caption{(a) A $\text{(15-to-1)}_{d,d/2,d/2}$ distillation protocol~\cite{Litinski2019a} has an active volume of 35 half-blocks, i.e., 17.5 blocks. It uses 15 half-distance $T$ states as inputs. Note that, in the 15-to-1 circuit, an $X$ Pauli correction must be applied to the output state to obtain the correct $|T\rangle$ state. (b) Because the $Z$ distance of some qubits is reduced, two blocks can be generated simultaneously by one workspace qubit. (c) The compressed ZX diagrams can be used to verify that the diagrams in (a) are identical. The diagrams correspond to the Tanner graph of a [15,1,3] Reed-Muller code.}
\label{fig:15to1}
\end{figure*}

\section{Magic state distillation}
\label{sec:distillation}

Many of the previously discussed operations consume $T$ states or CCZ states. These states need to be prepared via magic state distillation~\cite{Bravyi2005,Bravyi2012,Haah2018,Gidney2018a,Litinski2019a}. There exist (in principle, infinitely) many distillation protocols in the literature. Here, we consider only two protocols: 8-to-CCZ distillation which produces a distilled CCZ state from 8 noisy $T$ states, and 15-to-1 distillation which produces a distilled $T$ state from 15 noisy $T$ states. With surface codes, it is possible to prepare noisy $T$ states with an error rate proportional to the physical error rate using a protocol called state injection~\cite{Li2015}, which typically only requires physical $T$ gates in addition to standard surface-code operations. Since injected $T$ states typically have very high error rates, it is necessary to produce higher-quality $T$ states and CCZ states via magic state distillation.

\textbf{8-to-CCZ distillation.} Magic state distillation protocols can be constructed as quantum circuits consisting of $Z$-type $\pi/8$ rotations. An example is the 8-to-CCZ circuit in Fig.~\ref{fig:8toccz}a. It is a sequence of 7 PPRs which are executed using noisy $T$ states. An additional noisy $T$ state is used as an input qubit in the quantum circuit. The $X$ measurement at the end of the circuit is used to detect errors: If the outcome is $X = -1$, the output qubits are discarded. If the outcome is $X=+1$, the three output qubits constitute a distilled CCZ states with a quadratically suppressed error rate $\sim p^2$, where $p$ is the error rate of the noisy input $T$ states.

Because the input magic states are noisy and distillation protocols are error-detecting circuits, it is possible to significantly reduce the cost of distillation by reducing the code distances of various parts of the protocol~\cite{Litinski2019a}. The optimal choice of code distances depends on the physical error rate, target logical error rate, and the scaling of the logical error rate with the code distance. However, in Ref.~\cite{Litinski2019a}, it was observed that a reasonable operating regime is approximately the following: qubits corresponding to output magic states are encoded as $d_X \times d_X$ surface-code patches, input $T$ states in the circuit as $d_Z \times d_X$ patches, all measurements are performed with a temporal code distance of $d_m$, and input $T$ states that are used in PPRs as $d_m \times d_m$ surface-code patches. An $n$-to-$k$ distillation protocol with such parameters can then be labeled as $\text{($n$-to-$k$)}_{d_X,d_Z,d_m}$. When optimizing the code distances to reduce the volume of the distillation protocols, one often finds $d_X = d$, $d_Z \approx d/2$, $d_m \approx d/2$.

We first consider an $\text{(8-to-CCZ)}_{d,d,d/2}$ protocol. Because the temporal code distance is $d_m = d/2$, all logical blocks will be half-distance blocks as shown in Fig.~\ref{fig:8toccz}d, i.e., they will be executed in half of a logical cycle. The input magic states are half-distance qubits. Half-distance qubits are supported by the active-volume architecture, as we can store four half-distance qubits in each memory location, one in each quadrant of the full-distance patch. However, when we execute a PPR using a half-distance magic state, the stale half-distance $T$ state needs to participate in a reactive $Y$ measurement. Since we want to avoid storing half-distance $Y$ states in memory (as they have a significantly higher error rate), we use the block in Fig.~\ref{fig:8toccz}c to consume a magic state. This corresponds to an operation that consumes the magic state and applies an $X_{\pi/4}$ Clifford rotation to the stale $T$ state, changing the $X/Y$ measurement to an $X/Z$ measurement. (Note that this operation is identical to the auto-corrected $\pi/8$ rotations described in Ref.~\cite{Litinski2019}.) Furthermore, we break with our convention that all memory qubits are stored in the orientation of Fig.~\ref{fig:logicalblockintro}e where the logical $Z$ operators point in the north and south direction. The volume of distillation protocols can be reduced, if input magic states are stored in a rotated manner with the $Z$ operators pointing in the west and east direction. Therefore, the input qubits in a distillation protocol will feed into N-oriented $Z$-type blocks, rather than E-oriented $Z$-type blocks as would be required by the usual convention. Output qubits of distillation protocols, however, will follow the usual convention.

As shown in Fig.~\ref{fig:8toccz}, the 8-to-CCZ $\text{(8-to-CCZ)}_{d,d,d/2}$ protocol can be implemented by generating 25 half-distance blocks, i.e., with an active volume of 25/2. Note that, if necessary, the error resilience of the protocol can be increased by increasing the measurement distance $d_m$. Distilled CCZ states can also be converted to distilled $T$ states via the catalyzed CCZ-to-2T conversion of Ref.~\cite{Gidney2018a}, a modified version of which is shown in Fig.~\ref{fig:8toccz}b. The circuit corresponds to a 4-qubit $X$-type and qubit $Z$-type measurement, consuming a $Y$ state and a $T$ state with a reactive $Y$ measurement. Therefore, the CCZ-to-2T conversion has an active volume of 16.5 blocks.

\textbf{15-to-1 distillation.} In a similar way, we can construct a $\text{(15-to-1)}_{d,d/2,d/2}$ protocol shown in Fig.~\ref{fig:15to1}a. Here, we also reduce the $Z$ distance to $d_Z = d/2$, which means that some qubits will be encoded in rectangular $d \times d/2$ surface-code patches. A workspace qubit can generate two such qubits using multi-port half-distance blocks as shown in Fig.~\ref{fig:15to1}b. Therefore, multi-port blocks have pairs of ports on the east and west side of the block. Note that ports on the same side can be connected to different blocks, but they must be connected to blocks on the same side, e.g., the top east port of one block can only be connected to the top east port of another block, but not the bottom east port.

The resulting network of logical blocks in Fig.~\ref{fig:15to1}b is very complicated, but the compressed ZX diagrams of this network and of the original circuit can be used to verify that the operations are identical, as shown in Fig.~\ref{fig:15to1}c. Note that these compressed ZX diagrams correspond to the Tanner graph of a [15,1,3] Reed-Muller code, as the 15-to-1 distillation protocol is based on this code. The $\text{(15-to-1)}_{d,d/2,d/2}$ protocol can be implemented with 35 half-distance blocks, i.e., an active volume of 35/2. Remarkably, this protocol can be implemented with a range of $r=12$, as connected blocks are at most 6 workspace qubits apart.

\textbf{Multiple stages of distillation.} Typically, one stage of distillation will not be enough to produce sufficiently high-quality magic states. For example, if we need to produce Toffoli states with an error rate below $10^{-10}$, the input $T$ states in the 8-to-CCZ protocol need to have an error rate below $10^{-6}$. However, if noisy $T$ states produced by state injection have an error rate of $10^{-3}$, the input states to the 8-to-CCZ protocol need to be generated by an initial stage of distillation, e.g., via a 15-to-1 protocol using injected $T$ states as inputs.

The code distances used in the first stage of distillation can be reduced even further~\cite{Litinski2019a}, e.g., by using a $\text{(15-to-1)}_{d/2,d/4,d/4}$ distillation protocol. Here, all distances are halved compared to the protocol in Fig.~\ref{fig:15to1}a. In an active-volume architecture, we can use 35 workspace qubits to execute four instances of such a protocol simultaneously, i.e., one instance per quadrant of the workspace qubits. Because the measurement distance is now $d/4$, the 35 workspace qubits produce 8 distilled $T$ states every $d/2$ code cycles. These can then be used by an additional 25 workspace qubits to produce a CCZ state in the second stage of distillation. Therefore, the active volume of a $\text{(15-to-1)}_{d/2,d/4,d/4} \times \text{(8-to-CCZ)}_{d,d,d/2}$  protocol is 30 blocks.

Whether or not this protocol is suitable to distill sufficiently high-quality CCZ states depends on the physical error rate, target logical error rate and scaling behavior of the logical error rate. While detailed numerical simulations are required to determine the precise logical error rate of these distillation protocols, we can perform a very rough (and inaccurate) estimate using the method described in Ref.~\cite{Litinski2019a}. We can then estimate the output error rate of the 8-to-CCZ protocol as
\begin{equation}
	p_{\text{(8-to-CCZ)}_{d,d,d/2}} \approx 28 \cdot \left( 4 p\left(\frac{d}{2}\right) + p_{\rm in} \right)^2 + 2p(d) \, ,
\end{equation}
and of the 15-to-1 protocol as
\begin{equation}
	p_{\text{(15-to-1)}_{d/2,d/4,d/4}} \approx 35 \cdot \left( 4 p\left(\frac{d}{4}\right) + p_{\rm in} \right)^3 + 2p\left(\frac{d}{2}\right) \, .
\end{equation}
Here, $p_{\rm in}$ is the error rate of the input magic states, and $p(d)$ is the logical error rate of a surface-code spacetime block of size $d \times d \times d$.\footnote{This very rough estimate is obtained by observing that the logical operator of each PPM that is used to consume an input magic state is supported in 8 spacetime blocks of size $(d/2)^3$. In other words, the logical membrane (or correlation surface) encoding the PPM outcome has an error rate of approximately $8p(d/2)$. A flipped PPM outcome implies that we perform a $P_{-\pi/8}$ rotation instead of a $P_{+\pi/8}$ rotation. Such an $S$-gate error generates a $Z$ flip with a 50\% probability in the distillation protocol, hence the $4p(d/2)$ contribution in addition to the error of the input magic states. The 15-to-1 protocol suppresses such $Z$ flips with $p \rightarrow 35p^3$ and the 8-to-CCZ protocol with $p \rightarrow 28p^2$. In addition, each output qubit accumulates an idling error of $2p(d)$ that scales with the large code distance.}

As an example, we can assume that $p(d) = 10^{-d/2}$, as can be expected when operating at 10\% of the surface-code error threshold~\cite{Bombin2021a}. Suppose that we need to execute a quantum computation with $10^9$ Toffoli gates. If the computation primarily consists of adders, the active volume of the computation will be $\approx 10^{11}$ blocks. If we want to keep the probability of an error at the end of the computation below 1\%, we need to execute our computation with a full distance such that $p(d) < 10^{-13}$, e.g., we may use a quantum computer with $d=28$. We also need to generate CCZ states with an error rate of $\approx 10^{-11}$. With an injection error rate of $p_{\rm in} = 10^{-3}$, the first-stage 15-to-1 protocol produces $T$ states with an error rate of $6 \times 10^{-7}$ according to our rough estimate. The second-stage 8-to-CCZ protocol then produces CCZ states of $2.8 \times 10^{-11}$, which is close to the target error rate. 

We emphasize again that this is a very inaccurate estimate, and a detailed numerical study is required to obtain more accurate estimates. However, note that there is a lot of room for improvement, as distillation protocols can be optimized by tuning the distances and by considering different combinations of distillation protocols in addition to 15-to-1 and 8-to-CCZ protocols. In the context of this paper, the main goal is to obtain a rough estimate of the cost of a CCZ state. The active volume of the two-stage protocol described above is 30 blocks. Some extra volume will be required for state injection, and possibly to increase the measurement distance above $d/2$. We then estimate that it is reasonable to assume that a CCZ state can be distilled with a cost of $C_{|CCZ\rangle} \approx 35$. Furthermore, distilled CCZ states can be converted to two $T$ states with an extra cost of 16.5. Therefore, we also estimate that the cost of a $T$ state can be assumed to be $C_{|T\rangle} \approx 25$.

\section{Active-volume compilation}
\label{sec:compilation}

In the previous sections, we described how to translate various subroutines and distillation protocols into networks of logical blocks. In this section, we discuss how this can be used to translate full quantum computations into instructions for the workspace and memory modules of an active-volume quantum computer.

\begin{figure}[t]
\centering
\includegraphics[width=0.98\linewidth]{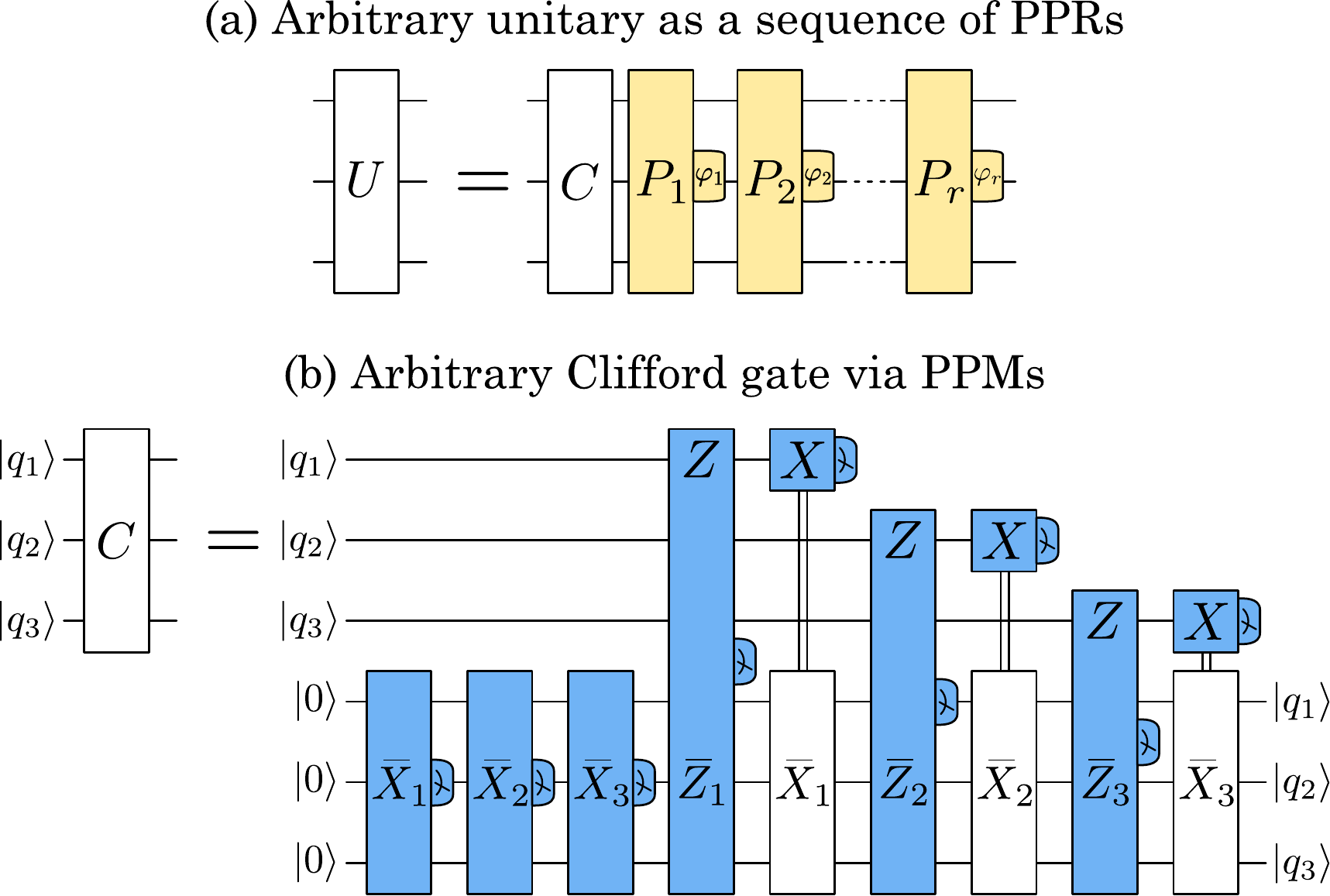}
\caption{(a) Any $n$-qubit quantum circuit consisting of Clifford gates and $n_r$ single-qubit rotations can be converted to $n_r$ arbitrary-angle PPRs and an $n$-qubit Clifford operation. (b) Any $n$-qubit Clifford operation can be implemented with $2n$ Pauli product measurements. Here, the Clifford operation maps $X_j \rightarrow \bar{X}_j$ and $Z_j \rightarrow \bar{Z}_j$.}
\label{fig:customsubroutine}
\end{figure}

\textbf{Generation of logical block networks.} If a quantum computation exclusively consists of standard subroutines such as adders, QROM reads, QFTs, PPRs and PPMs, then it can be translated into a sequence of operations described by networks of logical blocks, which can then be executed by workspace modules. However, an arbitrary quantum computation may contain custom subroutines in addition to such standard subroutines. Such custom subroutines can be translated into logical block networks using the general prescription shown in Fig.~\ref{fig:customsubroutine}a. Using the PPR conversion procedure described in Ref.~\cite{Litinski2019}, any $n$-qubit quantum circuit consisting of Clifford gates and $n_r$ single-qubit rotations can be converted into a sequence of $n_r$ PPRs and one $n$-qubit Clifford gate.

This $n$-qubit Clifford gate is an operation that maps $X_j \rightarrow \bar{X}_j$ and $Z_j \rightarrow \bar{Z}_j$, where $1 \leq j \leq n$, and $\bar{X}_j$ and $\bar{Z}_j$ are $n$-qubit Pauli operators. Such a Clifford gate can be implemented using $2n$ PPMs as shown in Fig.~\ref{fig:customsubroutine}b. Here, we first prepare an eigenstate of all $\bar{X}_j$, and then teleport the data qubits into that state via $Z_j \otimes \bar{Z}_j$ measurements, which applies the desired Clifford operation. Random $n$-qubit PPMs have an active volume of $\approx 1.5n$ blocks. Therefore, a random $n$-qubit Clifford gate has an active volume of $\approx 3n^2$ blocks. The arbitrary unitary in Fig.~\ref{fig:customsubroutine}a then has an active volume of $\approx 3n^2 + n_r \cdot (1.5n + C_{\rm rot})$, where $C_{\rm rot}$ is the cost of an arbitrary-angle $Z$ rotation.

Ideally, most operations in a quantum computation will be optimized standard operations such as adders, since custom operations with a cost of $\mathcal{O}(n^2)$ are costly in comparison. However, if small portions of the computation use such custom subroutines, they can be compiled into PPMs and PPRs using the prescription in Fig.~\ref{fig:customsubroutine}. If some unsupported custom subroutines are used often enough that their cost becomes significant, they can be optimized by translating them into block networks using the methods described in earlier sections.

\begin{algorithm}
 \Repeat{\rm all qubits in target memory locations}{
 Mark all memory locations as eligible for quickswaps in this layer \\
 \For{\rm each qubit $i$ with a target location $j$}{
    \eIf{\rm (qubit $i$ already in target location)}{
Mark location $j$ as ineligible}{
	If possible, perform a valid quickswap between qubit $i$ and a memory location $k$ as close to $j$ as possible. Mark both the initial and final locations as ineligible for quickswaps in this layer.}
   }
 }
 \caption{Algorithm 1: A greedy quickswap algorithm for a separation of $s \geq 2$.}
\label{algo:quickswap}
\end{algorithm}

\begin{figure}[t]
\centering
\includegraphics[width=\linewidth]{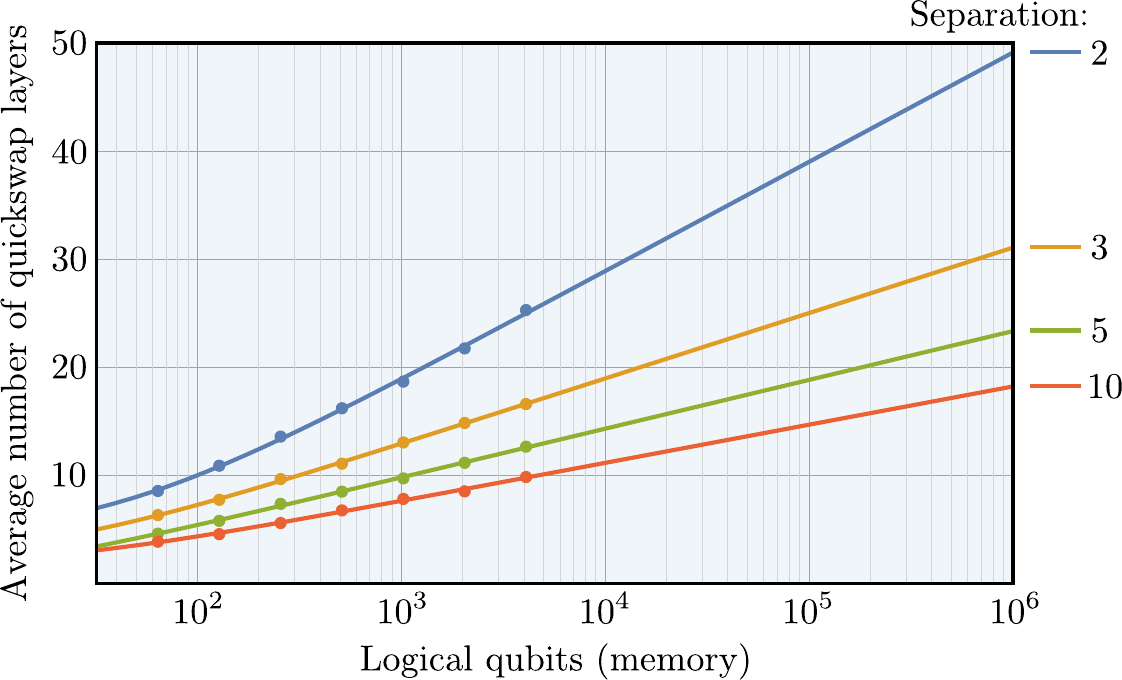}
\caption{Results of numerical simulations testing the efficiency of quickswaps in rearranging large memories. For a random configuration of an $n_q$-qubit memory, every $s$-th memory slot was assigned a target qubit that needs to be moved into this location, where $s$ is referred to as the separation. Layers of quickswaps were generated using the greedy algorithm described in the main text. The average number of quickswap layers required to move all relevant qubits into the target locations grows logarithmically with $n_q$. For $s=3$, even very large memories with a million logical qubits can be rearranged within 30 quickswap cycles on average, implying that, in this situation, the memory can be rearranged within a logical cycle, as long as the code distance is $d>30$.}
\label{fig:quickswapdata}
\end{figure}

\textbf{Management of resource states and catalysts.} In addition to the execution of logical blocks, workspace qubits are responsible for the generation of magic states and other resource states such as $Y$ states. The quantum computation should never stall due to a lack of available resource states. Therefore, a stockpile of resource states should be kept in memory. New resource states should be generated whenever the consumption of resource states from the stockpile is anticipated. In addition to data qubits and resource states, the memory also needs to store catalyst states such as phase-gradient states that are used to facilitate various logical operations. Such catalysts need to be generated at the very beginning of the computation when ``booting'' the quantum computer.

\textbf{Memory management.} In each logical cycle, workspace qubits execute logical blocks corresponding to logical operations and distillation protocols. Each configuration of logical blocks imposes certain conditions on the state of the memory at the end of the logical cycle, as explained in Sec.~\ref{sec:zmeasurements} and Fig.~\ref{fig:compstructure}. Some (but not all) data qubits, bridge qubits and resource states stored in memory will need to be moved to specific memory locations within one logical cycle. This can be done by generating sequences of quickswap layers as instructions for memory qubits. Quickswaps allow us to swap qubits stored in memory locations $i$ and $j$ within one code cycle, if $|i-j| = 2^{k}$, where $k$ is an integer.

In principle, a bitonic sorting network can rearrange an $n$-qubit memory via $\mathcal{O}(\log^2 n)$ quickswap layers, but this will typically not be fast enough. For example, a 1024-qubit memory would require $10\cdot 11/2 = 55$ layers of quickswaps, but this will be too slow if the code distance is $d < 55$. Fortunately, a complete rearrangement of the memory is rarely required, as only a subset of qubits have target memory locations. Consider a situation in which we are provided a randomly arranged $n$-qubit memory, and every $s$-th memory location is assigned a qubit that needs to be moved there by the end of the logical cycle. For example, in a computation consisting of adders, each adder segment has an active volume of $\approx 60$ blocks and has 6 input and 9 output qubits, in addition to the input and output qubits of the distillation protocol. In this situation, we can expect that the separation is $s \approx 3$. We can generate layers of quickswap operations via a simple greedy algorithm, as shown in Algorithm~\ref{algo:quickswap}.

By performing a numerical simulation of this algorithm for various sizes of $n$-qubit memories, we find that all relevant qubits can be moved into target locations within $\mathcal{O}(\log n)$ quickswap layers. The results are presented in Fig.~\ref{fig:quickswapdata}. The number of required quickswap layers increases with the separation. For a separation of $s=3$, a 2048-qubit memory can be rearranged in approximately 15 code cycles, implying that this strategy is sufficient for $d > 15$. In fact, extrapolating to extremely large memories, a million logical qubits can be rearranged in approximately 30 quickswap layers. Note that, if memories need to be rearranged faster, there are various possible improvements. Many target locations are required to store a resource state or be empty, in which case there is a freedom of choice that can be exploited when deciding which specific qubit to move into that location. Moreover, logical blocks of subsequent logical cycles are known ahead of time, which can be exploited by moving certain qubits close to target locations ahead of time. It is also possible to introduce additional quickswaps other than between qubits $i$ and $j$ with $|i-j| = 2^{k}$, e.g., quickswaps between qubits that are 3, 5 and 7 memory locations apart.

\textbf{Running out of memory.} Note that a significant fraction of qubits stored in memory may not be data qubits, but distilled magic states, stale magic states, catalysts and bridge qubits. For a quantum computation primarily consisting of adders, we can estimate that each 60-block adder segment is associated with 3 qubits of a distilled CCZ state, 6 qubits of stale CZ states, a bridge qubit for the carry and 2 qubits for the 8 half-distance $T$ states in the distillation, i.e., $\approx 12$ non-data qubits in memory. Thus, around 20\% of qubits stored in memory may be non-data qubits. Therefore, an active-volume quantum computer with $n$ qubits of memory may run out of memory, if a quantum computation has a memory requirement close to the maximum capacity.

However, since workspace qubits have the same functionality as memory qubits, we can use them as memory qubits whenever we run out of memory. Since this reduces the number of workspace qubits available for the generation of logical blocks, this will reduce the speed of the quantum computer. For example, if we need to borrow 20\% of workspace qubits as memory qubits, the speed of the quantum computer will be reduced by 20\%. Conversely, quantum computations with very low memory requirements may use unoccupied memory qubits as workspace qubits in order to speed up the quantum computation. Therefore, the estimate that an $n$-qubit active-volume quantum computer can execute a quantum computation with an active volume of $b$ blocks in $2b/n$ logical cycles is only an approximate estimate, as the precise number will depend on the state of the memory during the computation.

\textbf{Reactive measurements.} Finally, the only remaining task of memory qubits is the removal of stale $T$ states and CZ states via reactive measurements. Logical operations add such states to the memory, so they need to be removed as quickly as possible. However, since the choice of measurement basis depends on the outcome of previous measurements, the speed at which this can be done is governed by the reaction time $\tau_r$. This is the time that it takes to do all of the following steps: Perform a set of logical reactive measurements (single-qubit $X$ or $Z$, or two-qubit Bell-basis measurements), determine the measurement outcomes via a surface-code decoder, use the outcomes to determine the next set of reactive measurements, and send these instructions to the quantum computer. Therefore, it is impossible to execute a quantum computation with a reaction depth of $\delta_r$ faster than in a time $\delta_r \cdot \tau_r$. It is possible to execute logical blocks faster, but stale magic states will fill up the memory until no more memory is available. The final result of a quantum computation is unknown, until all stale magic states are removed via reactive measurements.

\section{Photonic active-volume interleaving modules}
\label{sec:photonicmodules}

In the previous sections, we have demonstrated that an active-volume quantum computer can execute quantum computations cost-efficiently, if it is capable of all the operations listed in Sec.~\ref{sec:overview}. In addition to standard operations required to store logical qubits in surface-code patches, these include quickswaps, lattice-surgery operations between between pairs of surface-code patches within range $r$, and transversal Bell-state preparations and measurements between pairs of surface-code patches within some range $r$.

These operations require physical two-qubit operations that are incompatible with a strictly 2D-local array of physical qubits, so some degree of non-locality is necessary. Remarkably, all these operations can be implemented in a photonic fusion-based~\cite{Bartolucci2021} quantum computer using the interleaving architecture of Ref.~\cite{Bombin2021} with small modifications. 

\textbf{Review of FBQC and interleaving.} Large million-physical-qubit fault-tolerant quantum computers are often thought of as two-dimensional arrays of static physical qubits with the ability to perform entangling two-qubit gates between nearest neighbors~\cite{Kitaev2003, Bravyi1998, Fowler2012}. Such a picture does not translate well to photonic qubits. While single photons can be used as qubits, a 2D array of one million single photons is not a million-qubit quantum computer, because photons have certain features that distinguish them from matter-based (non-photonic) qubits. Measurements of photons are destructive and entangling operations between photons are probabilistic when using linear optics. Therefore, rather than constructing a photonic quantum computer using static qubits and two-qubit gates, a different scheme is required. One such scheme is \textit{fusion-based quantum computing} (FBQC)~\cite{Bartolucci2021}. In this scheme, the two primary hardware components are \textit{resource-state generators} (RSGs) and \textit{fusion devices}. RSGs create identical, constant-size, few-photon resource states in periodic time intervals. Fusion devices are detectors for pairs of photons performing entangling two-photon Bell measurements (type-II fusions~\cite{Browne2005, GimenoSegovia2015}). All photons in the computation are created in an RSG and (ideally) destroyed in a fusion device. In FBQC, a surface-code quantum computation can be thought of as a 3D arrangement of resource states. Some photons in this arrangement participate in a fusion with a photon from a neighboring resource state, while others are part of a single-photon measurement, depending on the specific computation that is being executed. The FBQC-equivalent of a million-qubit fault-tolerant quantum computation is a 3D arrangement of a large number of resource states consisting of many 2D layers of \linebreak ${\sim}$1 million resource states each. Importantly, these resource states do not need to be generated all at once, but can be generated in multiple time steps. Photons that are produced in RSGs do not need to survive for the entire duration of the quantum computation, but only for as long as it takes to generate the next set of a few million photons. A naive approach is to use a network of 1 million RSGs to generate 1 million resource states at the same time, generating one 2D layer of the 3D arrangement after the other.

\begin{figure*}[t]
\centering
\includegraphics[width=\linewidth]{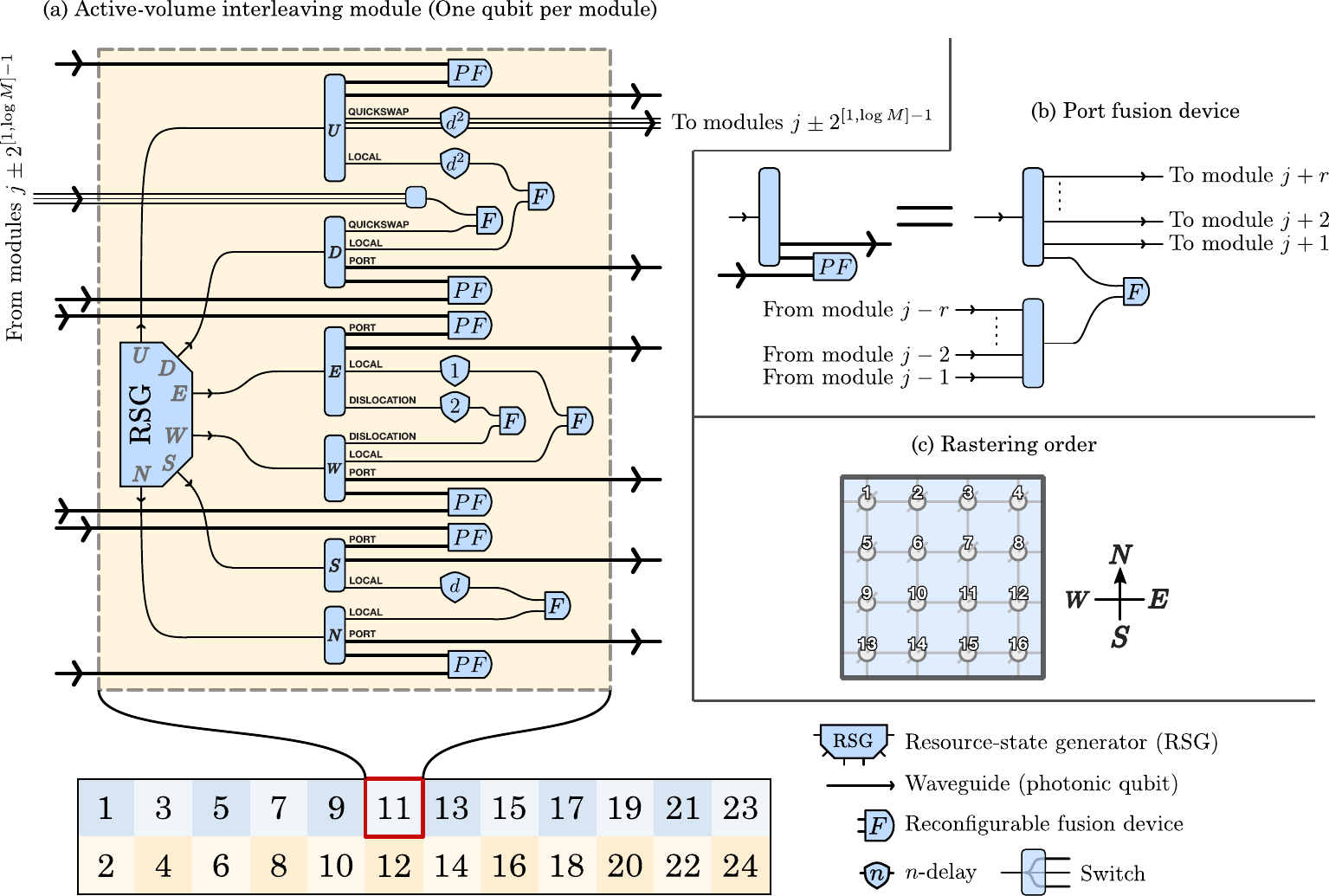}
\caption{(a) A photonic fusion-based active-volume quantum computer with $M$ qubit modules can be constructed by slightly modifying the interleaving modules presented in Fig. 7 of Ref.~\cite{Bombin2021}, as shown for the example of $M=24$. Each fusion device should be interpreted as a reconfigurable fusion device. Here, each interleaving module corresponds to one qubit module. (b) Port fusion devices connect modules to other modules within range. That is, module $j$ can send photons to modules $j +1\dots r$ and receive photons from modules $j - 1 \dots r$. (c) Each module generates logical qubits using the same rastering order, as shown for the example of $d=4$. Port fusions do not require any delays, as they fuse photons produced in the same cycle.} 
\label{fig:module1}
\end{figure*}

The naive approach requires a very large number of hardware components~--~millions of RSGs~--~to construct a useful fault-tolerant quantum computer. However, it is possible to exploit the existence of low-loss photonic delays in a photonic architecture. Commercially available optical fiber features loss rates below 0.2 dB/km (or 4.5\%/km) for wavelengths around 1550 nm. In other words, a photon entering a 1-km-long fiber has a $>$95\% probability of exiting the fiber 5 microseconds later. As shown in Ref.~\cite{Bombin2021}, such loss rates can be sufficient for fault-tolerant quantum computations with surface codes. If a photon enters this 1-km-long fiber every nanosecond, the fiber acts as a temporary memory for up to 5000 photons, albeit a very specific kind of memory: Photons can only be stored for exactly 5 microseconds and can only be accessed in the order in which they enter the fiber.  In an interleaving architecture, the aforementioned components are combined to form interleaving modules consisting of one RSG with its associated fusion devices and a few fiber delays (or other types of low-loss photonic delays). Essentially, each 2D layer of the 3D arrangement of resource states is not generated in a single step, but generated in multiple cycles with the help of long delay components~--~a technique which is referred to as \textit{interleaving}. For the specific example of 6-ring resource states consisting of 6 photonic qubits, a $d \times d$ surface-code patch can be generated by a single RSG with the help of a 1-delay, $d$-delay and a $d^2$-delay. We refer to a delay with $n$ time bins of size $\tau_{\rm RSG}$ as an $n$-delay, where $\tau_{\rm RSG}$ is the length of the cycle on which RSGs produce resource states (e.g., $\tau_{\rm RSG} = 1~\mathrm{ns}$ in the example above). 

\textbf{Active-volume interleaving modules.} Consider the interleaving module shown in Fig.~\ref{fig:module1}a. The module contains an RSG which produces a 6-ring resource state, where the 6 qubits of the resource state are labeled U (up), D (down), E (east), W (west), S (south), and N (north), which are the same labels that we use for the directions of surface-code spacetime diagrams. Each photon enters a switch which can be used to select between various output options labeled \textit{local}, \textit{dislocation}, \textit{port} and \textit{quickswap}. Notice that, if the port and quickswap options are removed, this module is identical to the module in Fig.~7 of Ref.~\cite{Bombin2021} without network options. The \textit{local} switch options are used to generate a $d\times d$ surface-code patch, as the resource states are generated in the order shown in Fig.~\ref{fig:module1}c for the example of $d=4$. In other words, E-photons must be delayed by 1 RSG cycle to be fused with W-photons, S-photons must be delayed by $d$ RSG cycles to be fused with N-photons, and U-photons must be delayed by $d^2$ RSG cycles to be fused with D-photons for interlayer fusions. The dislocation option is required to implement surface-code dislocations (e.g., for Hadamarded logical blocks).

Consider a quantum computer consisting of $M$ of the modules shown in Fig.~\ref{fig:module1}, where we label the modules 1 to $M$. Using the local and dislocation options, each module is capable of generating one logical distance-$d$ surface-code qubit. The \textit{port} options connect each module $j$ to modules $j \pm [1,r]$, where $r$ is the range of the active-volume quantum computer, as shown in Fig.~\ref{fig:module1}b. Each photonic qubit can be sent to modules $j + [1,r]$ via $r$ switch options, or fused with with an incoming photonic qubit from modules $j - [1,r]$ via $r$ additional switch options. Note that these port fusions do not involve any delays. These fusions are used to implement port connections between logical blocks. For example, the W-photons of resource states along the west boundary of surface-code patch $j$ can fuse with the W-photons of resource states of the same RSG cycle along the west boundary of a different patch $k$, which connects these patches in a lattice-surgery operation corresponding to a logical block connection via the E-port. Since the switch can be used to implement port fusions for any $|j-k| \leq r$, this implements the desired finite-range port connections in the N, E, S and W directions. Port fusions of U-photons correspond to transversal Bell-basis measurements, as fusions are physical Bell-basis measurements. Similarly, port fusions of D-photons correspond to the transversal preparation of Bell states.

Finally, quickswaps can be implemented via the additional switch options for U- and D-photons. These allow U-photons to be sent to modules $j \pm 2^{[1, \log M]-1}$ via a $d^2$-delay and $\log M$ switch options, and D-photons to fuse with incoming photons from modules $j \pm 2^{[1, \log M]-1}$. These fusions are identical to local up-down fusions, except that they involve photons produced in different modules. Therefore, these fusions can be used to swap the logical qubits of different modules.

\textbf{Variant with multiple qubits per module.} In the modules of Fig.~\ref{fig:module1}, the size of the longest delay, the $d^2$-delay, is determined by the code distance of the logical qubits (which is assumed to be fixed). However, in interleaving, we are typically interested in using delays that are as long as reasonably possible to generate as many logical qubits as possible with each RSG as long as photonic loss rates remain sufficiently low. If the maximum delay length that can be tolerated is $\lambda$ time bins of size $\tau_{\rm RSG}$, such that the longest delay stores photons for a time $T = \lambda \cdot \tau_{\rm RSG}$, then this maximum delay length may be $\lambda > d^2$. There are two ways in which we can take advantage of such long delays. The first option is to slow down the RSGs of the modules in Fig.~\ref{fig:module1} by redefining $\tau_{\rm RSG} \rightarrow \tilde{\tau}_{\rm RSG} = T/d^2$. The RSG is expected to be the costliest component of a photonic interleaving module. If RSGs with half the clock rate are half as expensive as RSGs operating at the full clock rate, then $\tilde{\tau}_{\rm RSG} / \tau_{\rm RSG} = \lambda/d^2$ modules with slow RSGs can be constructed with the same physical footprint as a single module with a fast RSG. In this construction, each interleaving module corresponds to a qubit module of the active-volume quantum computer, and each full-speed RSG contributes $\lambda/d^2$ qubit modules.

Alternatively, we can use the modules shown in Fig.~\ref{fig:module2}, where each interleaving module corresponds to $n$ qubit modules. They use $n$-delays, $dn$-delays and $\lambda$-delays instead of 1/$d$/$d^2$-delays, where $\lambda = nd^2$. These modules generate $n$ logical qubits at the same time, each offset by one time bin as shown in the rastering order of Fig.~\ref{fig:module2}b, such that all delay lengths are rescaled by a factor of $n$. Furthermore, port fusions and quickswaps may now involve logical qubits from the same module, so switchable delay lengths are introduced in Fig.~\ref{fig:memstate2}c, and port fusion devices are modified as shown in Fig.~\ref{fig:memstate2}d/e. In essence, these modules still use $\mathcal{O}(r + \log N)$ switch options for $N$ logical qubits and range $r$, and can generate $n = \lambda/d^2$ logical qubits for each full-speed RSG, as can be achieved by slowing down the RSGs of Fig.~\ref{fig:memstate1} as described in the previous paragraph.

\textbf{Performance metrics of a fusion-based active-volume quantum computer.} Assuming RSGs with a clock cycle of $\tau_{\rm RSG}$ and a maximum delay length of $\lambda = nd^2$ time bins of size $\tau_{\rm RSG}$, each RSG adds $n$ qubit modules to the quantum computer, i.e., $n/2$ memory modules and $n/2$ workspace modules. Each workspace module executes a logical block in $d \cdot \lambda \cdot \tau_{\rm RSG} = nd^3 \cdot \tau_{\rm RSG}$. Since each module can execute $n/2$ logical blocks in that time, each module increases the speed of the quantum computer by $\tau_{\rm RSG}/(2d^3)$ blocks per unit time. For the example of $d=32$, $\tau_{\rm RSG} = 1~\mathrm{ns}$ and a 1.6-km fiber delay with $\lambda = 8192$, each RSG increases the memory by 4 qubits and the speed by 15,000 blocks per second. Other examples of device parameters are shown in Fig.~\ref{fig:factsheet}g.

Note that the speed provided by each RSG is independent of $\lambda$, whereas the memory provided by each RSG scales linearly with $\lambda$. Therefore, the use of long delay lines is strictly advantageous and does not constitute a linear space-time trade-off in the overall performance of the quantum computer. However, because it takes $\lambda \cdot \tau_{\rm RSG}$ to perform a layer of reactive single-qubit or two-qubit measurements, the reaction time will scale with $\lambda$ in this architecture, although the reaction time may still be dominated by the time required for classical processing and feed-forward. Should delay lengths be long enough that the reaction time is a concern, an alternative architecture may be considered in which stale magic states are rerouted into interleaving modules with a shorter delay length in order to enable faster reactive measurements.

\section{Conclusion}
\label{sec:conclusion}

We have introduced an architecture for surface-code-based fault-tolerant quantum computing that takes advantage of non-local connections to reduce the cost of fault-tolerant quantum computations. When compiling quantum computations into surface-code operations, the spacetime volume cost (i.e., the number of logical qubits multiplied by the number of logical cycles required to finish the computation) consists of two contributions: active volume that is required to execute logical operations and idle volume that only contributes trivial operations to the computation without advancing the computation, but may be present due to inefficiencies in the architecture. In previous general-purpose architectures with 2D-local connections (i.e., baseline architectures), a vast majority of the spacetime volume cost of computations may be due to idle volume. Our newly proposed architecture uses non-local connections to eliminate almost all idle volume, such that only approximately half of the spacetime volume is idle volume.

\textbf{Active-volume quantum computers have simple performance metrics.} In the active-volume architecture, the performance of a quantum computer is characterized by the size of the memory, its speed in blocks per second, the per-block error rate, and the reaction time. The cost of a quantum computation is quantified by its memory requirement and its active volume, which is counted in units of blocks. The memory requirement determines whether a quantum computation can be executed on a specific device, while the runtime of a quantum computation is determined by its active volume in blocks and the speed of the device in blocks per second.

\textbf{The cost of a quantum computation depends on its subroutines, not just the Toffoli count.} We have computed the active volumes of various logical operations, as summarized in Tab.~\ref{tab:subroutines}. Such a table can be used to compute the active volume of large quantum computations by summing the active volumes of the constituent subroutines. We observe that some subroutines are cheaper than others, even if they use the same number of non-Clifford gates. When performing resource estimates for baseline (non-active-volume) architectures, an oft-used cost metric is the number of $T$ gates and Toffoli gates. However, the per-Toffoli active volume of a QROM read, for example, can be orders of magnitude higher than the per-Toffoli active volume of an adder. Therefore, when optimizing quantum computations for an active-volume architecture, it can be advantageous to rely on cheap operations such as quantum arithmetic, even if this does not minimize the Toffoli count. We emphasize that the active-volume architecture reduces the cost of both data loaders and arithmetic operations compared to a baseline architecture, but the cost reduction is significantly more pronounced in the case of arithmetic operations.

\textbf{Photonic qubits are a natural fit for active-volume quantum computing.} We showed how an active-volume architecture can be implemented using photonic fusion-based interleaving modules. However, it is also be possible to construct active-volume architectures with other types of photonic qubits, such as architecture based on GKP encoding~\cite{Gottesman2001,Bourassa2021}, or even with matter-based qubits. For example, interleaving-type schemes for superconducting qubits based on resonant cavities have been proposed in Ref.~\cite{Duckering2020}, and possible non-2D-local connectivity between surface-code patches has been discussed in for networks of ion traps~\cite{Webber2021} as well as superconducting qubits~\cite{Bravyi2022}. Still, the ease with which non-local connections can be implemented in a fusion-based architecture, and the synergy between active-volume architectures and long interleaving delays, are important advantages of photonic qubits.

\textbf{Outlook and future work.} While the tools provided in this paper can be used to determine the active volume of a large-scale quantum computation, detailed resource estimates for relevant algorithms are left for future work. Previous work on the optimization of fault-tolerant algorithms has mostly been focused on the minimization of the number of non-Clifford gates. The active-volume architecture opens up new subroutine-aware directions for the optimization of fault-tolerant algorithms beyond Toffoli counts. Furthermore, rigorous simulations of magic state distillation protocols need to be performed to validate the error rates of distilled magic states and provide a more accurate estimate of the cost of distillation. Moreover, the methods in this paper can be used to optimize other relevant subroutines for an active-volume architecture, such as arithmetic circuits beyond ripple-carry addition. One may also study the use of additional magic states from synthillation~\cite{Campbell2017} and the incorporation of classically-encoded PPMs~\cite{Chamberland2022} to reduce the cost of logical operations. It is also worthwhile investigating to what degree the advantages of an active-volume architecture can be exploited with strictly 2D-local connectivity, as, e.g., the AutoCCZ construction for ripple-carry addition~\cite{Gidney2018a} can outperform a baseline architecture while using only local connections. Another direction is the study of transversal gates beyond transversal Bell measurements, e.g., transversal CNOT gates. These can potentially reduce the volume of logical operations even further, but may negatively impact the error threshold, so rigorous numerical simulations are required. We hope that this paper can inspire future work into the construction of architectures for efficient fault-tolerant quantum computers that take advantage of non-local connectivity beyond nearest neighbors in two dimensions.

\section*{Author contributions}

Naomi Nickerson proposed the distinction between active and inactive logical volume in fault tolerant circuits and designed first methods of using switching and photonic delays to reduce the cost of the inactive volume. Daniel Litinski proposed the architecture presented in this manuscript and its photonic implementation, introduced the ZX-based description for networks of logical blocks, analyzed the impact on algorithmic subroutines and wrote the manuscript.

\section*{Acknowledgments}

The authors would like to thank Sam Pallister and William Pol for insightful discussions on the active volume of useful quantum algorithms, Fernando Pastawski for discussions on ZX diagrams, and Gabriel Mendoza for discussions on switch designs. We would also like to thank Jacob Bulmer, Axel Dahlberg, Daniel Dries, Matteo Lostaglio, Sam Roberts, Terry Rudolph, Vincent Russo and David Tuckett for detailed feedback on the draft, and our colleagues at PsiQuantum for useful discussions.

\appendix

\section{Example resource estimate}
\label{sec:resourceestimate}

This section explains in more detail how the numbers in Fig.~\ref{fig:examplealgorithm} are obtained. The algorithm to factor 2048-bit numbers presented in Ref.~\cite{Gidney2021} corresponds to a sequence of 500,000 lookup additions on around 6200 qubits. Each lookup addition consists of a 2048-qubit adder and a QROM that loads 1024 2048-bit numbers. Using Gidney adders, each adder consists of 8192 $T$ gates and has an active volume of around $1.2\times 10^5$ blocks. The QROM consists of 4096 $T$ gates and has an active volume around $1.62\times 10^6$ blocks. Therefore, the total circuit volume is $6200 \cdot 6.1 \times 10^9 = 3.8 \times 10^{13}$ and the active volume is $8.7 \times 10^{11}$ blocks.

\textbf{Baseline architecture.} The total spacetime volume of the algorithm in a baseline architecture is proportional to twice the circuit volume, i.e., $7.6 \times 10^{13}$ blocks of size $d \times d \times d$. With a logical error rate per $d \times d \times d$ space-time block of $p(d) = 10^{-d/2}$, a code distance of $d=28$ yields acceptable success probabilities. In a circuit-based quantum computer, each distance-$d$ surface-code patch consists of $2d^2$ physical qubits. In a baseline architecture, there are $2 \times 6200$ surface-code patches, so a total footprint of 19 million physical qubits. Each $T$ gate is executed in $d$ code cycles, so the entire algorithm is executed in $1.7 \times 10^{11}$ code cycles. With a code cycle time of 1 $\mathrm{\mu s}$, as could be the case for superconducting qubits, the entire duration of the computation is 48 hours. With a code cycle time of 1 ms, as could be the case for superconducting qubits, the entire duration of the computation is 5.4 years.

In a fusion-based photonic baseline architecture~\cite{Bombin2021}, the quantum computer is a network of interleaving modules, each containing a resource-state generator (RSG) and photonic delays with a maximum length of $\lambda$ time bins. With optical fiber delays, the physical length of the delay is $\lambda/5$ meters. With free-space delays, the physical length is $0.3\lambda$ meters. The vast majority of the physical device footprint is found in the RSGs, so the number of RSGs is a reasonable metric for the physical cost of the device. Each interleaving module increases the number of logical qubits by $\lambda/d^2$. With $\lambda=1000$, or 200 m fiber delays, around 9700 interleaving modules are required to encode $2 \cdot 6200$ logical qubits. Each interleaving module generates one resource state in every nanosecond. A total of $2 \cdot 3.8 \cdot 10^{13} \cdot 28^3$ resource states need to be generated to finish the computation, so it finishes in 48 hours. With 10 times longer delays, such as 2 km fiber delays with $\lambda = 10^4$, 10 times fewer interleaving modules are required, but the computation takes 10 times longer, so 20 days with  970 interleaving modules. Even longer delays may require lower-loss delays such as free-space delays. Using 30 km free-space delays with $\lambda = 10^5$, 97 interleaving modules finish the computation in 200 days. With 300 km free-space delays with $\lambda = 10^6$, 10 interleaving modules finish the computation in 5.4 years.

\textbf{Active-volume architecture.} Since the total spacetime volume is lower in an active-volume architecture, the code distance can be reduced from 28 to $d=26$. With 19 million physical qubits, there are around 7000 workspace modules executing 7000 logical blocks every $d$ code cycles. With a code cycle time of 1 $\mathrm{\mu s}$ and a total of $8.7 \times 10^{11}$ logical blocks, the computation finishes in 54 minutes. Note that the reaction depth of the computation may be as high as $2.6 \times 10^9$, depending on the amount of parallelization done using the methods discussed in Ref.~\cite{Gidney2021}. A 54-minute computation may then require a reaction time of around 1 $\mu$s, which is lower than the typical assumption of around 10 $\mu$s. This is not a concern with a 1 ms code cycle, where the computation finishes in 37 days.

In a fusion-based photonic active-volume architecture using modules as shown in Fig.~\ref{fig:module1}, delays with $\lambda=10^3$ can be used to construct a quantum computer with modules that each correspond to a distance-$d$ logical qubit and consume $d^2 / \lambda \approx 0.68$ resource states in every RSG cycle, i.e., every nanosecond. 9700 RSGs can therefore be used to feed around 14,000 modules, half of which are workspace modules. Each workspace module executes a logical block every $d\cdot \lambda$ RSG cycles, i.e., every 26 $\mu$s. Therefore, the computation finishes in 54 minutes, provided that the reaction time is short enough. Using 2 km fiber delays with $\lambda = 10^4$, the resource-state consumption rate of each interleaving module is 10 times lower, i.e., 0.068 resource states per RSG cycle. 14,000 interleaving modules can then be supplied by only 970 RSGs. Each workspace module executes one logical block every 260 $\mu$s for a total computational time of 8.9 hours. With $\lambda = 10^5$ and 97 RSGs, the computational time increases by another factor of 10 to 3.7 days. With $\lambda = 10^6$, 10 RSGs can supply 15,000 interleaving modules. Each of the 7,500 workspace modules executes one logical block every 26 ms. The computation finishes in 35 days. All these numbers can also be obtained by dividing the total number of resource states (which is active volume multiplied by $d^3$) by the rate of resource-state generation.

\textbf{A comment on Gidney adders and Cuccaro adders.} The above estimate assumes that ripple-carry addition is performed using the circuits proposed by Gidney in Ref.~\cite{Gidney2018}. However, their active-volume implementation in Fig.~\ref{fig:addercircuit} requires a number of stale CZ states to be stored in memory until an entire addition is executed, before these states can be removed via reactive measurements. Depending on the size of the memory of the quantum computer and the degree to which oblivious carry runways~\cite{Gidney2019a} are used to split large adders into independent pieces, the number of stale CZ states may be significant enough to temporarily slow down the quantum computation, as workspace modules may need to be borrowed as temporary memory. This can be avoided by using Cuccaro adders~\cite{Cuccaro2004} instead of Gidney adders. These adders use twice as many Toffoli gates and roughly twice as much active volume, so increase the cost of the algorithm, but avoid the storage problem. However, since the cost of the computation is dominated by the cost of QROMs, doubling the cost of adders to $2.4 \times 10^5$ blocks only increases the total volume by around 6\% to $9.2 \times 10^{11}$ blocks. As a side remark, the fact that QROMs dominate the cost also implies that a higher distillation cost that results in, e.g., CCZ states which are twice as costly will lead to a total volume increase by roughly the same amount.

\section{Active volume table}
\label{sec:table}

The active volumes of all the operations discussed in Secs.~\ref{sec:zmeasurements}-\ref{sec:compilation} are summarized in Tab.~\ref{tab:subroutines}.

\section{Additional figures}

\begin{figure*}[t!]
\centering
\includegraphics[width=0.8\linewidth]{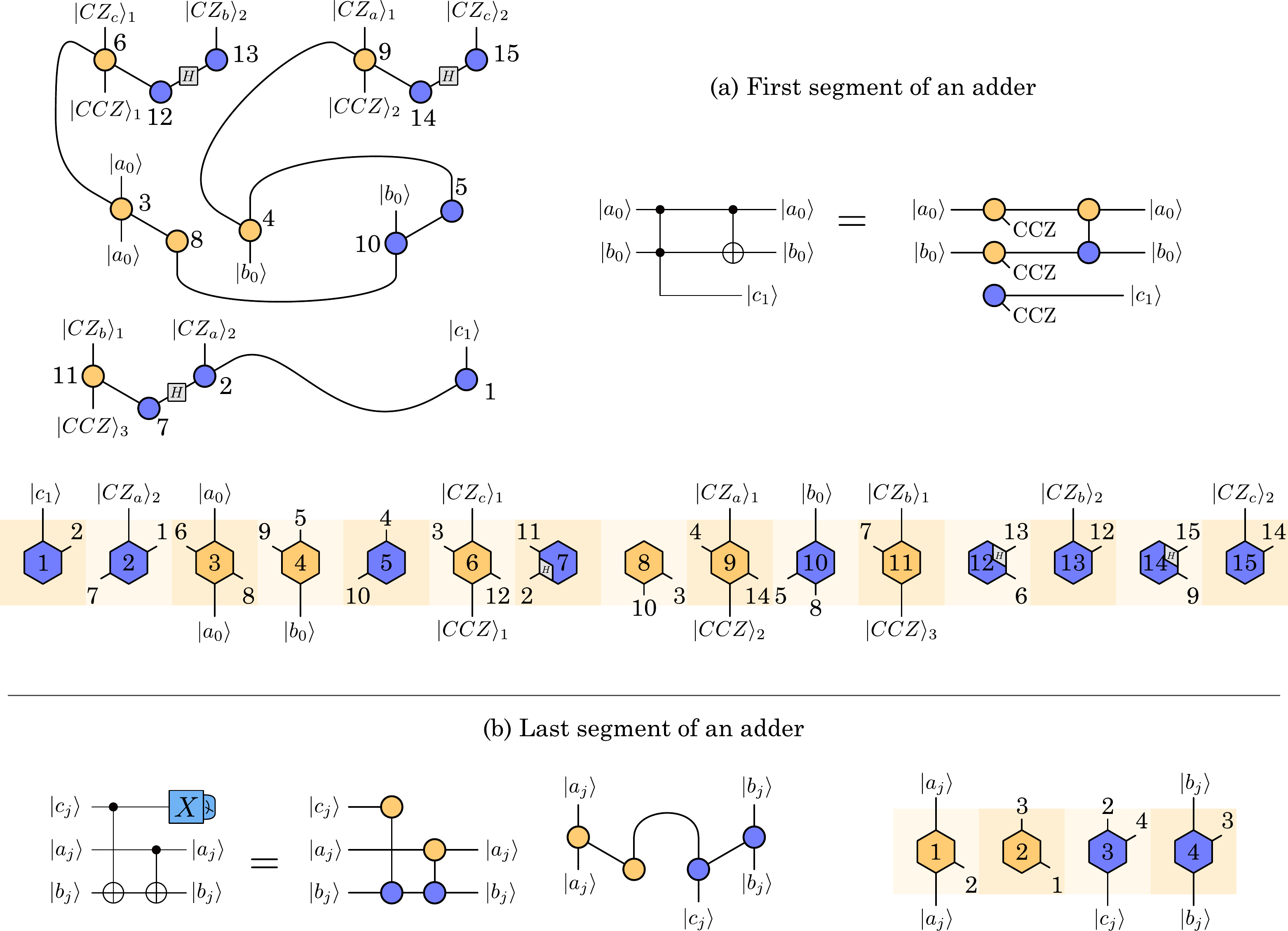}
\caption{The first (a) and last (b) segment of the adder in Fig.~\ref{fig:addercircuit} have a lower active volume of 15 and 4 blocks, respectively.}
\label{fig:adderfirstlast}
\end{figure*}

\begin{figure*}[t]
\centering
\includegraphics[width=\linewidth]{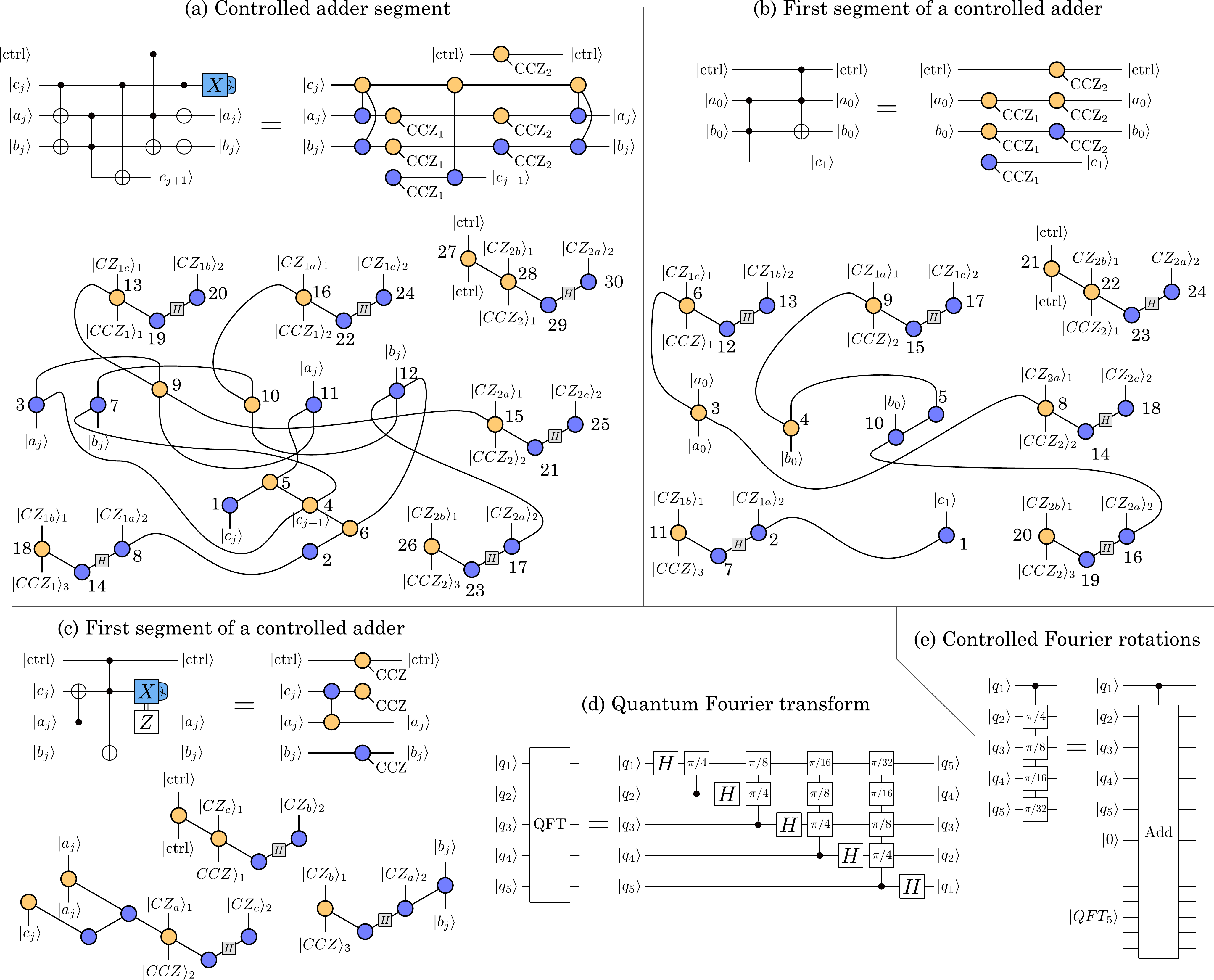}
\caption{(a-c) Controlled adders have a very similar construction, but a higher active volume compared to uncontrolled adders. (d) They can be used to implement a quantum Fourier transform consisting of Hadamard gates and controlled $\pi/2^n$ rotations. (e) These controlled rotations can be implemented using controlled additions into a phase-gradient register.}
\label{fig:ctrladder}
\end{figure*}

\begin{figure}[t]
\centering
\includegraphics[width=\linewidth]{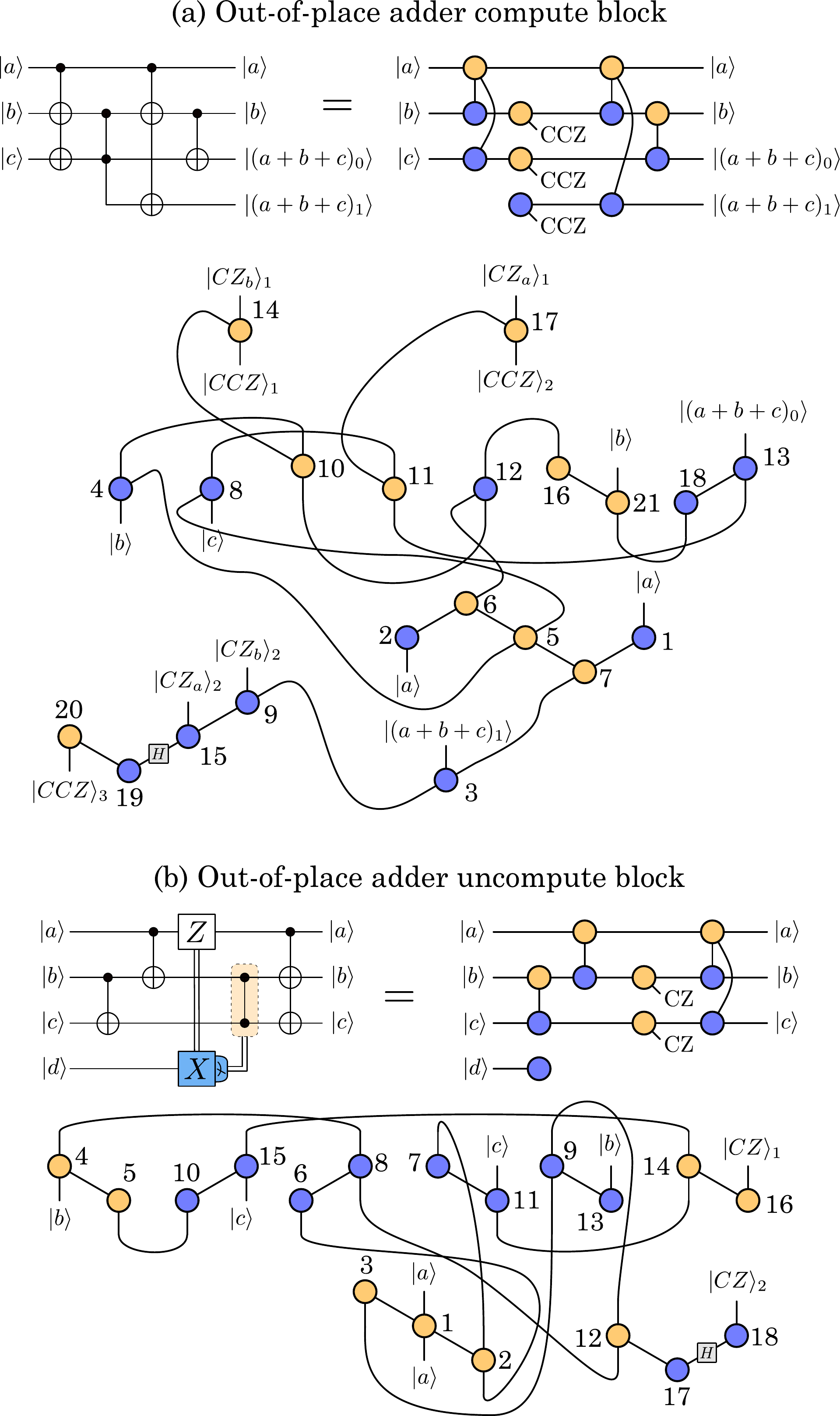}
\caption{The compute block of an out-of-place adder~\cite{Gidney2018} has an active volume of 21 blocks in addition to the volume of a CCZ state. The uncompute block has an active volume of 18 blocks.}
\label{fig:outofplaceadder}
\end{figure}

\begin{figure*}[t]
\centering
\includegraphics[width=0.8\linewidth]{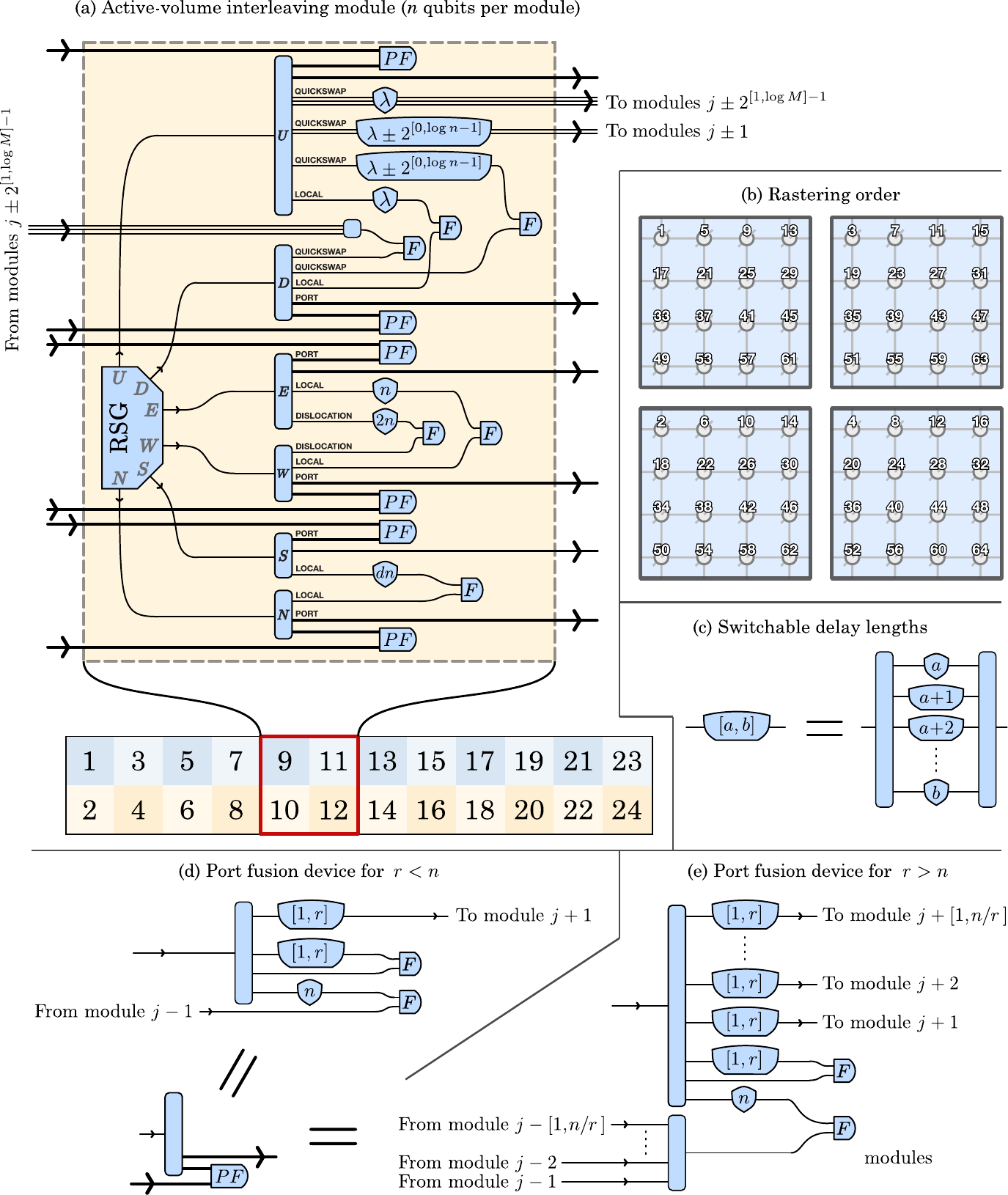}
\caption{Modified version of the modules in Fig.~\ref{fig:module1}, where each module generates $n$ logical qubits, i.e., each interleaving modules corresponds to $n$ qubit modules of the active-volume quantum computer.}
\label{fig:module2}
\end{figure*}

The additional figures show the first and last segment of a Gidney adder (Fig.~\ref{fig:adderfirstlast}), the block networks of a controlled adder (Fig.~\ref{fig:ctrladder}) and an out-of-place adder (Fig.~\ref{fig:outofplaceadder}), and a modified version of photonic active-volume interleaving modules (Fig.~\ref{fig:module2}).

\definecolor{col2}{HTML}{d2e6ea}
\def\arraystretch{1.2}
\begin{table*}[t!]
\scalebox{0.75}{
\begin{tabular}{p{0.55\linewidth}ccc}

\multicolumn{1}{c}{Subroutine} & Active volume (in blocks) & Reaction depth & Figure \\ 
\hline
\multicolumn{4}{c}{Elementary gates} \\
\hline
Hadamard & 3 & 0 & Fig.~\ref{fig:logicalblockintro}g  \\
CNOT & 4 & 0 & Fig.~\ref{fig:logicalblockintro}f \\
Two-qubit $Z \otimes Z$ (or $X \otimes X$) measurement & 2 & 0 & Fig.~\ref{fig:weighttwozmeas} \\
Conditional (reactive) CZ & 5 & 1 & Fig.~\ref{fig:toffoli}b \\
Toffoli & 12 + $C_{|CCZ\rangle}$ & 1 & Fig.~\ref{fig:toffoli}c \\
Controlled SWAP & 20 + $C_{|CCZ\rangle}$ & 1 & Fig.~\ref{fig:select}b \\
$Z_{c\cdot \pi/8}$ rotation with odd $c$ & 3.5 + $C_{|T\rangle}$& 1 &  \\
$Z_{c\cdot \pi/16}$ rotation with odd $c$ & $17.25 + \frac{1}{2}C_{|CCZ\rangle} + C_{|T\rangle}$ & 1.5 & Fig.~\ref{fig:adderpprs}c/d \\
Weight-$w$ $Z$-type (or $X$-type) PPM & $\lceil 1.5 w \rceil$ & 0 & Fig.~\ref{fig:zmeasurements}c \\
Weight-$(w_x,w_z)$ PPM & $\lceil 1.5 w_x \rceil + \lceil 1.5 w_z \rceil + 1$ & 0 & Fig.~\ref{fig:ppm} \\
Weight-$(w_x,w_z)$ $\pi/8$-angle PPR & $C_m + 1.5 + C_{|T\rangle}$ & 0 & Fig.~\ref{fig:ppm} \\
Weight-$(w_x,w_z)$ $b$-bit-precision PPR (variant 1~\cite{Ross2014}) & $C_m + 3b\cdot (4+ C_{|T\rangle}) + 1$ & $3b$ & Fig.~\ref{fig:pprs}e \\
Weight-$(w_x,w_z)$ $b$-bit-precision PPR (variant 2~\cite{Gidney2018a}) & $C_m + (b-1)(22.5+ C_{|CCZ\rangle}) - 3.5$ & $2b-3$ & Fig.~\ref{fig:adderpprs}a \\
Weight-$(w_x,w_z)$ $b$-bit-precision PPR (variant 3~\cite{Kliuchnikov2022}) & $C_m + \frac{1}{40}b \cdot (305 + 6 C_{|CCZ\rangle} + 24 C_{|T\rangle})$ & $0.75b$ &  \\
$n$ commuting equiangular PPRs with average weight $(w_x,w_z)$& $\approx (C_m + 39 + C_{|CCZ\rangle})\cdot n + \mathcal{O}(\log n \cdot C_{\rm rot})$ & $2n + \delta_r$ &  \\
Arbitrary $n$-qubit Clifford gate & $\approx 3n^2 $ & 0 & Fig.~\ref{fig:customsubroutine}b \\
Arbitrary $n$-qubit operation with $n_r$ rotations& $\approx 3n^2 + n_r\cdot(1.5n + C_{\rm rot})$ & 0 & Fig.~\ref{fig:customsubroutine}b \\
\hline
\multicolumn{4}{c}{Arithmetic and data loading} \\
\hline
$n$-qubit Gidney adder & $(n-1)\cdot (22+C_{|CCZ\rangle}) - 3$ & $2n-3$ & Fig.~\ref{fig:addercircuit}/\ref{fig:adderfirstlast} \\
Controlled $n$-qubit Gidney adder &  $(n-1)\cdot(30 + 2 C_{|CCZ\rangle}) + 9 + C_{|CCZ\rangle}$ & $4n-3$ & Fig.~\ref{fig:ctrladder} \\
Out-of-place adder (compute block) & $21 + C_{|CCZ\rangle}$ & 1 & Fig.~\ref{fig:outofplaceadder}a \\
Out-of-place adder (uncompute block) & $18$ & 1 & Fig.~\ref{fig:outofplaceadder}b \\
$n$-qubit quantum Fourier transform & $(n^2-1)\cdot (15 + C_{|CCZ\rangle}) - 3n + 1$ & $2n^2-n-1$ &Fig.~\ref{fig:ctrladder}d/e \\
SELECT of $n$ Pauli operators with average weight $(w_x,w_z)$ & $(n-1)\cdot(13 + C_m + C_{|CCZ\rangle}) $ & $n-1$ & Fig.~\ref{fig:select}a \\
QROM read of $n$ $b$-bit numbers, $\lambda$ numbers at a time & $(\frac{n}{\lambda} - 1) \cdot (15 + \frac{3}{4}b\lambda + C_{|CCZ\rangle}) $ & $n/\lambda + \log \lambda$  & \\
& $+~b\cdot(\lambda -1 ) \cdot (20 + C_{|CCZ\rangle} )$ & & \\
\hline
\multicolumn{4}{c}{Magic state distillation and management of resource states and catalysts} \\
\hline
Cloning a $Y$ state & 3 & 0 & Fig.~\ref{fig:pprs}c \\
Cloning two $\sqrt{T}$ states & 25.5 + $C_{|CCZ\rangle}$ + $C_{|T\rangle}$ & 0 & Fig.~\ref{fig:adderpprs}b \\
CCZ-to-2T conversion & 16.5 & 1 & Fig.~\ref{fig:8toccz}b \\
$\text{(8-to-CCZ)}_{d,d,d/2}$ distillation protocol & $25/2$ & 1 & Fig.~\ref{fig:8toccz}a \\
$\text{(15-to-1)}_{d,d/2,d/2}$ distillation protocol & $35/2$ & 1 & Fig.~\ref{fig:15to1}a \\
$\text{(15-to-1)}_{d/2,d/4,d/4} \times \text{(8-to-CCZ)}_{d,d,d/2}$ distillation protocol & $30$ & 2 & \\
Estimated cost of a $|T\rangle$ state & $C_{|T\rangle} \approx 25$ & & \\
Estimated cost of a $|CCZ\rangle$ state & $C_{|CCZ\rangle} \approx 35$ & & \\

\end{tabular}}
\caption{Active volumes of all operations described in this paper. In a weight-$(w_x,w_z)$ Pauli product operator, each $Z$ increases $w_z$ by 1, each $X$ increases $w_x$ by 1, each $Y$ increases both $w_x$ and $w_z$ by 1, and, if the total number of $Y$ operators is odd, $w_x$ and $w_z$ are again increased by 1. Here, $C_m = \lceil \frac{3}{2} w_x \rceil + \lceil \frac{3}{2} (w_z+1) \rceil + 1$, and $C_{\rm rot}$ is the cost to perform a single-qubit $Z$ rotation.}
\label{tab:subroutines}
\end{table*}

\cleardoublepage

\end{document}